\documentclass{elsarticle}
\usepackage[authormarkup=none,final]{changes}
\definechangesauthor[color=red]{1}
\definechangesauthor[color=red]{2} 
\definechangesauthor[color=red]{3} 

\usepackage{amsmath,amssymb}
\usepackage{eucal}  
\usepackage{mathrsfs}
\usepackage{placeins} 
\usepackage{czt}  
\usepackage{tikz}  %
\usetikzlibrary{tikzmark}
\usepackage{tzplot}
\DeclareFontFamily{U}{mathc}{}
\DeclareFontShape{U}{mathc}{m}{it}%
{<->s*[1.03] mathc10}{}
\DeclareMathAlphabet{\mathscr}{U}{mathc}{m}{it}
\usepackage{changes}
\usepackage{cancel}  
\usepackage{graphicx}
\usepackage{dcolumn}
\usepackage{bm}
\usepackage{color}
\usepackage{enumitem}
\usepackage{empheq} 
\usepackage{url}
\usepackage{gensymb}  
\newcommand{\ccdot}{%
	\hspace{-4pt}\cdot
	\hspace{-4pt}
}  
\newcommand{\conc}{%
	{\protect \rblot \ccdot \lblot}
}
\let\isout\sout 
\renewcommand{\sout}[1]{\ifmmode\text{\isout{\ensuremath{#1}}}\else\isout{#1}\fi}

\newcommand{\N}{\mathbb{N}}

\newtheorem{proposition}{Proposition}{\bf}{\it}
{\bf}{\it}
{\bf}{\it}
{\bf}{\it}
{\bf}{\it}

{\bf}{\it}
\newtheorem{lemma}{Lemma}{\bf}{\bf}

\begin{document}
	\begin{frontmatter}
		\title{Universal behaviors of the multi-time correlation functions   of random processes with renewal: the step noise case (the random velocity of a Lévy walk)}
		\author{Marco Bianucci*\corref{mycorrespondingauthor}}
		\address{Istituto di Scienze Marine, Consiglio Nazionale delle Ricerche (ISMAR - CNR),\\
			Forte Santa Teresa, Pozzuolo di Lerici, 19032 Lerici (SP), Italy}
		\ead{marco.bianucci@cnr.it}
		\author{Mauro Bologna}
		\address{Departamento de Ingenier\'ia El\'ectrica-Electr\'onica, Universidad de Tarapac\'a, Arica, Chile}
		\author{Daniele Lagomarsino-Oneto}
		\address{Istituto di Scienze Marine, Consiglio Nazionale delle Ricerche (ISMAR - CNR),
			Forte Santa Teresa, Pozzuolo di Lerici, 19032 Lerici (SP), Italy}
		\author{Riccardo Mannella}
		\address{Dipartimento di Fisica, Universit\`a di Pisa, 56100 Pisa, Italy}
		\date{\today}
		
		\begin{abstract} 
			Stochastic processes with renewal properties, also known as
			semi-Markovian processes, are powerful tools for modeling systems where memory effects and
			long-time correlations play a significant role. In this work, we study a broad class of
			renewal processes where a variable's value changes according to a prescribed Probability
			Density Function (PDF), $p(\xi)$, after random waiting times $\theta$. This model is
			relevant across many fields, including classical chaos, nonlinear hydrodynamics, quantum
			dots, cold atom dynamics, biological motion, foraging, and finance.
			
			We derive a general analytical expression for the $n$-time correlation function by
			averaging over process realizations. Our analysis identifies the conditions for
			stationarity, aging, and long-range correlations based on the waiting time and jump
			distributions. Among the many consequences of our analysis, two new key results emerge.
			First, for Poissonian waiting times, the correlation function quickly approaches that of
			telegraphic noise. Second, for power-law waiting times with $\mu>2$, , \emph{any $n$-time correlation
				function asymptotically reduces to the two-time correlation evaluated at the earliest and
				latest time points}.
			
			This second result reveals a universal long-time behavior where the system's full
			statistical structure becomes effectively two-time reducible. Furthermore, if the jump PDF
			$p(\xi)$ has fat tails, this convergence becomes 
			independent of the waiting time PDF and 
			is  significantly accelerated, requiring only
			modest increases in either the number of realizations or the trajectory lengths. Building
			upon earlier work that established the universality of the two-point correlation function
			(i.e., a unique formal expression depending solely on the variance of $\xi$ 
			and on the waiting-time PDF),
			the present study extends that universality to the full statistical description of a broad class
			of renewal-type stochastic processes. 
		\end{abstract}
		
		
	\end{frontmatter}
	
	\section{\label{sec:intro}Introduction} 
	A complete statistical characterization of a stochastic process requires knowledge of its
	$n$-time (or $n$-point) joint correlation functions.  
	In a previous paper~\cite{bblmCSF196} we investigated the two-time joint correlation function of
	stochastic processes with renewal $\xi[t]$\footnote{We use square brackets for the time
		argument of a stochastic process,  and parentheses for specific time-dependent realizations, e.g., 
		$\xi(u)$, with $t_0 \leq u \leq t$. }. 
	
	These processes are fundamental models in many scientific disciplines. As this article
	builds upon the findings of~\cite{bblmCSF196}, we refer the reader to that work for a
	comprehensive introduction to the subject's history and relevant literature.
	However, for clarity and self-containment, we summarize the essential background material
	in the initial sections of this manuscript.
	
	There are two different definitions of the stochastic process $\xi[t]$ arising from a random
	variable with renewal. 
	The first definition describes a shot-noise process (also called an intermittent process). 
	Here, a series of shots, or impulses, occur at random times $t_i$. 
	The intensity of the $i$-th shot is a random variable 
	$\xi_i$ drawn from a probability density function (PDF) $p(\xi)$, while the time interval between consecutive shots, 
	$\theta=t_i-t_{i-1}$, is a random variable with PDF 
	$\psi(\theta)$.
	
\begin{figure}[h]
		\centering
		\includegraphics[width=\textwidth]{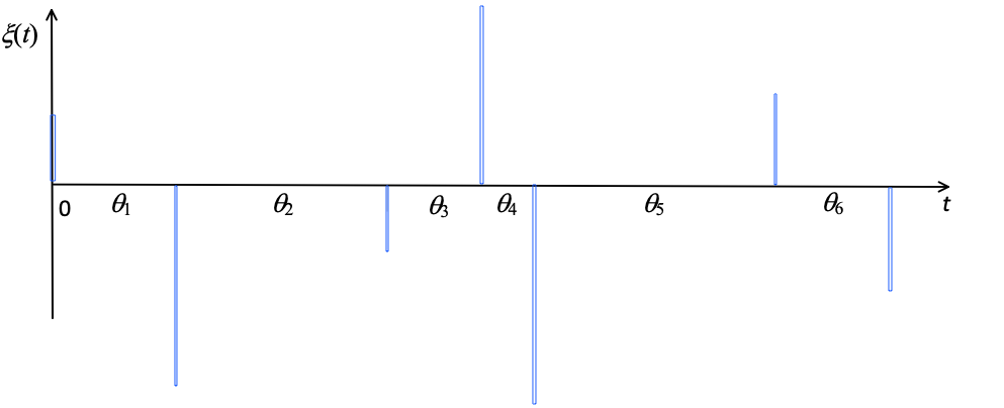}
		\caption{
			Schematic representation of a trajectory realization $\xi(t)$ for the noise of L\'evy
			flight-CTRW process (see text for details). In actual cases, the pulse heights are
			infinite, as the trajectory consists of a sum of shifted Dirac delta functions. Here, for
			visualization purposes, the pulses are depicted as very thin boxes of equal width, with
			heights determined by the random values of $\xi$.}
		\label{fig:trajectory_leapers}
	\end{figure}

	A schematic representation of a trajectory realization $\xi(t)$
	for this case is illustrated in Fig.~\ref{fig:trajectory_leapers}
	and can be formally written as a weighted sum of shifted Dirac delta functions of time, where the weight
	is the random value of $\xi$ and the time shift is the random number $\theta$ (a sort of non-stationary white noise).
	\begin{figure}[h]
		\centering
		\includegraphics[width=\textwidth]{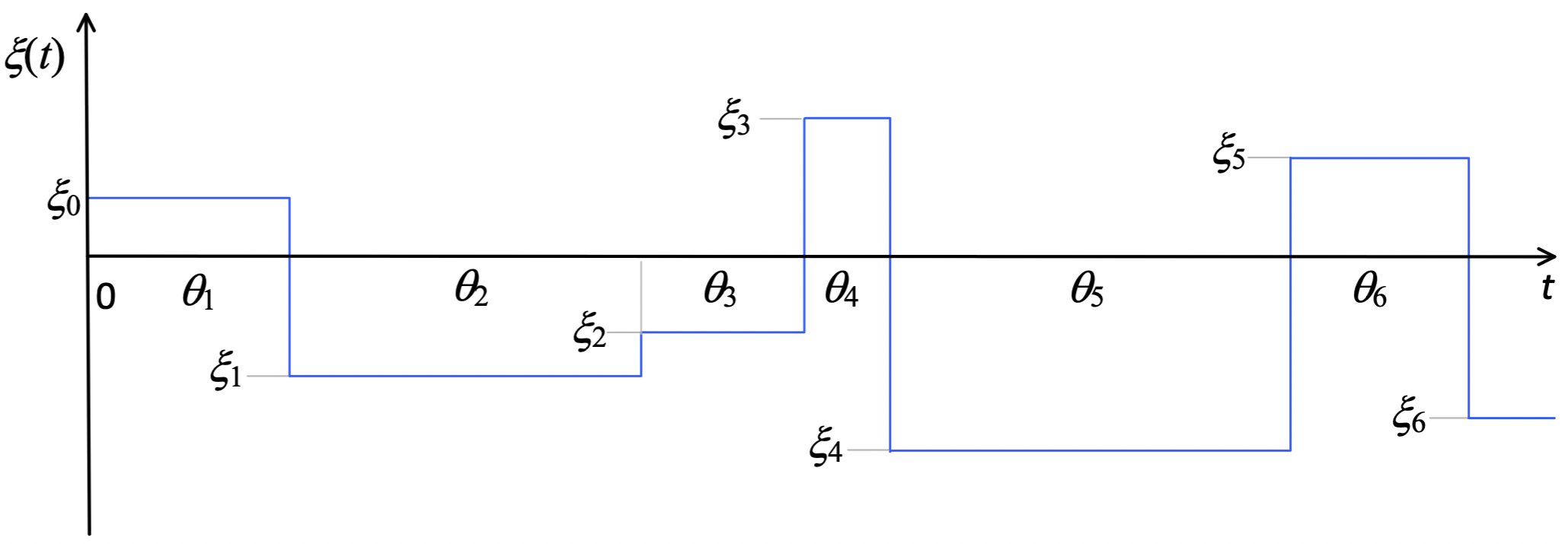}
		\caption{  Schematic representation of a trajectory realization $\xi(t)$
			for the noise of L\'evy walk random velocity
			(LWRV, see text for details).}
		\label{fig:trajectory_creepers}
	\end{figure}

	The second definition describes a step-like process. In this case, the process 
	$\xi[t]$ maintains a constant value, drawn from the PDF $p(\xi)$, for a random duration 
	$\theta$, after which it jumps to a new, independently drawn value. 
	A schematic of a trajectory is shown in Fig.~\ref{fig:trajectory_creepers}.
	
	In many important contexts, renewal processes constitute the primary phenomenon 
	of interest. For example, spike (or jump) noise, particularly with Poissonian statistics, has 
	been extensively studied and applied across various disciplines in a series of 
	works~\cite{cvsFNL05, cdvsPRE84, kcavsPRE92, vcdsJSTAT2015, acvsJSTAT2016, 
		acvksIJMPB30, cvCSF153}.
	
	However, as a process in its own right, the renewal process with a step-like 
	structure is often the central object of study due to its broad range of 
	applications. The well-known continuous-time random walk (CTRW) can itself be 
	considered a member of this family, as highlighted in our previous 
	work~\cite{bblmCSF196}. In fact, such processes occur in various physical 
	systems. Notable examples include blinking quantum dots~\cite{mbJCP121}; breath 
	figures, where patterns form from the growth and coalescence of water droplets 
	on a surface~\cite{bkPRL57, bdgEL27, dgyPRA44, m-mbbgyPA214}; spin systems 
	quenched from high to zero temperature (or, more generally, into the 
	low-temperature phase)~\cite{dbgJPA27, dhpPRL75, dhpJSP85}; and diffusion fields 
	evolving from random initial conditions~\cite{mbcsPRL77, dhzPRL77}, among many 
	others.
	
	Moreover, these processes are also relevant when $\xi[t]$ is considered as the 
	noise source acting on a Brownian particle, as in the following stochastic 
	differential equation (SDE):
	\begin{equation} 
		\dot{x} = -C(x) + I(x)\xi[t],
		\label{SDE} 
	\end{equation}
	where $-C(x)$ denotes a deterministic drift (velocity field), and $I(x)$ accounts 
	for a possible state-dependent noise intensity.
	
	Depending on how $\xi[t]$ is defined, whether as shot noise or step noise, this 
	SDE can describe different stochastic dynamics. Specifically, it can represent a 
	generalized L\'evy flight CTRW (when $-C(x) \ne 0$ and $I(x)$ is not constant) or 
	a L\'evy walk with random velocity (LWRV), respectively.
	
	In the shot-noise case, between two consecutive transitions of the random 
	variable $\xi[t]$, the Brownian particle is simply advected by the unperturbed 
	velocity field $-C(x)$, while it undergoes an instantaneous jump of size and sign 
	equal to $I(x)\xi(t)$ at each transition.
	
	In the step-noise case, i.e., the generalized LWRV setting, the stochastic 
	process $I(x)\xi[t]$ represents the state-dependent instantaneous velocity of 
	the variable $x$.

	It is worth emphasizing that the LWRV model integrates two fundamental properties: the
	ability to generate anomalously fast diffusion and a finite propagation speed for the
	random walker. Recent investigations in fields such as optics, Hamiltonian chaos, cold
	atom dynamics, biophysics, and behavioral science have demonstrated that the simple
	LWRV model provides significant insights into complex transport phenomena. For a
	comprehensive and self-contained introduction to L\'evy walks, including their
	theoretical foundations, a wide range of applications, recent advances, and future
	prospects, we refer the reader to the excellent review by Zaburdaev, Denisov, and
	Klafter~\cite{zdkRMP87}.
	
	As a key result of~\cite{bblmCSF196}, we established the universal nature of the
	two-time joint correlation function for this class of processes. More precisely,
	we derived two analytical expressions that apply to any stochastic process with
	renewal, whether of the spike or step type. These expressions depend solely on the
	waiting time (WT) probability density function (PDF) $\psi(t)$ and on the variance
	of the random variable $\xi$, but not on other specific features of its PDF,
	denoted by $p(\xi)$~\footnote{We denote by $\overline{f(\xi)}$ the average of any
		function $f(\xi)$ with respect to the PDF $p(\xi)$.}. This means that the result 
	remains the same for Gaussian, dichotomous, flat, and other cases alike.
	These results are exact and
	consistent with prior findings for dichotomous, multi-state, and subordinated
	Langevin
	processes~\cite{glJSP104,aagprwPRE71,abwgPRE83,nbkMMNP11,bsJSTAT2007,
		bmbProceedingNAS107}.

	The two-time joint correlation function is crucial for deriving the power spectrum of a 
	stochastic process, either in stationary or time-dependent regimes~\cite{kuboLRT, lbPRL115}. 
	When the process models noise acting on a Brownian particle, this function directly determines 
	the system’s time-dependent variance. In particular, the generalized Green-Kubo relation links 
	the diffusion properties to the scaling of the noise autocorrelation~\cite{dlkbPRX4}.
	Moreover, by using a perturbative approach, it is well known that the two-time correlation function of the noise $\xi[t]$ 
	for a dissipative system of interest $x[t]$, crucially determines its
	statistics, appearing in the state-dependent diffusion coefficient of the corresponding
	Fokker-Planck equation~\cite{bmJPC4,bbmJSP191}. When $\xi[t]$ has long memory ($\psi(t)\sim t^{-\mu}$,
	$1<\mu<2$), the asymptotic behavior of $x[t]$'s correlation mirrors that of $\xi[t]$~\cite{bsJSTAT2007}.
	Thus, the universality of $\xi[t]$'s two-time correlation function extends to generalized systems of interest,
	in particular under dissipation.

	However, in general, we may be interested in computing all the multi-time correlation 
	functions of the process $\xi[t]$. In particular, when we focus on the variable 
	$x$ in the SDE~\eqref{SDE}, with a nonzero drift term $-C(x)$ and/or a 
	non-constant function $I(x)$, the master equation (ME) for the reduced 
	probability density function (PDF) of $x$ depends on all the multi-time 
	correlation functions of the noise $\xi[t]$. Specifically, it involves the 
	corresponding multi-time (also called ``multi-point'') joint cumulants. For 
	further discussion, see~\cite{bmJPC4, bbmJSP191, bbJSTAT4, bCSF148, bCSF159}.
	Consequently, knowledge of the multi-time/multi-point joint correlation functions of $\xi[t]$ 
	(hereafter referred to simply as ``multi-time correlation functions'') is crucial for accurately 
	describing the statistical dynamics of the system~\eqref{SDE}.

	Beyond our previous work~\cite{bblmCSF196}, the existing literature provides 
	exact or asymptotic results, typically valid for large times, for the two-time 
	correlation function of stochastic renewal processes in specific systems~
	\cite{glJSP104, aagprwPRE71, abwgPRE83}, as well as general statistical 
	treatments of multi-state systems~\cite{nbkMMNP11} and occupation time 
	statistics~\cite{glJSP104}.
	
	Motivated by these considerations, and with the broader goal of deepening our 
	understanding of stochastic processes with renewal, we extend the analysis 
	carried out in~\cite{bblmCSF196} to the $n$-time correlation function of 
	$\xi[t]$, regardless of whether it acts as a noise source in a system governed 
	by Eq.~\eqref{SDE}, or whether it constitutes the primary variable of interest.
	
	Given the breadth of scenarios considered, and in order to maintain a concise and
	accessible presentation, we divide the analysis into two separate works: the
	present paper, where we are focused on the step-type case (i.e., random velocities
	in L\'evy walks), and a companion one addressing the case in which the stochastic
	process is of the spike type (i.e., the standard stochastic forcing in CTRW
	models).\\ In particular, we start from the definition of the $n$-time correlation
	function $\langle \xi(t_1) \xi(t_2) \dots \xi(t_n) \rangle_{t_0}$ as the ensemble
	average over trajectories of the product $\xi(t_1) \xi(t_2) \dots \xi(t_n)$, where
	$\xi(t)$ denotes a single realization of the process $\xi[t]$, with $n\in \N$ and
	where $t_0$ represents the initial time of the process. Then, we derive a general
	procedure for computing this correlation function in terms of the waiting time
	(WT) probability density function (PDF) $\psi(\theta)$ and the jump distribution
	$p(\xi)$.\\ We emphasize the term ``general'' to highlight that we obtain a
	unified analytical expression involving $\psi(\theta)$ and $p(\xi)$, which
	formally applies to any class of distributions, such as Gaussian, dichotomous, or
	power-law PDFs. To our knowledge, these findings are novel. The main element of
	novelty is to offer a complete and universal characterization of the statistics of
	$\xi[t]$: we impose no constraints on time scales or on the form of the PDF
	$p(\xi)$. Moreover, our results are presented directly in the time domain, rather
	than only in the Laplace domain, as is often the case in existing studies.
	
	This result provides a framework for a deeper understanding of the full
	statistical structure of a stochastic process $\xi[t]$. In the Poissonian case,
	for instance, Lemma~\ref{lem:1} and Proposition~\ref{prop:1} provide significantly
	more information than previously available in the literature. Specifically, all
	symmetric renewal stochastic processes become asymptotically indistinguishable
	from telegraph noise.\\ Another noteworthy application involves the stationary
	limit. When considering Eq.~\eqref{SDE}, aside from the trivial case of free
	diffusion, the statistical behavior of $x$ depends on the full set of multi-time
	correlation functions of $\xi[t]$. To determine whether $x$ attains an
	asymptotically stationary state, it is therefore necessary to understand the
	stationarity properties of all these correlation functions.
	Proposition~\ref{stationarity} sheds light on this aspect. 
	However, we argue that an even more significant result of the present paper concerns the case where 
	the WT PDF $\psi(\theta)$ exhibits a power-law decay with exponent $\mu > 2$, and the time 
	lags $t_i - t_{i-1}$ (for $i = 2, \dots, n$) are large compared to the average waiting time 
	$\tau$, with $t_1 - t_0 \gg \tau$. In this regime, we obtain a simple and universal behavior 
	of the $n$-time correlation function (see Proposition~\ref{prop:main}):
	\begin{equation}
		\label{gen_temp}
		\langle \xi(t_1) \xi(t_2) \dots \xi(t_n) \rangle_{t_0} \sim 
		\frac{\overline{\xi^n}}{\overline{\xi^2}} 
		\langle \xi(t_1) \xi(t_n) \rangle_{t_0}.
	\end{equation}
	It is evident that Eq.~\eqref{gen_temp} depends only on the first and last times, while 
	\emph{intermediate times do not appear}.
	
	It is worth emphasizing that the general expressions we  derive for the $n$-time 
	correlation function in this work involve the first $n$ moments of $\xi$. 
	One might therefore argue that these results hold only when the probability 
	density function (PDF) $p(\xi)$ decays sufficiently rapidly to ensure the 
	existence of these moments. However, in practical situations, the ensemble 
	over which averages are computed is finite. As a result, even for PDFs with 
	very heavy tails where moments like $\overline{\xi^n}$ may diverge 
	theoretically, the empirical average remains finite.\\
	Indeed, as confirmed by numerical simulations, the cases where $p(\xi)$ 
	exhibits heavy tails are precisely those in which the $n$-time correlation 
	function most closely approximates the universal two-time correlation function 
	evaluated at the extreme times. More specifically, in such cases,
	\emph{even if the WT PDF decays very slowly, i.e. with a power law with $\mu<2$,} increasing 
	the number of realizations in the ensemble leads to rapid convergence of the 
	normalized $n$-time correlation function towards the two-time counterpart, 
	\textit{regardless of the time lags}.\\
	This observation supports a key point of our approach: we do not impose any 
	restriction on the tails of either the waiting time PDF $\psi(\theta)$ or the 
	amplitude distribution $p(\xi)$. The generality of our framework accommodates 
	even those cases where classical assumptions about finite moments are 
	violated.
	
	Given the result of the previously mentioned paper, 
	about the universality of the two-time correlation functions, 
	this fact yields a far more universal result concerning the statistical behavior of 
	stochastic processes with renewal.
	
	We further perform numerical simulations of several relevant cases to verify our findings.
	
	For simplicity, throughout this work we assume that the waiting time $\theta$ 
	and the jump amplitude $\xi$ are independent random variables. Under this 
	assumption, their joint PDF factorizes as
	$\psi(\theta, \xi) = \psi(\theta) \, p(\xi)$.
	Nevertheless, we think that our method is sufficiently general and 
	straightforward to be extended to the more general case in which $\theta$ and 
	$\xi$ are statistically dependent and $\psi(\theta, \xi)$ cannot be written 
	as a product of marginals.	However, we have not yet addressed this situation.
	\section{Model and definitions}
	The graph of a trajectory realization $\xi(t)$ of the stochastic process $\xi[t]$ for the step-noise case
	is illustrated in 
	Fig.~\ref{fig:trajectory_creepers}. It consists of horizontal segments, which we refer to as laminar regions 
	(borrowing terminology from the dichotomous case, which is the noise of a standard L\'evy walk process). 
	The ordinates $\xi_1, \xi_2, \xi_3, \ldots, \xi_k, \ldots$ of each segment in 
	Fig.~\ref{fig:trajectory_creepers} are random numbers drawn from the
	PDF $p(\xi)$ and they last for random durations $\theta_1, \theta_2, \ldots, \theta_k, \ldots$ respectively, 
	according to the WT PDF $\psi(\theta)$. In formula, $\xi(t)$ is written as
	\begin{equation}
		\label{trajectory}
		\xi( t)=\sum_{q=0}^{\infty} \xi_q \; \Theta\left(t-t_0-\sum_{k=0}^q \theta_k\right) \Theta\left(\sum_{h=0}^{q+1} \theta_h-t+t_0\right),
	\end{equation}
	where $\Theta(t)$ is the Heaviside step function.
	We set $t_0$ as the initial time at which the stochastic process begins and we assume
	the time ordering defined by the notation $t_i \le t_j$ for $j > i$.
	The distance from the initial time $t_0$ to $t_1$ is important for measuring the aging of the
	process.
	The process of averaging over
	all the possible trajectories  $\xi(t)$ starting at the time $t_0$, is indicated by the angle brackets $\langle...\rangle_{t_0}$. 
	Thus, we can write the definition of $n$-times correlation function as:
	\begin{equation}
		\label{n_corr_def}
		\langle\xi( t_1)\xi( t_2)...\xi( t_n)\rangle_{t_0} =\int \xi( t_1)\xi( t_2)...\xi( t_n) P_{t_0}[\xi(t)]\delta\xi(t).
	\end{equation}
	where $ P_{t_0}[\xi(t)]\delta\xi(t)$ is the proper functional differential measure corresponding
	to a realization of the stochastic
	process $\xi[ t]$.
	Because all $\xi_k$ and  $\theta_k$ are independent random numbers, the PDF for the trajectory realization is 
	\begin{equation}
		\label{PDF_trajectory}
		P_{t_0}[\xi(t)]\delta\xi(t)=p_0(\xi_0)d\xi_0\prod_{q=1}^\infty\psi(\theta_q)d\theta_q\,p(\xi_q)d\xi_q
	\end{equation}
	where $\xi_0$ represents the value of $\xi$ at the initial time $t_0$, and 
	$p_0(\xi_0)$ is the PDF used to sample the initial value 
	of $\xi$ (i.e., at $t = t_0$). For example, if all trajectories start from the same initial 
	value $\xi'$ (e.g., $\xi' = 0$), then $p_0(\xi_0) = \delta(\xi_0 - \xi')$. 
	On the other hand, if the initial value $\xi_0$ is a random number with the same 
	PDF as the random variable $\xi$, then $p_0(\xi_0) = p(\xi_0)$.
	Of course, the influence of the initial PDF 
	$p_0(\xi_0)$ is particularly significant (i.e., persistent over time) when the aging 
	time of the process is long or infinite.
	
	The average of a function of the random number $\xi$ will be indicate with a bar over the same
	function, i.e.  
	$\int f(\xi)p(\xi) d\xi:= \overline{f(\xi)}$.
	Thus,  
	$\int \xi^np(\xi) d\xi:= \overline{\xi^n}$. With a slight abuse of notation, we also define  
	$\int \xi_0^{n} p_0(\xi_0) d\xi_0 := \overline{\xi_0^{n}}$.
	
	In addition to the overbar notation, for convenience we also introduce a bracket-style notation to denote the 
	average of a function of
	$\xi$ with respect to the PDF $p(\xi)$:
	\begin{equation}
		\label{mean}
		\lceil{f(\xi)}\rceil:=\overline{f(\xi)}.
	\end{equation}
	%
	
	Before presenting the main results, we first introduce some standard quantities commonly used 
	in the context of renewal processes.
	
	The probability density for an event to occur precisely at time $t$ is given by
	\begin{equation}
		\label{Rtilde}
		R(t-t_0) := \Theta(t-t_0) \sum_{n=1}^{\infty} \psi_n(t-t_0) \Rightarrow \hat{R}(s) = \frac{\hat{\psi}(s)}{1-\hat{\psi}(s)},
	\end{equation}
	where $\psi_n$ denotes the $n$-time convolution of $\psi(\theta)$
	and a hat over a function indicates its Laplace transform. 
	$R(t)$ is the rate function that appears in the master equation of the
	PDF of the CTRW. Another related quantity is
	\begin{equation}
		\label{R}
		\tilde{R}(t-t_0) := R(t-t_0) + \delta(t-t_0) \Rightarrow \hat{\tilde{R}}(s) = \frac{1}{1-\hat{\psi}(s)}.
	\end{equation}
	Note that by setting $\psi_0(\theta) := \delta(\theta)$,
	we can also write 
	$$\tilde{R}(t-t_0) = \Theta(t-t_0) \sum_{n=0}^{\infty} \psi_n(t-t_0).$$
	
	It is worth noting that in the case where the WT  PDF $\psi(\theta)$ is an exponential
	function, i.e., $\psi(\theta) =  \exp(-\theta/\tau)$, then $R(t-t_0) = 1/\tau$, i.e., is the
	usual rate of events. On the other hand, if the WT PDF decays with a  power law behaviour such
	as $\psi(\theta) \sim (T/\theta )^{-\mu}$, where $T$ is the time scaling 
	factor, then the rate function $R(t-t_0)$ depends on time. In this case, the  average WT, defined as

	\begin{equation}
		\label{tau}
		\tau := \int_0^\infty t\,\psi(t)\,dt,
	\end{equation}
	exists (i.e., is finite) for $\mu>2$, and for large times  
	$R(t) \sim 1 + (T/t)^{\mu-2}$. For
	$1 < \mu < 2$, there is not a finite average time and asymptotically  $R(t) \sim (T/t)^{2-\mu}$
	(e.g., \cite{bJSTAT2020}).
	
	Another
	standard quantity we will use is the survival probability $\Psi( t)$, i.e., the probability
	that after a time interval $ t$ from the last transition, the
	random variable $\xi$ has not changed value. Equivalently, $\Psi( t)$ is the probability
	that transitions occur only at times greater than or equal to $ t$ after the last transition.
	Thus, in terms of the WT  PDF, we have
	\begin{equation}
		\label{Psi}
		\Psi( t) := \int_{ t}^\infty \psi(u) \, du = 1 - \int_0^{ t} \psi(u) \, du 
		\Rightarrow \hat{\Psi}(s) = \frac{1 - \hat{\psi}(s)}{s}.
	\end{equation}
	Note that in the case of an exponential WT PDF, with deacay time $\tau$ (the Poissonian case), also the survival probability is
	exponential. On the other hand,
	for WT PDF with heavy tail $\sim (t/T)^{-\mu}$,  
	we have $\Psi( t) \sim  (t/T)^{-\mu+1}$.
	Finally, by substituting the expressions \eqref{PDF_trajectory} and \eqref{trajectory} in Eq.~\eqref{n_corr_def}, we obtain the analytic formula
	\begin{align}\label{n_corr_def_step_temp}
		&  \langle\xi( t_1)\xi( t_2)...\xi( t_n)\rangle_{t_0}\nonumber \\
		&=\int \left[\sum_{i_1=0}^{\infty} \xi_{i_1} \; \Theta\left(t_1-t_0-\sum_{k_1=0}^{i_1} \theta_{k_1}\right)  \Theta\left(\sum_{k_1=0}^{{i_1}+1} \theta_{k_1}-t_1+t_0\right)\right]\nonumber \\
		&\times\left[\sum_{i_2=0}^{\infty} \xi_{i_2} \; \Theta\left(t_2-t_0-\sum_{k_2=0}^{i_2} \theta_{k_2}\right) \Theta\left(\sum_{k_2=0}^{{i_2}+1}\theta_{k_2}-t_2+ t_0\right)\right]\times...\nonumber \\
		&...\times\left[\sum_{i_n=0}^{\infty} \xi_{i_n} \; \Theta\left(t_n-t_0-\sum_{k_n=0}^{i_n} \theta_{k_n}\right) \Theta\left(\sum_{k_n=0}^{{i_n}+1} \theta_{k_n}-t_n+t_0\right)\right]\nonumber \\
		&\times p_0(\xi_0)d\xi_0\prod_{q=1}^\infty\psi(\theta_q)d\theta_q\,p(\xi_q)d\xi_q.
	\end{align}
	
	Now, since the second Heaviside function after the $i$th jump resets the step trajectory $\xi(t)$ to zero
	(i.e.,  erasing the memory of previous jumps), it 
	follows that if $t_1\le t_2 \le t_3...\le t_n$, in the second, third and subsequent sums in Eq.~\eqref{n_corr_def_step_temp}, 
	the corresponding indices $i_2$, 
	$i_3$, ..., and so on, can start from the value of the preceding index, rather than from zero, namely,
	\begin{align}
		\label{n_corr_def_step}
		&  \langle\xi( t_1)\xi( t_2)...\xi( t_n)\rangle_{t_0}\nonumber \\
		&=\int \left[\sum_{i_1=0}^{\infty} \xi_{i_1} \; \Theta\left(t_1-t_0-\sum_{k_1=0}^{i_1} \theta_{k_1}\right)  \Theta\left(\sum_{k_1=0}^{{i_1}+1} \theta_{k_1}-t_1+t_0\right)\right]\nonumber \\
		&\times\left[\sum_{i_2=i_1}^{\infty} \xi_{i_2} \; \Theta\left(t_2-t_0-\sum_{k_2=0}^{i_2} \theta_{k_2}\right) \Theta\left(\sum_{k_2=0}^{{i_2}+1}\theta_{k_2}-t_2+ t_0\right)\right]\times...\nonumber \\
		&...\times\left[\sum_{i_n=i_{n-1}}^{\infty} \xi_{i_n} \; \Theta\left(t_n-t_0-\sum_{k_n=0}^{i_n} \theta_{k_n}\right) \Theta\left(\sum_{k_n=0}^{{i_n}+1} \theta_{k_n}-t_n+t_0\right)\right]\nonumber \\
		&\times p_0(\xi_0)d\xi_0\prod_{q=1}^\infty\psi(\theta_q)d\theta_q\,p(\xi_q)d\xi_q.
	\end{align}
	
	The expression on the right-hand side of Eq.~\eqref{n_corr_def_step} follows directly from the definition of the multi-time 
	correlation function of the renewal stochastic process $\xi[t]$.  
	It is not merely the outcome of an intuitive or ``common sense'' interpretation based on statistical reasoning, 
	which---especially in non-stationary cases like the present one---can be misleading.  
	Although Eq.~\eqref{n_corr_def_step} may appear complicated, it is actually straightforward to work with.
	
	\section{Intuitive explanation of the two and four time correlation functions cases\label{sec:corr_creepers_general}}
	The main focus of the previous work was how to deal  with  Eq.~\eqref{n_corr_def_step} in the $n=2$ case:
	the following \emph{universal expression} for the two-time correlation function
	was obtained (for simplicity, it was assumed that $\overline{\xi}=0$):
	\begin{align}
		\label{cor2t_}
		\langle\xi(t_1)\xi(t_2)\rangle_{t_0}
		& =  \overline{\xi_0^2} \,
		\Psi(t_2 - t_0) + 
		\overline{\xi^2} \int_{t_0}^{t_1} du_1 \, {R}(u_1 - t_0) \Psi(t_2 - u_1)\nonumber \\
		&
		=\left(\overline{\xi_0^2}- \overline{\xi^2}\right)
		\Psi(t_2 - t_0) +\overline{\xi^2} \int_{t_0}^{t_1} du_1 \, \tilde{R}(u_1 - t_0) \Psi(t_2 - u_1).
	\end{align}

	The  result~\eqref{cor2t_} is \textit{universal} in the sense that it depends on the variance of $\xi$ and
	on the WT PDF, but not on other specific details of the stochastic process.
	For example, it holds equally for Gaussian, dichotomous, or uniform PDFs (see  \cite{bblmCSF196} for details). 
	The statistical interpretation of expression \eqref{cor2t_} is straightforward: the correlation function is given by  
	$\overline{\xi_0^2}$ times the probability that $t_1$ and $t_2$ lie within the first laminar region (i.e., no  
	transitions  occur from $t_0$ to $t_2$),  
	plus $\overline{\xi^2}$ times the probability that $t_1$ and $t_2$ are, together,  in any subsequent laminar  
	region (i.e., after the last transition at $u_1$, no further transitions occur between  
	$u_1 \le t_1$ and $t_2$, summed over all possible $u_1 \le t_1$). 
	This result is expected: the correlation function  
	$\langle\xi( t_1)\xi( t_2)\rangle_{t_0}$ is zero otherwise, as the random values $\xi(t_1)$ and  
	$\xi(t_2)$ are independent when separated by a transition event, and $\overline{\xi}=0$.  
	
	If $\overline{\xi_0^2}=\overline{\xi^2}$, i.e., if the variance of the ensemble at initial time is the same variance of 
	the random variable $\xi$, then Eq.~\eqref{cor2t_} simplifies  as
	\begin{align}
		\label{bblmCSF196}
		\langle\xi(t_1)\xi(t_2)\rangle_{t_0}
		& =  
		\overline{\xi^2} \int_{t_0}^{t_1} du_1 \, \tilde{R}(u_1 - t_0) \Psi(t_2 - u_1).
	\end{align}
	From the definition given in Eq.~\eqref{n_corr_def_step}, the general $n$-point case 
	can be derived by using a formal but straightforward procedure, as detailed in 
	\ref{app:A_A}.

	However, prior to rigorously addressing the general case, it is useful to first consider the case $n=4$, while maintaining the additional simplifying assumption that the odd moments of $\xi$ vanish.
	The result is justified using the same reasonable statistical arguments used previously for the case $n=2$: 
	\begin{align}
		\label{corr4}
		&\added[id=1]{ \langle \xi(t_1)\xi(t_2)\xi(t_3)\xi(t_4)\rangle_{t_0}=
			\overline{\xi_0^4} \,
			\Psi(t_4 - t_0) + 
			\overline{\xi^4} \int_{t_0}^{t_1} du_1 \, {R}(u_1 - t_0) \Psi(t_4 - u_1)}
		\nonumber\\
		&
		\added[id=1]{+\left[\left(\overline{\xi_0^2} -\overline{\xi^2}\right) \,
			\int_{t_2}^{t_3}du_2\,\psi(u_2-t_0)
				+\overline{\xi^2}
				\int_{t_0}^{t_1}du_1\tilde{R}(u_1-t_0) \int_{t_2}^{t_3}du_2\,\psi(u_2-u_1)
				\right]}\nonumber\\
			&\added[id=1]{\times	\overline{\xi^2} \int^{t_3}_{u_2} du_3 \tilde{R}\left(u_3-u_2\right)\int_{t_4}^\infty du_4\,\psi(u_4-u_3).}
		\end{align}
		In the case  where $\overline{\xi_0^2} =\overline{\xi^2}$ and also using the  
		result \eqref{cor2t_} 
		for the 2-time  correlation function with initial time $t_0=u_2$,  Eq.~\eqref{corr4} can be written as
		\begin{align}
			\label{corr4_}
			& \langle \xi(t_1)\xi(t_2)\xi(t_3)\xi(t_4)\rangle_{t_0}=
			\frac{\overline{\xi^4}}{ \overline{\xi^2}}\langle \xi(t_1)\xi(t_4)\rangle_{t_0}
			\nonumber\\
			&
			+
			\overline{\xi^2}
			\int_{t_0}^{t_1}du_1\tilde{R}(u_1-t_0) \int_{t_2}^{t_3}du_2
			\,\psi(u_2-u_1) \langle \xi(t_3-u_2)\xi(t_4-u_2)\rangle.
		\end{align}

		The interpretation of the result in Eq.~\eqref{corr4}, or the equivalent in 
		Eq.~\eqref{corr4_}, for the four-time correlation function, in terms of the probability of 
		times lying in the same laminar region, 
		as for the case  $ n=2 $,  is as  follows.
		In the time range between two transitions of the random variable $ \xi $, we cannot have 
		an odd number (i.e., 1 or 3) of subsequent times, taken from the four times 
		$ (t_1,t_2,t_3,t_4) $ of the correlation function 
		$ \langle \xi(t_1)\xi(t_2)\xi(t_3)\xi(t_4)\rangle $. Thus, we have only two possibilities:  
		1) all four times lie in the same laminar region, or  
		2) $ t_1 $ and $ t_2 $ are in one laminar region, while $ t_3 $ and $ t_4 $ are in a later 
		laminar region. 
		
		Since the times are ordered, the first case is equivalent to the condition that the two extreme 
		times, $ t_1 $ and $ t_4 $, are in the same laminar region. This is precisely the same 
		condition as that previously explored for the two-time correlation function. This leads to the first term 
		on the right-hand side of Eqs.~\eqref{corr4} and \eqref{corr4_}. 
		
		The second case is more subtle: the probability of having $ t_1 $ and $ t_2 $ in one laminar 
		region, while $ t_3 $ and $ t_4 $ are in a later laminar region, is not simply  
		$ \langle \xi(t_1)\xi(t_2)\rangle\langle \xi(t_3)\xi(t_4)\rangle $. This is evident from the 
		definition of the survival probability 
		$ \Psi(t_2 - u_1):=\int_{t_2}^\infty \psi(u_2-u_1)du_2 $ in Eq.~\eqref{cor2t_}. 
		In fact, $ \langle \xi(t_1)\xi(t_2)\rangle $ is evaluated under the assumption that after  
		$ t_2 $, a transition of $ \xi $ can occur at \textit{any} time $ u_2>t_2 $. However, in the 
		present case, we also have the constraint that a transition occurs between $ t_2 $ and $ t_3 $, 
		that is, $ u_2<t_3 $. 
		
		This leads to the part enclosed in the square brackets 
		in Eqs.~\eqref{corr4} (corresponding to the leftmost part of the last line in \eqref{corr4_} in the case in which $\overline{\xi_0^2} =\overline{\xi^2}$):
		$$\left(\overline{\xi_0^2} -\overline{\xi^2}\right) \,
		\int_{t_2}^{t_3}du_2\,\psi(u_2-t_0)
		+\overline{\xi^2}
		\int_{t_0}^{t_1}du_1\tilde{R}(u_1-t_0) \int_{t_2}^{t_3}du_2\,\psi(u_2-u_1).
		$$
		However, the integral over $ u_2 $ cannot be evaluated unless we consider that once the 
		transition of $ \xi $ has occurred at time $ u_2 $, the same $ u_2 $ serves as the initial
		time for the two-time correlation function related to the remaining times $ t_3, t_4 $. 
		This is exactly captured by the remaining terms in Eqs.~\eqref{corr4} and \eqref{corr4_}, which 
		close the integral over $ u_2 $.
		
		%
		The reader should observe that, if the WT PDF decays as $(T/t)^{\mu}$ with $\mu>2$, where $T
		$ is the WT time scale,  that implies   
		$R(t) \sim 1 +a(\mu)(T/t)^{(\mu-2)}$~\cite{bJSTAT2020}, for  
		$t_1-t_0\gg T$, $t_2-t_1\gg T$, $t_3-t_2\gg T$ and $t_4-t_3\gg T$, the dominant term in Eqs.~\eqref{corr4} and 
		Eq.~\eqref{corr4_} is that in the first line, it does not depend on the intermediate times and decays as 
		\textit{the two-time  correlation function evaluated at the two extreme times} (the first and
		the last times, respectively). Thus, in the simplified case in which $\overline{\xi_0^2} =\overline{\xi^2}$ we have: 
		\begin{align}
			\label{cor4t_theta_power_slow}
			&\langle\xi( t_1)\xi( t_2)\xi( t_3)\xi( t_4)\rangle_{t_0}
			\sim  
			\frac{\overline{\xi^4}}{\overline{\xi^2}}\langle\xi( t_1)\xi( t_4)\rangle_{t_0}
			\nonumber\\
			&
			= 
			\overline{\xi^4} \int_{t_0}^{t_1} du_1 \, {R}(u_1 - t_0) \Psi(t_4 - u_1).
		\end{align}

		\section{The \added[id=2]{general case of }$n$-time correlation function\added[id=2]{s}\label{sec:ntimes_creepers}}
		%
		%
		In the field of renewal processes, the Laplace transform has been a widely used tool. 
		Indeed, a rigorous approach to obtain the $n$-time correlation function from the 
		general definition in Eq.~\eqref{n_corr_def_step} is to directly apply the Laplace 
		transform to Eq.~\eqref{n_corr_def_step}, and simplify the resulting expressions by 
		exploiting the properties of the Heaviside functions. 
		However, for large $n$, this method for evaluating the multi-time 
		correlation function of $\xi[t]$ quickly becomes cumbersome. Our alternative approach, which is detailed 
		in \ref{app:A_A}, although still rigorous, keeps the calculation more tractable and compact.
		This approach confirms and generalizes the 
		statistical interpretation of the correlation function in terms of the joint probability 
		that groups of  times fall within the same laminar region. It is essentially based on 
		the following key points:
		\begin{enumerate}[label=(\roman*)]
			\item The time-ordering assumption, i.e., $t_i \leq t_j$ for $i < j$.
			
			\item A variable transformation that converts waiting times (i.e., the time 
			intervals between successive events) into the absolute times of events. 
			Integration over these intervals leads to the $i_k$-fold convolution of the 
			WT PDF, and consequently, to the rate functions $R$.
			
			\item The two steps above, followed by integration over all remaining 
			waiting times and over the random number $\xi$ of all the events, yield a 
			formal expression equivalent to the summation over all possible compositions 
			of the ordered times $t_1 \le t_2 \le \dots \le t_n$ into groups lying within 
			the same laminar region, separated by any number of transition events.
		\end{enumerate}

		To rephrase the exact procedure outlined in~\ref{app:A_A} in a way more close to the intuitive
		statistical interpretation we have used
		for the $n=2,4$ cases, it is useful to introduce a specific
		notation  that, given a composition of the $n$ ordered times 
		$t_1\le t_2 \le ...\le t_n$,
		it represents the joint probability 
		that the times in each group of this composition lie in the same laminar region. To 
		introduce this notation by steps,
		we start with the probability that
		we have a single time $t_i$ in a
		laminar region, bounded by any  time $u_i\ge u_L$ and  any time $u_j\le t_R$. 
		We indicate this  probability by 
		enclosing these times between the ``bra'' symbol ``$\lblot$'' and the 
		``ket'' symbol ``$\rblot$'', which together we call ``closed angle 
		brackets''.
		Thus,   from the definition of \added[id=3]{the} rate function $R$ and \added[id=3]{the} WT PDF, we have
		\begin{align}
			\label{closed_corr_1}
			\lblot t_{i}\rblot:=
			\int^{t_{i}}_{u_L} du_{i} \tilde{R}\left(u_{i}-u_L\right)\int_{t_{i}}^{t_R} du_{i}^\prime\, \psi(u_{i}^\prime-u_{i}).
		\end{align}
		The probability that in the same laminar region we have only  two times $t_i$ and
		$t_j>t_i$ is %
		\begin{align}
			\label{closed_corr}
			\lblot t_{i}\,t_{j}\rblot:=
			\int^{t_{i}}_{u_L} du_{i} \tilde{R}\left(u_{i}-u_L\right)\int_{t_{j}}^{t_R} du_{i}^\prime\, \psi(u_{i}^\prime-u_{i}).
		\end{align}
		In fact,  by definition, 
		$\tilde{R}\left(u_{i}-u_L\right)$ is the probability for unit of time
		to have a transition event at  the time $u_i$ after the  ``initial'' time $u_L$
		and  
		$\int_{t_{j}}^{t_R} du_{i}^\prime\, \psi(u_{i}^\prime-u_{i})$ is the probability that
		the next transition event, after that in $u_i$, will happen after $t_j$, but
		before $t_R$. Integrating over all possible $u_i\in [u_L,t_i]$ gives
		the joint probability that $t_i$ and $t_j$ are in the same laminar region.
		
		For example, in the Poissonian case, i.e. when $\psi(t)=e^{-t/\tau}/\tau$,
		we have $\tilde R(t)=\delta(t)+1/\tau$ and from \eqref{closed_corr} we obtain:
		\begin{align}
			\label{closed_corr_poiss}
			\lblot t_{i}\,t_{j}\rblot&=
			\int^{t_{i}}_{u_L} du_{i} 
			\left(\delta\left(u_{i}-u_L\right)+\frac{1}{\tau}\right) \int_{t_{j}}^{t_R} du_{i}^\prime\, 
			e^{-(u_{i}^\prime-u_{i})/\tau}/\tau
			\nonumber \\
			&=
			e^{-\frac{t_j-t_i}{\tau }}-e^{-\frac{t_R-t_i}{\tau }}=
			e^{-\frac{t_j-t_i}{\tau }}(1-e^{-\frac{t_R-t_j}{\tau }}).
		\end{align}
		As expected in this classical case, we do not have any dependence on the 
		initial time $u_L$, while the joint probability depends on the maximum possible length of the laminar region, 
		controlled by the time $t_R$.
		
		For the probability that in the same laminar region there are only the
		$m$ ordered times $t_{i+1}\,t_{i+2}...\,t_{i+m}$, 
		the  projection property below clearly holds.
		\begin{align}
			\label{closed_corr_proj}
			\lblot  t_{i+1}\,t_{i+2}...\,t_{i+m}\rblot= 
			\lblot t_{i+1}\,t_{i+m}\rblot.
		\end{align}
		We call ``two-sides normalized
		closed correlation function'' the definitions in Eqs.~\eqref{closed_corr_1}-\eqref{closed_corr} and \eqref{closed_corr_proj}. 
		The reason for the adjective ``normalized'' is because comparing the definition
		in Eq.~\eqref{closed_corr} with Eqs.~\eqref{cor2t_}-\eqref{bblmCSF196}, we see that 
		$\lblot t_{i}\,t_{j}\rblot$  is 
		similar to a correlation function (we will go deeper in that hereafter) but is divided by the moment of the 
		random variable $\xi$. 
		The term ``closed'' correlation function is introduced because, as it is clear from Eq.~\eqref{closed_corr}, it is similar 
		to a correlation function, but where the times are constrained in a (closed)
		range, delimited by $u_L\ge 0$ on the left side and 
		$t_R\le \infty$ on the right side.

		The left opening of the closed correlation function, denoted using the standard left 
		angle bracket, such as $\langle t_i t_j\rblot$, is obtained, as in 
		Eq.~\eqref{closed_corr_prod}, by extending $u_L$, which represents the minimal 
		starting point of the laminar region, down to $t_0$, the initial time of the stochastic 
		process $\xi[t]$: 
		\begin{align}
			\label{left_closed_corr}
			&\langle t_i \,t_j\rblot:=
			\int^{t_i}_{t_0} du_i \tilde{R}\left(u_i-t_0\right) \int_{t_j}^{t_R} du_j\,\psi(u_j-u_i);
		\end{align}
		while the right opening, indicated by the standard right angle bracket, is obtained by setting $t_R=\infty$ 
		(i.e., extending up to infinity the 
		maximum possible value for the right side of the
		laminar region): 
		\begin{align}
			\label{right_closed_corr}
			&\lblot t_i \,t_j\rangle:=\int^{t_i}_{u_L} du_i \tilde{R}\left(u_i-u_L\right) \int_{t_j}^{\infty} du_j\,\psi(u_j-u_i).
		\end{align}
		Now we are ready to introduce the notation that associate to any composition
		of the $n$ ordered times $t_1,t_2,...,t_n$, the probability that
		the times in each group of this composition lie in the same laminar region. 
		This is done by separating the groups of
		the composition by using the centered dot symbol between  the ``ket'' and ``bra'' brackets 
		(i.e., $\conc$:). Thus, a composition made of $p$ groups is represented as
		\begin{align}
			\label{closed_corr_prod_pre}
			&	\langle  t_{1},t_{2},...,t_{i_1}
			\conc t_{i_1+1},t_{i_1+2},...,t_{i_2}\conc
			t_{i_2+1},t_{i_2+1},...,t_{i_3}\rblot
			\hspace{-4pt}\cdot...\cdot\hspace{-4pt}
			\lblot t_{i_{p-1}+1},
			t_{i_{p-1}+2},...,t_{i_p}\rangle 
			\nonumber\\
			=&	\langle  t_{1}\,t_{i_1}
			\conc t_{i_1+1}\,t_{i_2}\conc
			t_{i_2+1}\,t_{i_3}\rblot\hspace{-4pt}\cdot...\cdot\hspace{-4pt}\lblot t_{i_{p-1}+1} \,t_{i_p}\rangle 
		\end{align}
		with $i_p=n$. 
		In equation~\eqref{closed_corr_prod_pre}, the use of the same bracket symbols 
		introduced in~\eqref{closed_corr} is, of course, not accidental. It reflects 
		the same rationale as applied in that earlier case. Consequently, within each group, we retain 
		only the first and last times (which may coincide), as shown on the right-hand 
		side of~\eqref{closed_corr_prod_pre}.
		
		However, as previously noted in the discussion of the case $n=4$, the joint probability that the first group of times 
		$t_{1}, t_{2}, \ldots, t_{i_1}$ lies within the same laminar region \emph{and} 
		that the second group of times $t_{i_1+1}, t_{i_1+2}, \ldots, t_{i_2}$ lies 
		within a subsequent laminar region etc., is \emph{not} simply the product of the  
		corresponding normalized closed correlation functions defined 
		in~\eqref{closed_corr}.
		
		Indeed, as is easily verified by considering the definition of the rate function 
		$R(u)$, this joint probability is instead given by a kind of integral 
		convolution of these closed correlation functions---an operation we refer to as 
		\emph{concatenation}. This is the reason for placing a centered dot symbol 
		between the ``ket'' and ``bra'' brackets.
		More precisely, we have (recalling that $i_p = n$, i.e., $t_{i_p} = t_n$):
		%
		%
		\begin{align}
			\label{closed_corr_prod}
			\langle  t_{1}\,t_{i_1}&
			\conc t_{i_1+1}\,t_{i_2}\conc
			t_{i_2+1}\,t_{i_3}\rblot\hspace{-4pt}\cdot...\cdot\hspace{-4pt}\lblot t_{i_{p-1}+1} \,t_{i_p}\rangle\nonumber \\
			:=&
			\int^{t_{1}}_{{t_0}} du_{1} \tilde{R}\left(u_{1}-{t_0}\right) \int_{t_{i_1}}^{t_{i_{1}+1}} du_{1}^{\prime}\,\psi(u_1^{\prime}-u_{1})\nonumber
			\\
			&\times
			\int^{t_{i_{1}+1}}_{u_{1}^{\prime}} du_{2} \tilde{R}\left(u_{2}-u_{1}^{\prime}\right)\int_{t_{i_2}}^{t_{i_2+1}} du_{2}^{\prime}\, \psi(u_{2}^{\prime}-u_{2})\nonumber
			\\
			&\times
			\int^{t_{i_2+1}}_{u_{2}^{\prime}} du_{3} \tilde{R}\left(u_{3}-u_{2}^{\prime}\right) 
			\int_{t_{i_3}}^{t_{i_3+1}} du_{3}^{\prime}\,\psi(u_{3}^{\prime}-u_{3})...\nonumber
			\\
			&\times...\times
			\int^{t_{i_{p-1}}}_{u_{p-1}^{\prime}} du_{p} \tilde{R}\left(u_{p}-u_{p-1}^{\prime}\right)\int_{t_{n}}^{\infty} du_{p}^\prime\,
			\psi(u_{p}^\prime-u_p)
		\end{align}

		Finally, for later convenience, we give a bilinear property to the
		closed angle brackets:
		\begin{equation}
			\label{angle_bilinear}
			\xi^n\lblot t_i,t_{j+k}\rblot=\lblot\xi^n\, t_i,t_{j+k}\rblot=\lblot t_i,t_{j+k}\,\xi^n\rblot.
		\end{equation}
		With these definitions and exploiting also the alternative way, given
		in Eq.~\eqref{mean}, to indicate the average over the random 
		variable  $\xi$,   the 2-time and the 4-time correlation functions in Eqs.~\eqref{cor2t_}
		and \eqref{corr4}, corresponding to the simplified case in which $\overline{\xi_0^2} =\overline{\xi^2}$, can be written
		in a compact form as:
		\begin{align}
			\label{corr2t_2}
			\langle \xi(t_1)\xi(t_2)\rangle&=
			\overline{\xi^2}\langle  t_1 \,t_2\rangle
			=  \lceil \langle \xi^2 t_1 \,t_2\rangle \rceil
		\end{align}
		\begin{align}
			\label{corr4_2}
			&	\langle \xi(t_1)\xi(t_2)\xi(t_3)\xi(t_4)\rangle=
			\overline{\xi^4}\langle  t_1 \,t_4\rangle
			+\overline{\xi^2}^2\langle  t_{1} \,t_{2}\conc   t_{3\,} t_{4}\rangle
			\nonumber \\ &
			\added[id=1]{=	\lceil \langle \xi^4 t_1 \,t_4\rangle\rceil
				+\lceil \langle  \xi^2t_{1} \,t_{2}{\protect \rblot 
					\hspace{-2pt}\rceil \ccdot
					\lceil\hspace{-2pt}\lblot}  \xi^2 t_{3\,} t_{4}\rangle\rceil}
			\nonumber \\ &
			\added[id=1]{=	\lceil\langle\xi^4  t_1 \,t_2\,t_3\,t_4\rangle\rceil
				+\lceil \langle  \xi^2t_{1} \,t_{2}{\protect \rblot 
					\hspace{-2pt}\rceil \ccdot
					\lceil\hspace{-2pt}\lblot}  \xi^2 t_{3\,} t_{4}\rangle\rceil}
			\nonumber \\ &
			=\lceil\langle   {\xi}^{2} t_{1}\,t_{2}
			\big(1+{\protect \rblot 
				\hspace{-2pt}\rceil \ccdot
				\lceil\hspace{-2pt}\lblot}
			\big) 
			{\xi}^{2} t_3\,t_{4} 
			\rangle\rceil
		\end{align}
		respectively. 
		\added[id=1]{Note that in the second line of Eq.~\eqref{corr4_2} we have used the bilinear property introduced in Eq.~\eqref{angle_bilinear}, while the third line is obtained by exploiting, in the first term, the projection property of
			Eq.~\eqref{closed_corr_proj}.}
		We observe that in the case in which $\overline{\xi_0^2} \ne\overline{\xi^2}$, Eqs.~\eqref{corr2t_2} becomes
		\begin{align*}
			\added[id=1]{ \langle \xi(t_1)\xi(t_2)\rangle =
				\left(\overline{\xi_0^2}- \overline{\xi^2}\right)
				\Psi(t_2 - t_0) +\lceil \langle \xi^2 t_1 \,t_2\rangle \rceil}.
		\end{align*}
		\added[id=1]{Moreover, Eq.~\eqref{corr4_2} becomes even more involved. Therefore, for the sake of simplicity,
			hereafter we will assume, unless explicitly stated otherwise, that the PDF of the initial state of the system
			is identical to the PDF of the random variable $\xi$.}

		Now, we are in a position to state the following Proposition, which outlines a 
		procedure to generalize the ``intuitive'' statistical (though not rigorous) approach used in the 
		$n=2$ and $n=4$ cases to the general $n$-point case. As such, the method presented 
		in this Proposition suffers from the same limitations as those previous cases: while reasonable, 
		it is susceptible to pitfalls because it deals with non-stationary statistics. 
		For this reason, in~\ref{app:A_A}, we provide a rigorous proof of the final result obtained 
		from this intuitive approach. 
		
		\added[id=1]{Finally, we emphasize that, unlike the proposition stated below, the formal proof presented in~\ref{app:A_A}
			does not rely on the simplifying assumption that the initial ensemble is prepared such that the PDF of $\xi_0$
			is the same as that of $\xi$.}
		
		\begin{proposition}
			\label{prop:main}
			Assuming that the PDF of the system's initial state
			coincides with that of the random variable $\xi$,  the $n$-time  correlation function for the stochastic process defined as a random step function
			with renewal (the noise for the L\'evy walk with random 
			velocity) is obtained through the following four steps procedure:
		\end{proposition}
		\begin{enumerate}[label=(\roman*)]
			\item \label{itema2}
			Write a sequence of $n$ ordered times between angle brackets:
			$\langle t_1\,t_2\,t_3...t_n\rangle$.
			
			\item \label{itemb2}
			Take any composition of this sequence (i.e., a partition where the order matters), made
			of subsequences (or blocks $\left\{ m_i\right\}$) of  terms by inserting the concatenated 
			``ket$\cdot$bra'' (i.e., 
			$\conc$) separators between these ordered times.
			In other words, partition the ordered sequence into $p \leq n$ blocks 
			$\langle\left\{ m_1\right\} 
			\conc
			\left\{ m_2\right\}\conc...\conc\left\{ m_p\right\}\rangle$
			where $m_i$ is the  number of elements of the $i$-th block. 
			Of course,  $\sum_{i=1}^p m_i=n$. For example, a partition made of $p$ blocks could be
			$$\langle\underbrace{t_1\,t_2\,t_3\,t_4\,t_5\,t_6)}_{m_1=6}\conc
			\underbrace{t_7\,t_8}_{m_2=2}\conc\,...\,\conc
			\underbrace{t_{n-3}\,t_{n-2}\,t_{n-1}\,t_n}_{m_p=4}\rangle.$$
			The number of possible such compositions is $2^{n-1}$, as at each position (excluding the two extrema) 
			the separation term ``$\conc$'' can be inserted or not inserted.  
			\item \label{itemc2} Assign a factor of $\overline{\xi^{m_i}}$ to each block consisting of $m_i$ ordered terms. 
			\item \label{iteme2} $\langle\xi(t_1)\xi(t_2)\dots\xi(t_n)\rangle$ is obtained summing all   
			$2^{n-1}$  compositions 
			obtained in this way.
		\end{enumerate}
		\added[id=1]{Since the sum over all compositions of $n$ ordered objects $a_1 a_2 \ldots a_n$ 
			into $p$ blocks ($1 \leq p \leq n$) can be formally written as  
			$$\{a_1\Big(1+\}\{\Big)a_2\Big(1+\}\{\Big)...a_{n-1}\Big(1+\}\{\Big)a_n\},$$}
		a \added[id=1]{convenient} way to 
		\added[id=1]{express compactly} 
		the sum of all compositions \added[id=1]{as prescribed in point~\ref{iteme2}},
		is to exploit both the bi-linear property 
		of the closed angle brackets given in \eqref{angle_bilinear} and the alternative definition of average
		over the PDF of $\xi$ given in Eq.~\eqref{mean}: 
		\begin{align}
			\label{corrGen_step_fin}
			\langle\xi(t_1)\xi(t_2)\xi(t_3)\xi(t_4)...\xi(t_n)\rangle_{t_0}
			=& 
			\added[id=1]{
				= \lceil\langle   {\xi} t_{1}
				\big(1+{\protect \rblot 
					\hspace{-2pt}\rceil \ccdot
					\lceil\hspace{-2pt}\lblot}
				\big) 
				{\xi} t_2 
				\big(1+{\protect \rblot 
					\hspace{-2pt}\rceil \ccdot
					\lceil\hspace{-2pt}\lblot}
				\big)...}
			\nonumber \\
			& .. 
			{\xi}t_{n-1}
			\big(1+
			{\protect \rblot 
				\hspace{-2pt}\rceil \ccdot
				\lceil\hspace{-2pt}\lblot}
			\big)
			{\xi}t_{n}\rangle
			\rceil.
		\end{align}%
		Another way is to first sum all the compositions corresponding to
		a fixed number $p$ of blocks  (they are 
		$N(p)=\frac{\left(n-1\right)!}{(p-1) !\left[n-p\right] !}$) and then summing
		for all $p=1, 2,...n$ (of course we have
		$\sum_{p=1}^{n} N(p)=2^{n-1}$), thus
		\begin{align}
			\label{corrGen_gen}
			&\langle\xi(t_1)\xi(t_2)\dots\xi(t_n)\rangle \nonumber \\
			&= \sum_{p=1}^{n} 
			\bigg[
			\sum_{\{m_i\in\N\}:\sum_{i=1}^p m_i = n}  	
			(\overline{\xi^{m_1}})(\overline{\xi^{m_2}})
			(\overline{\xi^{m_3}})
			\dots(\overline{\xi^{m_p}})\times \nonumber \\
			&\langle t_1\,t_2...t_{m_1}\conc t_{m_1+1}\,t_{m_1+2}...\,t_{m_1+m_2}\conc
			t_{m_1+m_2+1}\,t_{m_1+m_2+2}...\,t_{m_1+m_2+m_3}\conc
			\dots \nonumber \\
			&\dots\conc t_{m_1+m_2+\dots+m_{p-1}+1}\, t_{m_1+m_2+\dots+m_{p-1}+2}...\,\underbrace{t_{m_1+m_2+\dots+m_{p-1}+m_p}}_{t_n}\rangle
			\bigg].
		\end{align}
		%
		%
		Finally, by exploiting the projection property of the closed correlation function, defined in Eq.~\eqref{closed_corr_proj},
		the results \eqref{corrGen_gen} can  also be written as:
		\begin{align}
			\label{corrGen_}
			&\langle\xi(t_1)\xi(t_2)\dots\xi(t_n)\rangle \nonumber \\
			&=  \sum_{p=1}^{n} 
			\bigg[
			\sum_{\{m_i\in\N\}:\sum_{i=1}^p m_i = n}  
			(\overline{\xi^{m_1}})(\overline{\xi^{m_2}})
			(\overline{\xi^{m_3}})\dots(\overline{\xi^{m_p}})\times \nonumber \\
			&\times\langle t_1\, t_{m_1})\conc t_{m_1+1}\, t_{m_1+m_2}\conc
			t_{m_1+m_2+1}\, t_{m_1+m_2+m_3}\conc
			\dots \nonumber \\
			&\dots\conc t_{m_1+m_2+\dots+m_{p-1}+1}\,\underbrace{t_{m_1+m_2+\dots+m_{p-1}+m_p}}_{t_n}\rangle.
			\bigg]
		\end{align}
		
		\added[id=1]{We note that in~\ref{app:A_A} a generalization of Eq.~\eqref{corrGen_} to an arbitrary initial ensemble is derived
			without relying on the ``bra'' and ``ket'' notation introduced in Eq.~\eqref{closed_corr}.
			This notation was originally introduced merely as a shorthand for the joint PDF of a set of times
			within the same laminar region. By contrast, the rigorous derivation in~\ref{app:A_A}
			relies solely on algebraic manipulations of the multi-time correlation function
			defined in Eq.~\eqref{n_corr_def_step}.
		}
		
		\added[id=1]{This manipulation is organized into several steps. Among them, the most relevant ones, directly connected to the 
			procedure in Proposition~\ref{prop:main}, are steps~(b) and~(c). 
			In step~(b), we rearrange the multiple sum in \eqref{n_corr_def_step} and show that the Heaviside functions 
			constrain the times 
			$t_1,t_2,\ldots,t_n$ of the multi-time correlation function to lie, in groups, within the same laminar region
			(see also the remarks, organized into three points, at the end of step~(b) in~\ref{app:A_A}).  }
		
		\added[id=1]{In step~(c), we  establish the formal equivalence between the rearranged multiple sum and
			the sum over compositions of  $n$ times (partition in groups, where the order matter). 
			At the same time, we also show that the summand can be directly related to the joint probability of such grouping events.  }
		
		\added[id=1]{ The remainder of the appendix consists of formal manipulations of the integrals over the waiting times, aimed at introducing the 
			rate function explicitly and thereby making manifest the equivalence with the expression in Eq.~\eqref{corrGen_}.
		}

		The expression \eqref{corrGen_gen} for the $n$-time  correlation function is formally 
		equal to the closed-form formula that expresses multi-time  correlations 
		(or multivariate moments) as sums of products of $G$-cumulants 
		(see also~\cite[Section 4.4.3, Eq.~(94)]{bbJSTAT4}). 
		
		In other words: 
		\begin{itemize} 
			\item the expressions \eqref{corrGen_gen}-\eqref{corrGen_} resemble the exponential formula in 
			combinatorial mathematics, which provides the exponential generating function in 
			terms of set partitions (from which the cumulants are defined); 
			\item in \eqref{corrGen_gen}-\eqref{corrGen_} the partitions preserve the ordering of the elements, 
			making them compositions, consistent with the definition of $G$-cumulants;
			\item the two points above allow to obtain the  master equation for the PDF of $x$ of Eq.~\eqref{SDE},
			by using the result \eqref{corrGen_} and by exploiting the generalized cumulant approach formulated in the already
			cited series of papers~\cite{bbJSTAT4,bCSF148,bCSF159}. However, we refrain from doing so, 
			as it falls outside the scope of the present study.
			
		\end{itemize}

		It is particularly relevant to observe that the result \eqref{corrGen_} involves only the
		moments of the random variable $\xi$ and the concatenation of the closed two-time
		correlation functions.
		\subsection{\added[id=2]{Some examples}}
		The explicit examples of the non symmetric two-time and the symmetric sixth and eighth-time 
		correlation functions may serve for illustration.
		
		In the non symmetric two-time correlation function we have $\overline{\xi}\ne0$, therefore, from \eqref{corrGen_} we get 
		\begin{align*}
			&\langle t_1\,t_2\rangle,&(p=1)\\
			&\langle t_1\conc t_2\rangle &(p=2)
		\end{align*}
		from which
		\begin{equation}
			\label{corr2_gen}
			\begin{array}{lll}
				\langle\xi(t_1)\xi(t_2)\rangle =& &
				\\
				\;&\;\\
				\overline{\xi^2}\langle  t_1\, t_2\rangle &(p=1)\\
				\;&\;\\
				+\overline{\xi}^2
				\langle  t_1\conc t_2\rangle
				&(p=2).
			\end{array}
		\end{equation}
		Eq.~\eqref{corr2_gen} generalizes to the non symmetric PDF case the result of 
		Eq.~\eqref{corr2t_2}.
		The contribution corresponding to $p=1$ does not need any further manipulation, while the  contributions
		corresponding to $p=2$ can be made explicit by exploiting the definition of the concatenation
		of closed  correlation function given in \eqref{closed_corr_prod}.
		Thus we get 
		\begin{align}
			\label{corr2_gen2} 
			&\langle\xi(t_1)\xi(t_2)\rangle 
			=\overline{\xi^2}\langle  t_1\, t_2\rangle 
			+\overline{\xi}^2
			\int^{t_1}_0 du_1 \tilde{R}\left(u_1\right) \int_{t_1}^{t_2} du_1^{\prime}\,\psi(u_1^{\prime}-u_1)%
			\nonumber\\
			&\times	\int^{t_2}_{u_1^{\prime}} du_2 \tilde{R}\left(u_2-u_1^{\prime}\right) \int_{t_2}^{\infty} du_2^{\prime}\,\psi(u_2^{\prime}-u_2).
		\end{align}
		Now the case $n=6$ and symmetric $\xi$ PDF .
		
		The compositions with even elements are 
		\begin{align*}
			&\langle t_1\,t_2\,t_3\,t_4\,t_5\,t_6\rangle,&(p=1)\\
			&\langle t_1\,t_2\,t_3\,t_4\conc
			t_5\,t_6\rangle\text{ and }  \langle t_1\,t_2\conc
			t_3\,t_4\,t_5\,t_6\rangle &(p=2)\\
			&\langle t_1\,t_2\conc t_3\,t_4\conc
			t_5\,t_6\rangle, &(p=3)
		\end{align*}
		from which
		\begin{equation}
			\label{corr6_}
			\begin{array}{ll}
				\langle\xi(t_1)\xi(t_2)\xi(t_3)\xi(t_4)\xi(t_5)\xi(t_6)\rangle =& 
				\\
				\;&\;\\
				\overline{\xi^{6}}\langle  t_1\, t_6\rangle \;\;\;
				\left[=\frac{\overline{\xi^{6}}}{\overline{\xi^{2}}}\langle \xi(t_1)\xi(t_6)\rangle\right]
				\;\;\;&(p=1) \\
				\;&\;\\
				+\overline{\xi^{4}}\,\overline{\xi^{2}}
				\big(\langle  t_1\, t_4\conc
				t_5\, t_6\rangle
				+\langle t_1 \,t_2\conc t_3\, t_6\rangle\big)
				\;\;\;&(p=2)
				\\
				\;&\;\\
				+\left(\overline{\xi^{2}}\right)^3\langle  t_1\, t_2\conc  t_3\, t_4\conc t_5\, t_6\rangle
				\;\;\;&(p=3).
			\end{array}
		\end{equation}
		Again, the contribution corresponding to $p=1$ does not need any further manipulation, while the  contributions
		corresponding to $p=2$ and $p=3$ is made explicit by exploiting the definition of the concatenation
		of closed  correlation function given in \eqref{closed_corr_prod}. 
		Thus, for $p=2$ we have
		\begin{align}
			\label{b2_corr6}
			&\overline{\xi^{4}}\,\overline{\xi^{2}}
			\left(\langle t_1 \,t_2\conc t_3\, t_6\rangle
			+\langle  t_1\, t_4\conc
			t_5\, t_6\rangle\right)\nonumber
			\\
			&=\overline{\xi^{4}}\,\overline{\xi^{2}}
			\bigg( 
			\int^{t_1}_0 du_1 \tilde{R}\left(u_1\right) \int_{t_2}^{t_3} du_1^{\prime}\psi(u_1^{\prime}-u_1)%
			\nonumber\\
			&\times	\int^{t_3}_{u_1^{\prime}} du_3 \tilde{R}\left(u_2-u_1^{\prime}\right) \int_{t_6}^{\infty} du_2^{\prime}\psi(u_2^{\prime}-u_2)
			\nonumber \\
			&+
			\int^{t_1}_0 du_1 \tilde{R}\left(u_1\right) \int_{t_4}^{t_5} du_1^{\prime} \psi(u_1^{\prime}-u_1)%
			\nonumber\\
			&\times	\int^{t_5}_{u_1^{\prime}} du_2 \tilde{R}\left(u_2-u_1^{\prime}\right) \int_{t_6}^{\infty} du_2^{\prime}\psi(u_2^{\prime}-u_2)
			\bigg),
		\end{align}
		and for $p=3$:
		\begin{align}
			\label{6times}
			%
			&\left(\overline{\xi^{2}}\right)^3 \langle  t_1\, t_2\conc  t_3\, t_4\conc t_5\, t_6\rangle &\,
			\nonumber 
			\\
			&=\left(\overline{\xi^{2}}\right)^3
			\bigg( 
			\int^{t_1}_0 du_1 \tilde{R}\left(u_1\right) \int_{t_2}^{t_3} du_1^{\prime}\psi(u_1^{\prime}-u_1)
			\nonumber\\
			&\times	\int^{t_3}_{u_1^{\prime}} du_2 \tilde{R}\left(u_2-u_1^{\prime}\right)\int_{t_4}^{t_5} du_2^{\prime} \psi(u_2^{\prime}-u_2)%
			\nonumber 
			\\
			&\times
			\int_{u_2^\prime}^{t_5} du_3 
			du_3 \tilde{R}\left(u_3-u^\prime_2\right) \int_{t_6}^{\infty} du_3^{\prime}\psi(u_3^{\prime}-u_3)
			\bigg).
		\end{align}

		Then the 8-time  correlation function for the symmetric $\xi$ PDF : the compositions of the ordered times are
		\begin{align*}
			&\langle t_1\,t_2\,t_3\,t_4\,t_5\,t_6\,t_7\,t_8\rangle \text{;}&(p=1)\\
			&\langle t_1\,t_2\,t_3\,t_4\,t_5\,t_6\conc
			t_7\,t_8\rangle \text{ , }
			\langle t_1\,t_2\,t_3\,t_4\conc
			t_5\,t_6\,t_7\,t_8\rangle\text{ , }
			\langle t_1\,t_2\conc
			t_3\,t_4\,t_5\,t_6\,t_7\,t_8\rangle\text{;}&(p=2)\\ 
			&\langle t_1\,t_2\,t_3\,t_4\conc 
			t_5\,t_6\conc
			t_7\,t_8\rangle\text{ , }
			\langle t_1\,t_2\conc 
			t_3\,t_4\,t_5\,t_6\conc 
			t_7\,t_8\rangle\text{ , }
			\langle t_1\,t_2\conc  
			t_3\,t_4\conc  
			t_5\,t_6\,t_7\,t_8\rangle\text{;}&(p=3)
			\\&\langle t_1\,t_2\conc  
			t_3\,t_4\conc  
			t_5\,t_6\conc  
			t_7\,t_8\rangle.&(p=4)
		\end{align*}
		from which,
		\begin{equation}
			\label{corr8_}
			\begin{array}{ll}
				\langle\xi(t_1)\xi(t_2)\xi(t_3)\xi(t_4)\xi(t_5)\xi(t_6)\xi(t_7)\xi(t_8)\rangle & 
				\\
				\;&\;\\
				=\overline{\xi^{8}}\langle \xi(t_1)\xi(t_8)\rangle/\overline{\xi^{2}}
				\;\;\;&(p=1) \\
				\;&\;\\
				+\left(\overline{\xi^{6}}\,\overline{\xi^{2}}\langle  t_1\, t_6)\conc  t_7\, t_8)\rangle 
				+\left(\overline{\xi^{4}}\right)^2\langle  t_1\, t_4
				\conc  t_5\, t_8\rangle\right.&
				\\
				\left.+\overline{\xi^{2}}\,\overline{\xi^{6}}\langle  t_1\, t_2\conc  t_3\, t_8\rangle\right)
				\;\;\;&(p=2)
				\\
				\;&\;\\
				+\overline{\xi^{2}}\,\overline{\xi^{4}}{}\left(
				\langle  t_1,t_4\conc  
				t_5\, t_6\conc  
				t_7\, t_8\rangle 
				+\langle  t_1\, t_2\conc  t_3\, t_6\conc 
				t_7\, t_8\rangle
				\right.&
				\\
				\left.+\langle t_1\, t_2\conc  t_3,t_4\conc 
				t_5    , t_8\rangle
				\right)
				\;\;\;&(p=3)
				\\
				\;&\;\\
				+\left(\overline{\xi^{2}}\right)^4\langle t_1\, t_2\conc 
				t_3\, t_4\conc  t_5\, t_6\conc 
				t_7\, t_8\rangle
				\;\;\;&(p=4)
			\end{array}
		\end{equation}
		From this expression, by using again the definition of closed correlation function given in \eqref{closed_corr_prod}, 
		it is straightforward to obtain the explicit expressions of  the  contributions,
		for $p=1,2,3$ and $4$ to the 8-time  correlation function in \eqref{corr8_}, in terms of waiting
		time PDF, but, for the sake of simplicity, we do not report here the result.
		\section{\added[id=2]{Further consequent results}}
		
		\subsection{The general dichotomous case\label{sec:dicho_creepers}}
		The dichotomous case is easily treated by using the compact way, given in \eqref{corrGen_step_fin}, to write the
		$n$-time correlation function. 
		
		In the case of symmetric dichotomous random variable
		$\xi$, with values $\pm 1,$  we can  replace ${\xi^{n}}$ ($n$ even) with $1$ and then replace with the identity the 
		``brackets'' $\lceil...\rceil$. Doing that
		in  \eqref{corrGen_step_fin} implies to  set $\rceil\lceil=1$ there. Thus we get (note that in
		this case we can safely set 
		$\lblot t_{3}\,t_{4}\rblot=\lblot \xi(t_{3}) \xi(t_{4})\rblot):$  %
		\begin{align}
			\label{corrGen_step_fin_}
			& \langle\xi(t_1)\xi(t_2)\xi(t_3)\xi(t_4)...\xi(t_n)\rangle_{t_0}
			=\langle  t_{1}\,t_{2}
			\big(1+\conc
			\big) 
			t_3,t_{4} 
			\big(1+\conc 
			\big)...
			\nonumber \\
			&... 
			t_{n-3}\,t_{n-2}
			\big(1+
			\conc 
			\big)
			t_{n-1}\,t_{n}\rangle\nonumber \\
			&=\langle  \, \xi(t_1)\xi(t_{2}) 
			\left(1+
			\conc
			\right) 
			\xi(t_{3}) \xi(t_{4})
			\left(1+\conc\right)... 
			\nonumber \\& 
			\dots 
			\xi(t_{n-3}) \xi(t_{n-2})\left(
			1+\conc  
			\right)\xi(t_{n-1}) \xi(t_n) \rangle
		\end{align}
		Eq.~\eqref{corrGen_step_fin_} shows that in general, any kind of factorization property, for the multi-time
		correlation function for the general dichotomous noise, does not hold.
		As we will see hereafter, only in the Poissonian case, in which the WT PDF decays exponentially, 
		in Eq.~\eqref{corrGen_step_fin_} we can replace 
		``$\conc $'' with ``$\rangle\langle-1''$, leading to the well known factorization property of the multi-time 
		correlation function of the telegraph noise. 
		%
		
		%
	
	\subsection{The symmetric Poissonian cases\label{sec:poiss_telegraph}}
	
	The Poissonian stochastic process with renewal is characterized by the following WT  PDF 
	\begin{equation}
		\label{WTpoisson}
		\psi(t)=\frac{1}{\tau} \exp(-t/\tau),
	\end{equation}
	consequently (see Eqs.~\eqref{Rtilde}-\eqref{R}) $R(t)=1/\tau$ and $\tilde R(t)=1/\tau+\delta(t)$. 
	It represents a particular, but really important, case of the broad class of  stationary stochastic processes. 
	Given its significance, we will start by examining this scenario%
	\footnote{To be more precise and to consider a more general situation, we specify, even if it
		should not be necessary, that all the results we obtain hereafter, concerning the case where
		the WT  PDF decays as in \eqref{WTpoisson}, are trivially extended to the situation in which the time
		behavior of the WT  PDF is more complex, as long as the function \eqref{WTpoisson} represents
		the ``large'' time behavior of the envelope of $\psi(t)$.}.%

	For the sake of simplicity, we assume in this Poissonian case that the $\xi$ PDF  is 
	symmetric, that is, that all odd moments of $\xi$ vanish. Under this assumption, the 
	maximum value of ``$p$'' in Eq.~\eqref{corrGen_} is $n/2$. 
	
	Inserting \eqref{WTpoisson} into \eqref{bblmCSF196}, we obtain an 
	exponential behavior, that is standard for Poissonian processes~\cite{bCSF159,Cox_Renewal_Theory}
	and that for this case was already  reported in~\cite{bblmCSF196}:
	\begin{align}\label{poisson_}
		\langle\xi(t_1)\xi(t_{2})\rangle =
		\overline{\xi^2}\,
		e^{-\frac{1}{\tau}(t_2-t_1)} +
		\left(\overline{\xi_0^2} - \overline{\xi^2}\right)
		e^{-\frac{1}{\tau}t_2}.
	\end{align}
	Eq.~\eqref{poisson_} shows that if $\overline{\xi_0^2} = \overline{\xi^2}$, the correlation function is 
	always stationary. Otherwise, stationary is asymptotically achieved for $t_2 \gg \tau$. 
	In the latter case, it is worth noting that no conditions are required for the
	time lag  $t_2-t_1$. 
	
	A less trivial result, that for the best of our knowledge is not reported in literature, is the explicit expression 
	for the general $n$-time  correlation function red in the Poissonian case.
	For that,  we exploit the result \eqref{corrGen_} of
	Proposition~\ref{prop:main}, according to which the key quantity we need to evaluate is
	the closed correlation function defined  in Eqs.~\eqref{closed_corr}-\eqref{closed_corr_prod}.
	Starting the multiple integrations from the right side of the concatenated closed correlation functions, 
	it is easy to show that from   \eqref{WTpoisson}, for a generic term of the sum in Eq.~\eqref{corrGen_} 
	we have (note: $t_{i_p}=t_n$)
	\begin{align}
		\label{pois} 
		&...\lblot t_{i_{p-3}+1}\, t_{i_{p-2}}\conc  t_{i_{p-2}+1}\, t_{i_{p-1}}\conc  t_{i_{p-1}+1}\, t_{i_{p}}\rangle\nonumber
		\\
		=&...
		\left(e^{-\frac{1}{\tau}(t_{i_{p-2}}-t_{i_{p-3}+1})}-e^{-\frac{1}{\tau}(t_{i_{p-2}+1}-t_{i_{p-3}+1})}\right)
		\nonumber
		\\
		&\times \left(e^{-\frac{1}{\tau}(t_{i_{p-1}}-t_{i_{p-2}+1})}-e^{-\frac{1}{\tau}(t_{i_{p-1}+1}-t_{i_{p-2}+1})}\right)e^{-\frac{1}{\tau}(t_{i_{p}}-t_{i_{p-1}+1})}
		\nonumber \\
		=&...e^{-\frac{1}{\tau}(t_{i_{p-2}}-t_{i_{p-3}+1})}\left(1-e^{-\frac{1}{\tau}(t_{i_{p-2}+1}-t_{i_{p-2}})}\right)\nonumber
		\\
		&\times e^{-\frac{1}{\tau}(t_{i_{p-1}}-t_{i_{p-2}+1})}\left(1-e^{-\frac{1}{\tau}(t_{i_{p-1}+1}-t_{i_{p-1}})}\right)e^{-\frac{1}{\tau}(t_{i_{p}}-t_{i_{p-1}+1})}
	\end{align}
	%
	%
	%
	%
	that explicitly shows its stationary nature. Moreover, 
	if the ``observation'' time-scale is much larger than the ``microscopic'' time $\tau$, i.e., if 
	$\forall i\in[0,1,...,n-1]$ we have $t_{i+1} - t_i \gg\tau$,
	then 
	\begin{equation}
		\label{pois:approx}
		\left(1 - e^{-\frac{1}{\tau}(t_{i+1} - t_i)}\right) \approx 1.
	\end{equation}
	Therefore, in this case, we can safely set %
	\begin{equation}
		\label{poiss:opencorr}
		\lblot t_i\, t_{i+h}\rblot\approx e^{-\frac{1}{\tau}(t_{i+h}-t_i)}=\langle\xi(t_1)\xi(t_{2})\rangle /
		\overline{\xi^2}
	\end{equation}
	%
	and the concatenation of closed correlation functions in Eq.~\eqref{pois} reduces to a product of ``normalized'' 
	correlation functions: 
	\begin{align}
		\label{pois2} 
		&...\lblot t_{i_{p-3}+1}\, t_{i_{p-2}}\conc  t_{i_{p-2}+1}\, t_{i_{p-1}}\conc  
		t_{i_{p-1}+1}\, t_{i_{p}}\rangle	
		\nonumber \\
		\approx&...e^{-\frac{1}{\tau}(t_{i_{p-2}}-t_{i_{p-3}+1})}
		e^{-\frac{1}{\tau}(t_{i_{p-1}}-t_{i_{p-2}+1})}
		e^{-\frac{1}{\tau}(t_{i_{p}}-t_{i_{p-1}+1})}, 
	\end{align}
	When inserting Eq.~\eqref{pois2} into the general result \eqref{corrGen_}, we obtain the $n$-time correlation 
	function for the Poissonian case.
	For example, for $n=6$, by combining Eq.~\eqref{pois2} with \eqref{corr6_} we  get
	\begin{equation}
		\label{corr6_Pois}
		\begin{array}{ll}
			\langle\xi(t_1)\xi(t_2)\xi(t_3)\xi(t_4)\xi(t_5)\xi(t_6)\rangle & 
			\\
			\;&\;\\
			=\overline{\xi^{6}}
			e^{-\frac{1}{\tau}(t_6-t_1)} &(p=1) \\
			\;&\;\\
			+\overline{\xi^{4}}\,\overline{\xi^{2}}
			\big(	e^{-\frac{1}{\tau}(t_4-t_1)}
			e^{-\frac{1}{\tau}(t_6-t_5)}
			+	e^{-\frac{1}{\tau}(t_2-t_1)}	e^{-\frac{1}{\tau}(t_6-t_3)}\big)
			\;\;\;&(p=2)
			\\
			\;&\;\\
			+\left(\overline{\xi^{2}}\right)^3	e^{-\frac{1}{\tau}(t_2-t_1)} 	e^{-\frac{1}{\tau}(t_4-t_3)}
			e^{-\frac{1}{\tau}(t_6-t_5)}
			\;\;\;&(p=3).
		\end{array}
	\end{equation}
	It is apparent that
	for $t_{i+1} - t_i \gg\tau$ the dominant term 
	in \eqref{corr6_Pois} is that with $p=3$. Looking at the
	more general result \eqref{corrGen_},  it is also clear that  Eq.~\eqref{pois2} implies that for  large 
	time scale separation between the observation times and 
	the decay time of the WT PDF, the dominant term in the sum of compositions given in  \eqref{corrGen_}, is 
	the last one,  i.e., the one corresponding to $p = n/2$ (the maximum 
	value for the symmetric case). Thus, we have
	\begin{align}
		\label{poi}
		&\langle \xi(t_1) \xi(t_2) \ldots \xi(t_n) \rangle \approx 
		\left(\overline{\xi^{2}}\right)^{\frac{n}{2}} e^{-\frac{1}{\tau}(t_2 - t_1)} e^{-\frac{1}{\tau}(t_4 - t_3)}
		\ldots e^{-\frac{1}{\tau}(t_n - t_{n-1})}+ O\left(n\Delta t_m/\tau\right)\nonumber \\
		&=\langle\xi(t_1)\xi(t_{2})\rangle\langle\xi(t_3)\xi(t_{4})\rangle\ldots\langle\xi(t_{n-1})\xi(t_{n})\rangle
		+ O\left(n\Delta t_m/\tau\right),
	\end{align}
	where $\Delta t_m/\tau:=\min[t_{i+1}-t_i, 0\le t\le n-1]$ is the minimum  time lag. 
	Therefore, in this asymptotic case, only the variance of the random variable $\xi$ is relevant and 
	\textit{the factorization property holds}\footnote{Note that in the special case of a dichotomous Poissonian stochastic process , 
		i.e., for $\overline{\xi^{2}}=1$,
		comparing the large time behavior of Eq.~\eqref{poisson_} with Eq.~\eqref{pois}, it is easy to verify 
		that  we  can safely replace ``$\rblot     
		\lblot$'' with ``$\rangle\langle-1$'', and, when this replacement is done in  Eq.~\eqref{corrGen_step_fin_}, it leads
		to the exact factorization property, as anticipated in Section~\ref{sec:dicho_creepers}.}.

	
	%


	Comparing the limit result \eqref{poi}, holding for general renewal 
	Poissonian stochastic process $\xi[t]$, with the exact factorization property of the telegraph noise, we are led to  the following
	\begin{proposition}
		\label{prop:1}
		Given a symmetric stochastic process $\xi[t]$ with renewal and WT  PDF given by 
		$\psi(t)=(1/\tau)\exp[-t/\tau]$, for   $\vert t_i-t_j\vert \gg \tau$, with $i,j\in [1,2,...,n]$, 
		if the first $n$ moments of the random variable $\xi$ are finite, 
		the $n$-time  correlation function $\langle \xi(t_1) \xi(t_2)...\xi(t_n)\rangle$ tends to the same multi-time  
		correlation function of the telegraph noise.
	\end{proposition} 
	From Proposition~\ref{prop:1}, with some abuse of the term ``indistinguishable'', it directly follows
	\begin{lemma}
		\label{lem:1}
		Under the conditions holding for Proposition~\ref{prop:1}, the stochastic process $\xi[t]$ becomes
		indistinguishable from the telegraph noise.
	\end{lemma} 
	Proposition~\ref{prop:1} and Lemma~\ref{lem:1},  which hold when there is a 
	separation between the observation time scale (regarding the times $t_1,t_2,...,t_n$ of the
	multi-time correlation functions) and the intrinsic time scale of the 
	stochastic process $\xi[t]$,  are consistent with the generalized central limit 
	theorem (GCLT, also known as Operator Central Limit Theorem - OCLT) presented in~\cite{bCSF148}, and more 
	specifically with \cite[Section 7.2]{bCSF159}, but   represent novel and stronger 
	results.
	
	In fact,  
	the GCLT concerns the limit behavior of the sum (the integral,  here) of the random process
	$\xi[t]$, (i.e., it regards the statistics of the Brownian
	variable $x$, fulfilling the equation  $\dot x=\xi[t]$), while the limit result  
	\eqref{poi} concerns the full statistics of $\xi[t]$,
	a much stronger statement.  For that reason we use the word ``indistinguishable''. 
	The added value of this result can be appreciated, for example, considering the  
	general SDE \eqref{SDE}, with  $\xi[t]$ a Poissonian stochastic process with renewal. 
	In fact, if the drift dynamics is slow compared to that
	of $\xi[t]$, then, according to  Lemma~\ref{lem:1},  $\xi[t]$ can be replaced by a telegraph
	noise with the same correlation function.
	On the other hand, it is not difficult to show that, when the noise $\xi[t]$ 
	is a telegraph process, the dynamics of the PDF of $x$ governed by 
	Eq.~\eqref{SDE},  assuming the initial preparation of the ensemble of $x$ does not depend on the 
	state of the noise (or that the time $t$ 
	is enough greater than $\tau$ to make irrelevant the initial
	condition), exactly satisfies the following master equation (see~\ref{app:ME_dicho})
	\footnote{note that Eq.~\eqref{MEG_cnumber_PDF_appro_2} is derived from Eq.~\eqref{MEG_cnumber_PDF_appro}
		by simply observing that, by using the Hadamard formula, the following series of equalities holds 
		(for details, see~\cite{bJMP59}):
		$e^{\partial_x C(x) u} \partial_x I(x)...=e^{\partial_x C(x) u} \partial_x I(x)e^{-\partial_x C(x) u}e^{\partial_x C(x) u}...
		= \partial_x C(x)\; e^{\partial_x C(x) u}\frac{I(x)}{C(x)}e^{-\partial_x C(x) u}e^{\partial_x C(x) u}...
		=\partial_x C(x)
		\frac{I\left(x_0(x;-u)\right)}{C\left(x_0(x;-u)\right)} e^{\partial_x C(x) u}...$}:
	\begin{align}
		&\partial_t P(x;t)
		= \partial_x \big[ C(x) P(x;t) \big] \nonumber \\
		& + \overline{\xi^{2}} \partial_x I(x)\int_0^t e^{-u / \tau} 
		e^{\partial_x C(x) u} \partial_x I(x) P(x;t-u) d u
		\label{MEG_cnumber_PDF_appro}
		\\
		=& \partial_x \big[ C(x) P(x;t) \big] \nonumber \\
		& + \overline{\xi^{2}} \partial_x I(x)\partial_x C(x)
		\int_0^t e^{-u / \tau} \frac{I\left(x_0(x;-u)\right)}{C\left(x_0(x;-u)\right)} 
		e^{\partial_x C(x) u} P(x;t-u) du,
		\label{MEG_cnumber_PDF_appro_2}
	\end{align}
	where $x_0(x; -u) := \left(e^{\partial_x I(x) u} \, x \, 
	e^{-\partial_x C(x) u}\right)$ represents the backward time evolution of the 
	variable $x$ in the absence of external perturbations, that is, under the sole 
	influence of the drift field $-C(x)$.
	
	Now, if $\xi(t)$ in Eq.~\eqref{SDE} is not a telegraph process but a more 
	general Poissonian renewal noise, and if there is enough time-scale 
	separation between the fluctuations of the noise and the deterministic 
	dynamics induced by $-C(x)$, to satisfy Lemma~\ref{lem:1}, 
	Eq.~\eqref{MEG_cnumber_PDF_appro} still holds.

	We do not delve into this matter, which concerns the ME equivalent 
	to the SDE~\eqref{SDE}, because in this paper we focus our attention on the 
	stochastic process $\xi[t]$.

	\subsection{The stationary condition for the case of power law WT\label{sec:stationary}}
	
	According to our previous works~\cite{bCSF159,bblmCSF196}, as well as standard 
	results from renewal theory~\cite{Cox_Renewal_Theory}, we have shown that when 
	$\xi[t]$ is a Poissonian process, it is stationary, meaning that its correlation 
	functions remain invariant under a uniform time shift of all arguments. 
	
	Furthermore, Proposition~\ref{prop:1} and Lemma~\ref{lem:1} establish that, for 
	time intervals much larger than $\tau$, all multi-time correlation functions of 
	any Poisson renewal process converge to those of telegraph noise. This is indeed a new result.
	
	\added[id=2]{It is now natural to ask whether the stationarity of $\xi[t]$ persists in the more general case of a waiting time (WT) PDF does not decay exponentially.}
	\deleted{It is now natural to ask which of these results persist in the more general 
		case where $\xi[t]$ remains stationary, but the waiting time (WT) PDF no longer 
		decays exponentially.} Before addressing this question, we first verify whether the 
	$n$-time correlation functions obtained via Procedure~\ref{prop:main} reproduce 
	the well-known results for Brownian motion with Lévy walks and random velocities.
	
	Indeed, it is well known that in the case of a free Brownian particle governed by 
	$\dot{x} = \xi[t]$, in the limit $t \gg \tau$, with $\tau$ being the first moment 
	of the WT PDF, as defined in Eq.~\eqref{tau}, provided it is finite, the process 
	$x[t]$ becomes stationary. This classical result pertains to the statistics of the 
	time integral of the noise realization, $x(t) = \int_0^t \xi(u)\,du$. In the 
	following, we present a similar but more general result that characterizes the 
	detailed statistics of the process $\xi[t]$ itself, when the observation time scale is 
	much larger than the characteristic time scale of the WT PDF.

	More precisely, we have the following:
	\begin{proposition}
		\label{stationarity}
		Let $\xi[ t]$ be a stochastic process with renewal with finite mean time $\tau$,  where $\tau$
		is precisely defined by Eq.~\eqref{tau}, then, if for any  $n\in\N$ the ordered times 
		$t_0\le t_1\le t_2\le....\le t_n$ are such that $t_1\gg t_{j+1} - t_{j} \gg\tau$ then all its  
		multi-time  correlation functions are asymptotically stationary (or invariant by time translations). 
	\end{proposition}

	Note that, in addition to the trivial Poissonian case, the finite mean time condition, i.e., 
	$\tau < \infty$, where $\tau$ is defined by
	Eq.~\eqref{tau} and assumed as necessary in Proposition~\ref{stationarity},
	is also satisfied for WT PDF that asymptotically
	decays as a power law, with an exponent greater than 2: $\psi(t) \sim (T/t)^{\mu}$, 
	with $\mu >2$. 
	
	Proposition~\ref{stationarity} further implies that, under coarse-grained time resolution, the
	stochastic process with renewal $\xi(t)$ behaves as a stationary process.

	In   \cite{bblmCSF196} we have shown that  if  $\tau$ is finite, then the two-time correlation function is stationary when $t_1\gg\tau$.  
	To demonstrate the more general Proposition~\ref{stationarity} we start observing that the general multi-time  correlation function 
	$\langle \xi(t_1)\xi(t_2)...\xi(t_n)\rangle$, is expressed in terms of  concatenate closed correlation functions (see 
	Eq.~\eqref{closed_corr_prod} and Proposition ~\ref{prop:main}).
	Thus, considering the composition with $p$ groups 
	$\langle  t_{1}\,t_{i_1}
	\conc t_{i_1+1}\,t_{i_2}\rblot\hspace{-4pt}\cdot... \cdot\hspace{-4pt}\lblot t_{i_{k-1}+1} t_{i_{k}}
	\conc  t_{i_{k}+1} t_{i_{k+2}}\rblot\hspace{-4pt}\cdot ...
	\cdot\hspace{-4pt}\lblot  t_{i_{p-2}+1} t_{i_{p-1}}
	\conc  t_{i_{p-1}+1} t_p\rangle$
	we have
	\begin{align}
		\label{closed_corr_stazionario_2t}
		&	... \conc t_{i_{k-1}+1} t_{i_{k}}\conc...\conc  t_{i_{p-1}+1} t_p\rangle\nonumber \\
		&=...
		\int^{t_{i_{k-1}+1}}_{u_{k-1}^{\prime}} du_{k} \tilde{R}\left(u_{k}
		-u_{k-1}^{\prime}\right) \int_{t_{i_k}}^{t_{i_k+1}} du_{k}^{\prime}\,\psi(u_{k}^{\prime}-u_{k})...
		\nonumber \\
		&	... \int^{t_{i_{p-1}+1}}_{u_{p-1}^{\prime}} du_{p} \tilde{R}\left(u_{p}
		-u_{p-1}^{\prime}\right)
		\int_{t_{n}}^{\infty} du_{p}^\prime\,
		\psi(u_{p}^\prime-u_p).
	\end{align}
	%
	As shown in~\cite{bblmCSF196}, to demonstrate the asymptotic stationarity of the term \eqref{closed_corr_stazionario_2t}, we can apply a change of 
	integration variables to make the limits of integration depend only on the time lags. This transformation reveals that, for large times, if the WT PDF $\psi(t)$ decays as $(T/t)^{\mu}$ with $\mu > 2$, which corresponds to the case where the average waiting time $\tau$ is 
	finite, then a suitable approximation for $\tilde{R}$ is given by (see~\cite{bJSTAT2020})\footnote{With an appropriate choice of $\psi(t)$, 
		Eq.~\eqref{stat.1} can be 
		exact (see~\cite{bJSTAT2020}).}:
	\begin{eqnarray}\label{stat.1} 
		\tilde{R}(t)\approx\delta(t)+\frac{1}{\tau}\left[1+\left(\frac{T}{t}\right)^{\mu-2}\right].
	\end{eqnarray}
	For sufficiently large times, this expression can be safely approximated as a constant. Under this approximation, the concatenation of closed correlation functions in Eq.~\eqref{closed_corr_stazionario_2t} reduces to a simple product of closed correlation functions.
	
	By adopting the following standard form for the power-law WT PDF:
	\begin{equation}
		\label{WTManneville}
		\psi(t)=\frac{(\mu -1)}{T}\left(\frac{T}{t+T}\right)^{\mu},
	\end{equation}
	which corresponds to an idealization of the Manneville map~\cite{abghimprsvyCSF15} and by directly integrating Eq.~\eqref{closed_corr_stazionario_2t}, starting from the last term (which extends to infinity), we obtain, after a tedious but straightforward algebraic manipulation, the following result for $t_1 \gg t_i - t_j \gg T$:
	\begin{align}
		\label{stat2}\nonumber
		&...\conc k \to (k-1) \conc t_{i_{k-1}+1}t_{i_{k}}\conc k\to (k+1)\conc...\nonumber \\
		&
		\sim...\left\{k \to (k-1) \right\}\nonumber \\
		&\times
		T^{\mu-2}
		\left[
		(t_{i_{k}}-t_{i_{k-1}+1})^{-(\mu-2) }-(t_{i_{k}+1}-t_{i_{k-1}+1})^{-(\mu-2) }\right]
		\nonumber \\
		&\times \left\{k \to (k+1) \right\}...
	\end{align}
	The expressions ${k \to (k-1)}$ and ${k \to (k+1)}$ on either side of the central term indicate the same expression as in the
	central term, but with the index $k$ replaced by $k-1$ and $k+1$, respectively.
	
	Eq.~\eqref{stat2} explicitly shows that for large time lags the concatenation of two-time closed correlation
	functiosn depends only on the  time lags and this ends the demonstration of Proposition~\ref{stationarity}.

	Beyond implying stationarity, we observe that the result in Eq.~\eqref{stat2} is analogous to 
	Eq.~\eqref{poiss:opencorr}, but it applies to power-law decays of the WT PDF rather than 
	exponential ones. However, while \textit{exponential} decay of the closed correlation function 
	leads to the asymptotic validity of the factorization property, the \textit{power-law} decay 
	described in Eq.~\eqref{stat2} indicates that this property no longer holds.
	
	In other words, and as expected, for power-law decays of $\psi(t)$, the stochastic process 
	with renewal $\xi(t)$ is not asymptotically equivalent to telegraph noise. That is, 
	Proposition~\ref{prop:1} and Lemma~\ref{lem:1} do not apply.

	Nevertheless, from the asymptotic expression \eqref{stat2} for the closed correlation function, we can still 
	deduce crucial information about the general (universal) limit behavior of the stochastic process $\xi(t)$. 
	This is, in fact, the central result of this work, to which is devoted the next section.

	\section{The universal limit behavior of the $n$-time  correlation functions for power law WT  decays.\label{sec:limt_result}}
	Given the result of Eq.~\eqref{stat2}, the following important fact is obtained:
	\begin{proposition}
		\label{prop:universal}
		If the WT  PDF exhibits a power-law decay, then, under the same assumptions and conditions
		as in   Proposition~\ref{stationarity}, the $n$-time  correlation function asymptotically 
		reduces to the two-time correlation function evaluated at the two extreme times:
		\begin{empheq}[box=\fbox]{align}
			\label{corrnt}
			\langle \xi(t_1)\xi(t_2)\dots\xi(t_n)\rangle/ \overline{\xi^n} \sim \langle t_1\,t_n\rangle=
			\langle \xi(t_1)\xi(t_n)\rangle/\overline{\xi^2}
		\end{empheq}
		which means it does not depend on the intermediate times.
	\end{proposition}
	
	By comparing Eq.~\eqref{corrnt} with the procedure described in Proposition~\ref{prop:main}
	and the corresponding Eq.~\eqref{corrGen_}, Eq.~\eqref{corrnt} implies that the dominant term
	in the sum of Eq.~\eqref{corrGen_} corresponds to the case where all times lie within the same
	laminar region ($p=1$).
	
	By combining Proposition~\ref{prop:universal} with the universal result for the two times correlation function 
	found in \cite{bblmCSF196},  and also reported in Eq.~\eqref{bblmCSF196}, we arrive at
	the following
	
	\begin{lemma}
		\label{lem:2}
		Under the conditions outlined in Proposition~\ref{prop:universal},
		the common asymptotic expression for the $n$-time 
		correlation functions for
		any stochastic process with renewal of the type considered in this paper
		(the
		noise for the L\'evy walk with random velocity), is given by Eq.~\eqref{cor2t_},
		and it depends only on the average of $\xi^n$ and the WT  PDF $\psi(u)$:
		\begin{empheq}[box=\fbox]{align}
			\label{corrnt-2t}
			\langle \xi(t_1)\xi(t_2)\dots\xi(t_n)\rangle /\overline{\xi^n}
			\to 
			\int_{t_0}^{t_1} du_1 \, \tilde{R}(u_1 - t_0) \Psi(t_n - u_1).
		\end{empheq}
	\end{lemma}

	Proposition~\ref{prop:universal}, together with Lemma~\ref{lem:2}, represents one of the main results
	of this paper.  
	\added[id=1]{The    universal property  stated by Lemma~\ref{lem:2},
		implies a corresponding universal statistical behavior of any Brownian variable with drift, 
		perturbed by renewal-type  noise   as in the model of Eq.~\eqref{SDE}, provided that $\mu>2$.}
	
	\added[id=1]{The reader should note that Proposition~\ref{prop:universal} and Lemma~\ref{lem:2} are 
		derived under the same assumptions and conditions as Proposition~\ref{stationarity}. 
		Consequently, they pertain to the stationary regime, where the initial preparation of the 
		system is irrelevant. In this context, the r.h.s. of~\eqref{corrnt-2t} may be replaced by the 
		stationary limit expression valid at large times, as given in~\cite[Eq.~(40)]{bblmCSF196}, that explicitly does not depend on $t_0$:}
	\added[id=1]{
		\begin{equation}
			\label{corr2t_aged}
				\langle \xi(t_1)\xi(t_2)\dots\xi(t_n)\rangle /\overline{\xi^n}
				\to \frac{1}{\tau}
				\int_{t_n-t_1}^{\infty} (u - t_n+t_1)\psi(u)\,du.
		\end{equation}
	}
	\added[id=1]{The right-hand side of the above limit result coincides with the expression for the two-time 
		correlation function in the dichotomous case (see, e.g.,~\cite{gnzPRL54}). However,  
		this same result now extends to any $n$-time correlation function, without requiring any 
		specific assumptions on the probability density function of the random variable $\xi$.
		.}

	To prove Proposition~\ref{prop:universal} 
	%
	we insert the result from Eq.~\eqref{stat2} into the general 
	expression for the $n$-time correlation function given in Eq.~\eqref{corrGen_}. 
	This yields a sum of products of power-law decay functions, each evaluated at the 
	time differences between pairs of intermediate times. 
	When all these time differences are much larger than $T$, the dominant 
	contribution arises from the first term in the sum-corresponding to $P = 1$,
	which involves only the two extreme times.
	An  example serves to illustrate the point:
	let us consider the case of eight-time correlation 
	function  where the $\xi$ PDF  is symmetric. This corresponds to the case of  Eq.~\eqref{corr8_}.
	According to Eq.~\eqref{corr8_}, the  partition  in which we group  all the eight times together (the case $p=1$) 
	gives the contribution
	\begin{equation}
		\label{8_part1}
		\overline{\xi^{8}} \langle t_1 t_8\rangle
	\end{equation}
	\added[id=1]{that, by using \eqref{stat2} yields}:
	\added[id=1]{
		\begin{equation}
			\label{8_part1_}
			\overline{\xi^{8}} \langle t_1 t_8\rangle
			\sim	\overline{\xi^{8}}
			T^{\mu-2}
			\left(
			t_{1}^{-(\mu-2) }-t_{8}^{-(\mu-2) }\right).
		\end{equation}
	}
	On the other hand, one of the three terms of the partition made of two groups ($p=2$) gives  the contribution 
	\begin{equation}
		\left(\overline{\xi^{4}}\right)^2\langle  t_1\, t_4
		\conc  t_5\, t_8\rangle
	\end{equation}

	\added[id=1]{	that, by using \eqref{stat2}, yields:}
		\begin{align}
			&	\left(\overline{\xi^{4}}\right)^2\langle  t_1\, t_4
			\conc  t_5\, t_8\rangle
			\sim	\left(\overline{\xi^{4}}\right)^2
			T^{2(\mu-2)}
			\left(
			t_{1}^{-(\mu-2) }-t_{4}^{-(\mu -2)}\right)
			\nonumber \\
			&\times
			\left[
			(t_{5}-t_4)^{-(\mu-2) }-(t_{8}-t_4)^{-(\mu-2) }\right].
		\end{align}
		Thus, for $t_1\ll t_2...\ll t_8$ we get
		\begin{align}
			\label{es_es}
			&\frac{\left(\overline{\xi^{4}}\right)^2}{\overline{\xi^{8}} }
			\frac{\langle  t_1\, t_4
				\conc  t_5\, t_8\rangle}{\langle t_1 t_8\rangle}
			\nonumber \\
			\sim&
			\frac{\left(\overline{\xi^{4}}\right)^2}{\overline{\xi^{8}} }
			\frac{
				T^{2(\mu-2)}
				\left(
				t_{1}^{-(\mu -2)}-t_{4}^{-(\mu-2) }\right)
				\left[
				(t_{5}-t_4)^{-(\mu-2) }-	(t_{8}-t_4)^{-(\mu-2) }\right]}
			{
				T^{\mu-2}
				\left(
				t_{1}^{-(\mu-2) }-t_{8}^{-(\mu-2) }\right)}
			\nonumber \\
			\sim	&
			\frac{\left(\overline{\xi^{4}}\right)^2}{\overline{\xi^{8}} }
			\frac{
				T^{\mu-2}
				\left(
				t_{1}^{-(\mu-2) }-0\right)
				\left[
				(t_{5}-t_4)^{-(\mu-2) }-0\right]}
			{
				\left(
				t_{1}^{-(\mu-2) }-0\right)}\nonumber \\
			=&	
			\frac{\left(\overline{\xi^{4}}\right)^2}{\overline{\xi^{8}} }
			T^{\mu-2}	(t_{5}-t_4)^{-(\mu-2) }\xrightarrow{ \Delta t\gg T } 0
		\end{align}

	\added[id=1]{
		It is important to emphasize that  in Eq.~\eqref{es_es}, the dominance of the denominator 
		over the numerator is further 
		reinforced by the coefficient  
		$\frac{\left(\overline{\xi^{4}}\right)^2}{\overline{\xi^{8}} }$. 
		In fact, by  H\"older's inequality
		we have }
	\begin{equation}
		\label{holder}
		\added[id=1]{			\frac{\left|\overline{\xi^{m_1}}\right|\,\left|\overline{\xi^{m_2}}\right|\dots\left|\overline{\xi^{m_p}}\right|}
			{\overline{\xi^n}}\le 1}
	\end{equation}
	\added[id=1]{	with $\sum_{i=1}^p m_i=n$, where the  equality holds only for the symmetric dichotomous 
		case. Moreover, the left-hand side of \eqref{holder} becomes smaller the heavier the tails of the PDF.
		As an illustration, consider the case of a  Gaussian PDF with unitary variance.  
		For $n=6$, the coefficients associated with the contributions $p=1,2,3$ in Eq.~\eqref{corr6_}
		are: 
		$$\overline{\xi^6} = 15 \quad (p=1), \qquad 
		\overline{\xi^4}\,\overline{\xi^2} = 3 \quad (p=2), \qquad 
		\left(\overline{\xi^2}\right)^3 = 1 \quad (p=3).$$
		This illustrates that even in the light-tailed Gaussian case, the $p=1$ term prevails.}
	
	\added[id=1]{Extending this argument to the general case of Eq.~\eqref{corrGen_}, we observe that, 
		in the summation of Eq.~\eqref{corrGen_}, the coefficients 
		$\left[(\overline{\xi^{m_1}})(\overline{\xi^{m_2}})\dots(\overline{\xi^{m_p}})\right]$, 
		with $\sum_{i=1}^p m_i = n$, are typically dominated by the case $p=1$. 
		That is, the leading contribution comes from the single-term coefficient $\overline{\xi^{n}}$.}

	\added[id=1]{In the case of a power-law PDF for $\xi$, i.e.\ $p(\xi) \sim \xi^{-\beta}$, 
		it is straightforward to verify that the left-hand side of \eqref{holder} 
		decays increasingly rapidly as $n$ grows. 
		Consequently, the convergence expressed by Eq.~\eqref{corrnt-2t} 
		is further strengthened for larger values of $n$.
	}

	\added[id=1]{The dominance of the partition with a single group ($p=1$) in Eq.~\eqref{corrGen_} 
		becomes particularly evident when $n > \beta - 1$. 
		In this case, the $n$-th moment of $\xi$ does not exist (it diverges), and 
		Eq.~\eqref{corrnt} formally loses its meaning. 
		However, in practical applications, averages are computed over a finite number $N$ of realizations of $\xi$, 
		corresponding to a finite initial ensemble and a finite trajectory length (i.e., a finite number of transitions per trajectory). 
		As a result, the average of $\xi^n$ appears large (and increases with $N$), but remains finite.}
	
	This leads to a more general result than Proposition~\ref{prop:universal}. 
	Indeed, in such cases, the convergence of the $n$-time correlation function to the two-time correlation function 
	occurs independently of $\mu$, and in fact holds regardless of the form of the waiting time probability density function (WT PDF).

	More precisely we have the following:
	\begin{proposition}
		\label{prop:PDF_power}
		In the case of power-law PDF for the random variable $\xi$, i.e.,  $p(\xi) \sim \xi^{-\beta}$, 
		and   $\beta < n+1,$ and $n$ a given integer, let us redefine $\overline{\xi^m}$ for $m \leq n$ as 
		the empirical average of $\xi^m$ computed over a large but finite number $N$ of instances. 
		In this formulation, $\overline{\xi^n}$ increases with $N$, and for any fixed $n$, 
		the convergence described by Eq.~\eqref{corrnt-2t} is achieved simply by increasing $N$, 
		regardless of the value of the time lags.
	\end{proposition}   
	Proposition \ref{prop:PDF_power},  very well supported by the numerical simulations, is a  strong statement, 
	implying the universal behavior for
	all the $n$-time correlation functions, of any stochastic process with renewal, regardless
	of the time lags, provided that the $\xi$ PDF  decays as a power law with    $n>\beta -1$.  
	
	The demonstration of
	this proposition is trivial: it follows directly, by inspection,  from the  general 
	expression for the $n$-time correlation function given in Eq.~\eqref{corrGen_}. In fact, the
	coefficient of the partition with $p=1$, is just  $\overline{\xi^n}$, and, increasing $N$,  
	becomes dominant respect to the coefficients of the other partitions.  
	
	\added[id=1]{The reader should note that, whereas in Proposition~\ref{prop:universal} the initial condition 
		(i.e., the preparation of the ensemble) is irrelevant, since that proposition pertains to the case $\mu > 2$, 
		where the system's statistics become asymptotically stationary, Proposition~\ref{prop:PDF_power} 
		also applies to situations in which the WT PDF decays very slowly, with $1 < \mu < 2$. 
		In this regime, the system cannot attain a stationary state, and the aging time diverges. 
		Consequently, for a generic WT PDF, the convergence described by Proposition~\ref{prop:PDF_power} 
		is more accurately expressed as}
	\added[id=1]{
		\begin{empheq}[box=\fbox]{align}
			\label{corrnt-2t_power}
			\langle \xi(t_1)\xi(t_2)\dots\xi(t_n)\rangle 
			\xrightarrow{N \text{large}}
			\overline{\xi_0^n} \,
			\Psi(t_n - t_0) + 
			\overline{\xi^n} \int_{t_0}^{t_n} du_1 \, {R}(u_1 - t_0) \Psi(t_n - u_1).
		\end{empheq}
	}

	\section{Comparison with numerical simulations.\label{sec:numerics}}
	In this section, we use numerical simulations of various stochastic processes 
	with renewal to verify
	Proposition~\ref{prop:1}/Lemma~\ref{lem:1},  Proposition~\ref{prop:universal}/Lemma~\ref{lem:2} and Proposition~\ref{prop:PDF_power}. 
	
	All codes were written in Fortran 90. \added[id=3]{The random number generator used is RAN2~\cite{press_etal:1992} from Numerical Recipes. Most simulations were done on multicore 
		machines using the OPENMPI library.
		When not indicated otherwise, correlation functions were
		obtained averaging $24 \times 4 \times 10^7 = 9.6 \times 10^8$ stochastic trajectories.
		In practice, for each parameter and PDF considered, 24 statistically independent correlation functions were computed, each one obtained averaging $4 \times 10^7$ trajectories. 
		The final correlation function shown in the different figures was obtained as the average of these 24 correlation functions, with an associated error on the average shown as error bars
		in some figures.}
	We adopt the notation $\Phi^{(n)}(t_1,\dots,t_n) := \langle \xi(t_1)\dots\xi(t_n)\rangle$, 
	where the initial time $t_0$ is always set to zero and it is therefore not explicitly indicated.
	
	As in  \cite{bblmCSF196}, 
	we consider four different PDF's for the variable $\xi$, all with unit variance:
	\begin{enumerate}
		\item
		symmetric two state PDF (dichotomous case):
		\begin{equation}
			\label{dichotomous}
			p(\xi)=\frac{1}{2}\delta(\xi+1)+\frac{1}{2}\delta(\xi-1);
		\end{equation}
		\item 
		Normal PDF:
		\begin{equation}\label{pNormal}
			p(\xi) = \frac{1}{\sqrt{2\pi}} e^{-\frac{\xi^2}{2}};
		\end{equation}
		\item
		flat PDF
		\begin{equation}\label{pFlat}
			p(\xi) =\frac{1}{2\sqrt{3}} \Theta(\sqrt{3}- \xi)\Theta(\xi+\sqrt{3});
		\end{equation}
		\item
		power law decaying PDF:
		\begin{equation}
			\label{pPowerLaw}
			p(\xi)=\frac{\sqrt{2}}{\pi}\frac{1}{\left(\xi ^4+1\right)}.
		\end{equation}
	\end{enumerate} 
	\begin{figure*}[htb!]  
		\includegraphics[width=\textwidth]{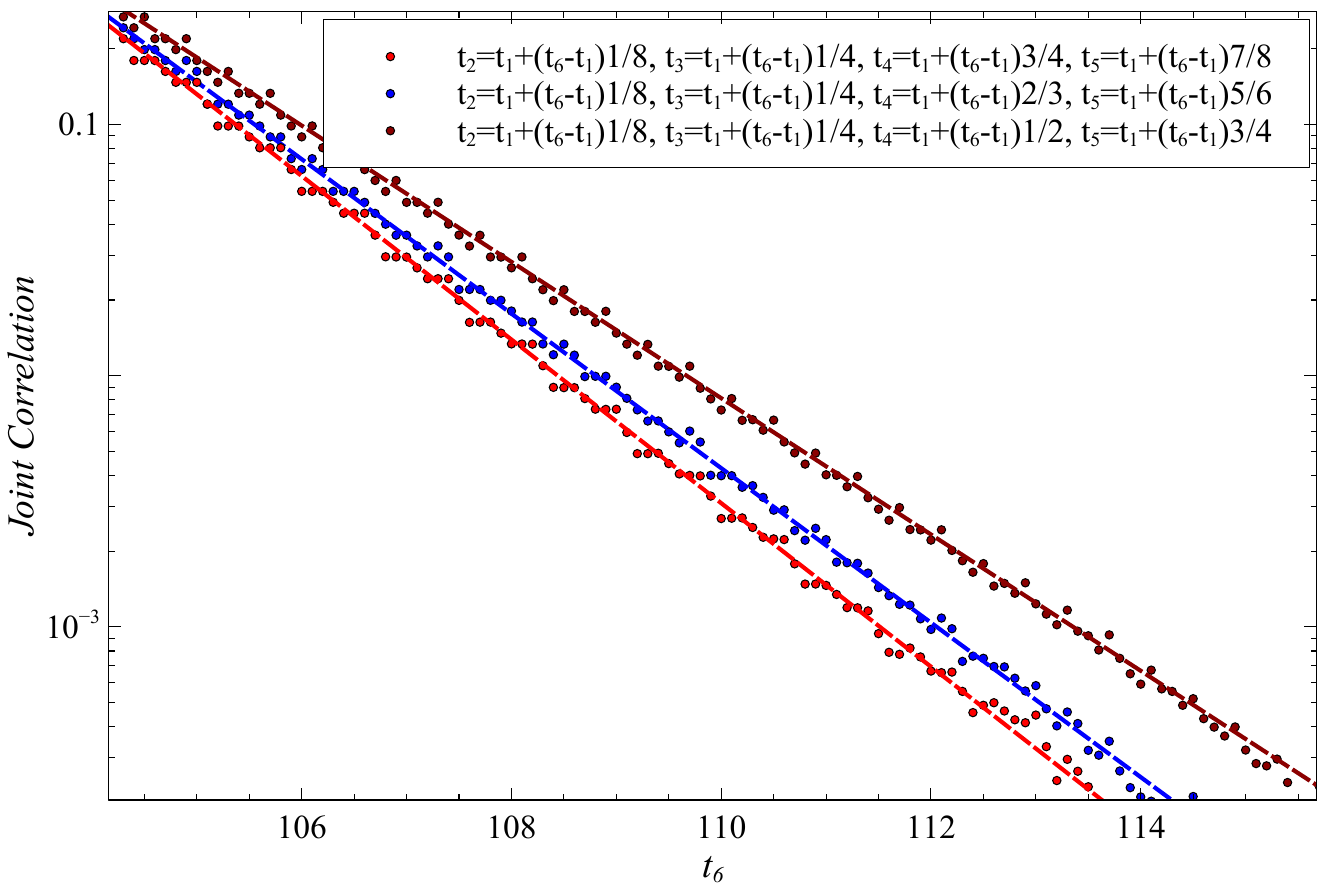}
		\caption{
			Log-plots of the 6-time correlation function for $t_1 = 100$, in the case of exponential 
			waiting times (i.e., a Poisson process) with $\tau = 1$, for the PDF of Eq.~\eqref{dichotomous} (dichotomous case). 
			This figure is meant as an example of the tests carried out to validate the numerical simulations.
			Dots represent the results of numerical simulations, while dashed lines correspond to the factorized expression 
			$\exp[-(t_2 - t_1)/\tau] \exp[-(t_4 - t_3)/\tau] \exp[-(t_6 - t_5)/\tau]$, which is exact in this case. 
			Different colors show correlations computed considering different intermediate time values, 
			as indicated in the legend. The agreement between simulations and theory is excellent.
		}
		\label{mo6dice_T1_t1_100}
	\end{figure*}
	
	\begin{figure*}[h!]  
		\centering
		\includegraphics[width=\textwidth]{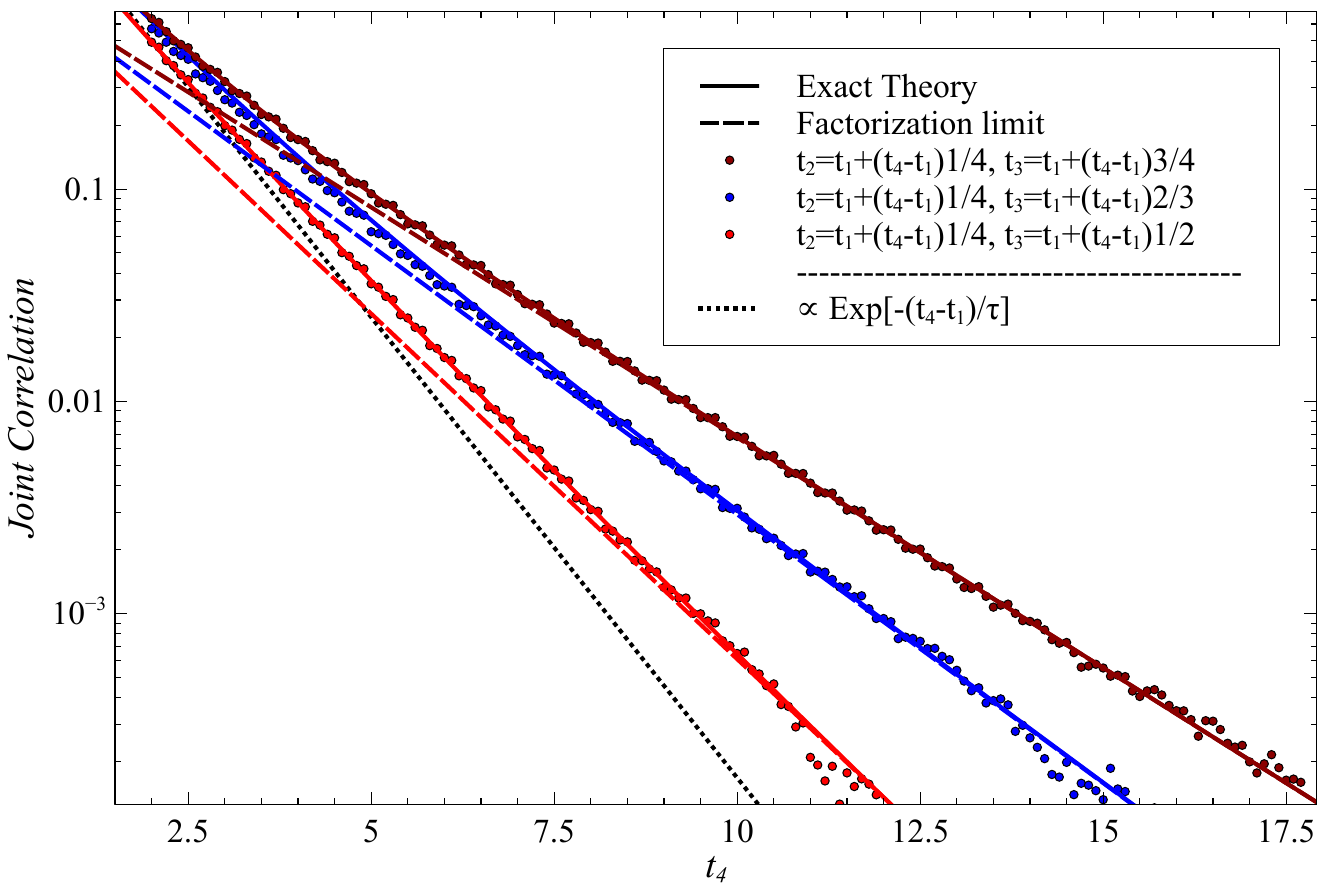}
		\caption{
			Log-plots of the 4-time correlation function for $t_1 = 0$, in the case of exponential waiting times with $\tau = 1$, for the PDF of Eq.~\eqref{pNormal} (Normal PDF). 
			Dots represent the results of numerical simulations.
			Solid lines, which are nearly indistinguishable from the numerical simulations, represent the exact theoretical 
			result obtained by inserting Eq.~\eqref{pois} into Eq.~\eqref{corr4_2}. 
			Dashed lines are the
			functions 
			$\exp[-(t_2 - t_1)/\tau] \exp[-(t_4 - t_3)/\tau]$ and represent the factorization limit:
			asymptotically, they tend to the exact theoretical results.
			The dotted line shows the 2-time correlation function, illustrating the so-called “universal limit result” which, however, fails in Poissonian cases when the PDF lacks heavy tails.
		}
		\label{gaue_T1_t1_0}
	\end{figure*}
	\begin{figure*}[h!]   
		\includegraphics[width=\textwidth]{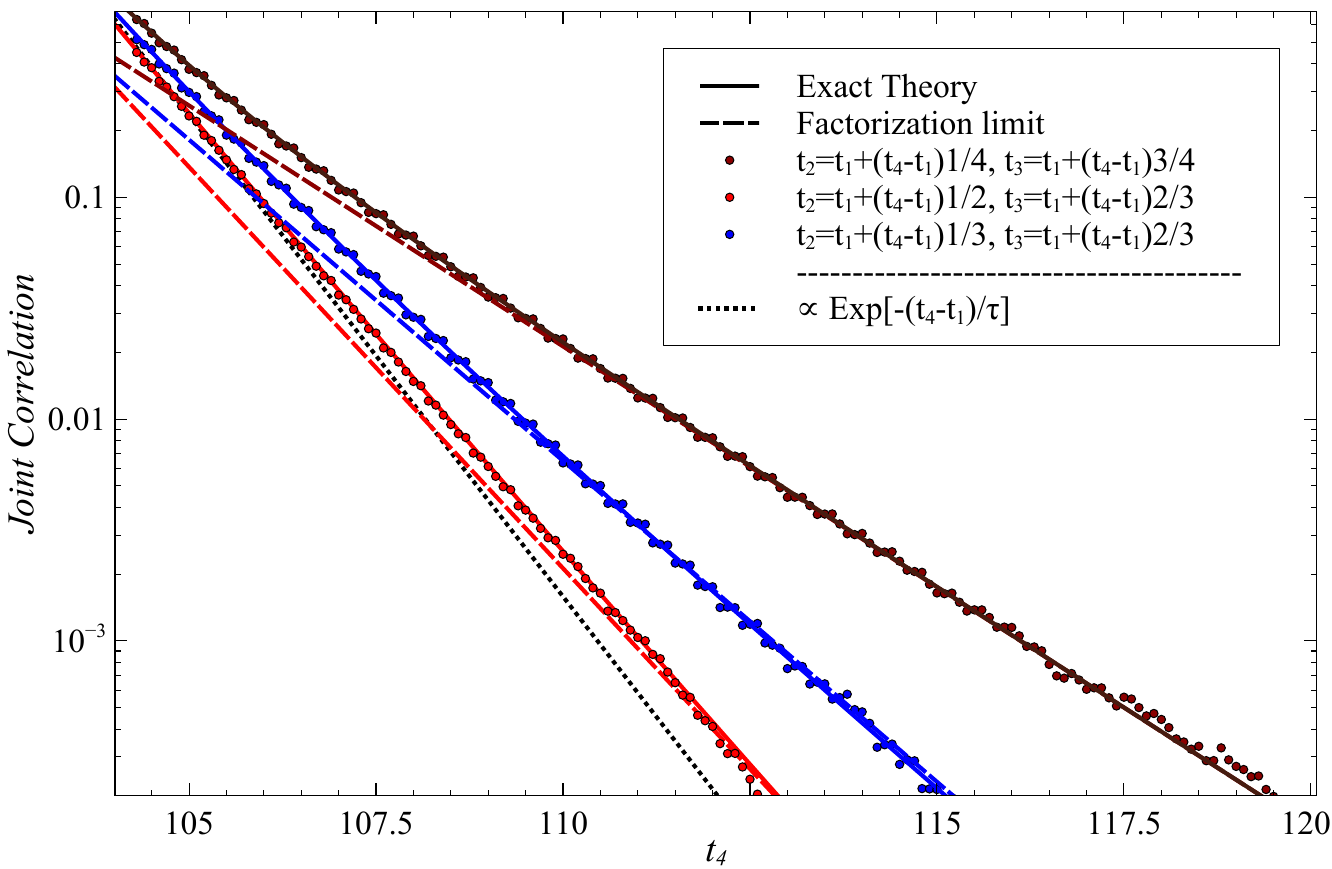}
		\caption{
			Log-plots of the 4-time correlation function and $t_1 = 100$, in the case of exponential 
			waiting times with $\tau = 1$, for the PDF of Eq.~\eqref{pNormal} (Normal PDF). 
			Dots represent the results of numerical simulations.
			Solid lines, which are nearly indistinguishable from the numerical simulations, represent the exact theoretical 
			result obtained by inserting Eq.~\eqref{pois} into Eq.~\eqref{corr4_2}. 
			Dashed lines are the
			functions 
			$\exp[-(t_2 - t_1)/\tau] \exp[-(t_4 - t_3)/\tau]$ and represent the factorization limit:
			asymptotically, they tend to the exact theoretical results.
			The dotted line shows the 2-time correlation function, illustrating the so-called “universal limit result” which, however, fails in Poissonian cases when the PDF lacks heavy tails.}
		\label{gaue_T1_t1_100}
	\end{figure*}
	\begin{figure*}[h!]  
		\includegraphics[width=\textwidth]{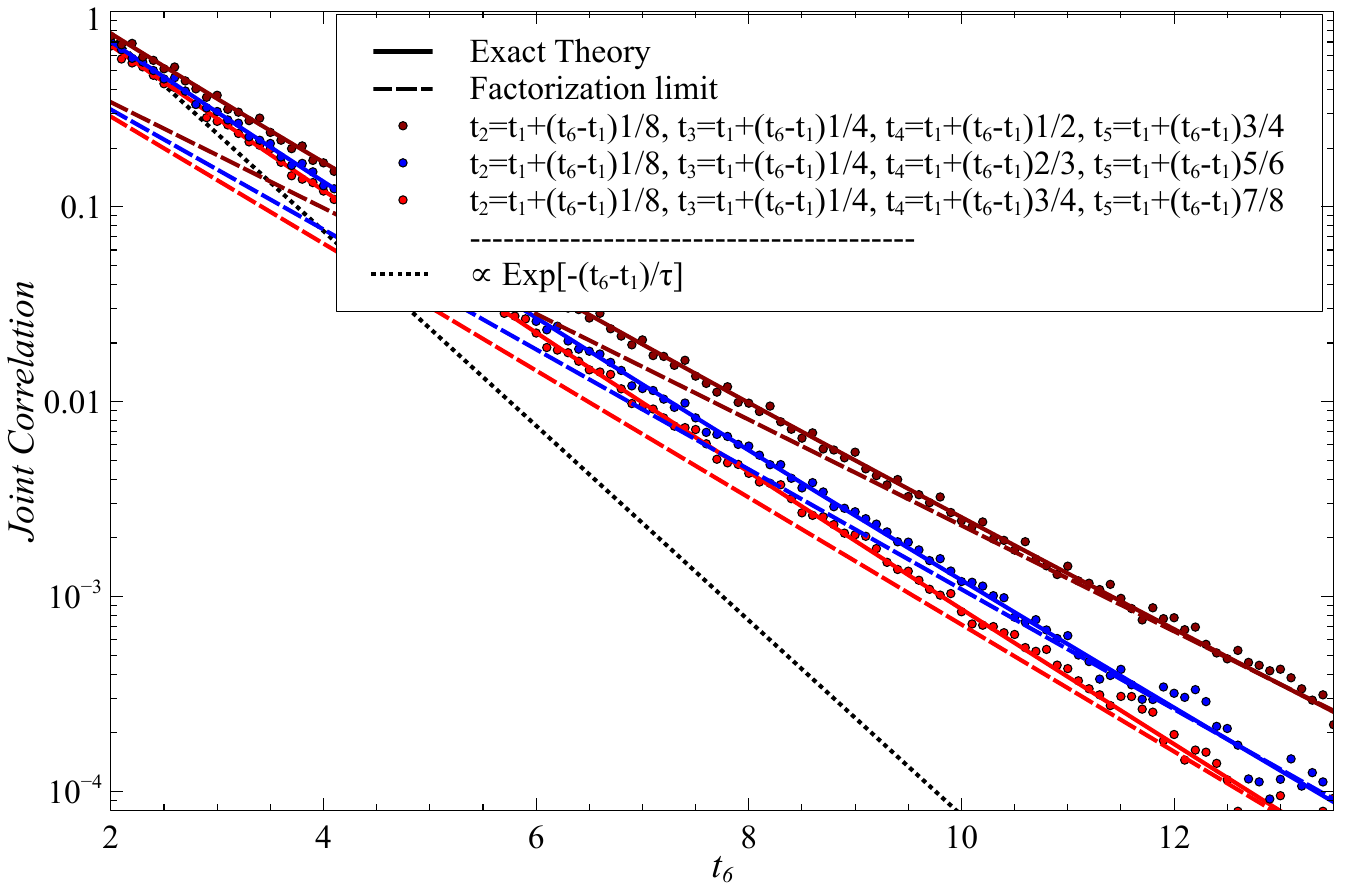}
		\caption{
			Log-plots of the 6-time correlation function and $t_1 = 0$, in the case of exponential 
			waiting times with $\tau = 1$, for the PDF of Eq.~\eqref{pFlat} (Flat PDF). 
			Dots represent the results of numerical simulations.
			Solid lines, which are nearly indistinguishable from the numerical simulations, represent the exact theoretical 
			result obtained by inserting Eq.~\eqref{pois} into Eq.~\eqref{corr6_}. 
			Dashed lines are the
			functions $\exp[-(t_2-t_1)/\tau]\exp[-(t_4-t_3)/\tau]\exp[-(t_6-t_5)/\tau]$ and represent the factorization limit:
			asymptotically, they tend to the exact theoretical results.
			The dotted line shows 2-time correlation function, illustrating the so-called “universal limit result” which, however, fails in Poissonian cases when the PDF lacks heavy tails.}
		\label{mo6flae_T1_t1_0}
	\end{figure*}
	\begin{figure*}[h!]  
		\includegraphics[width=\textwidth]{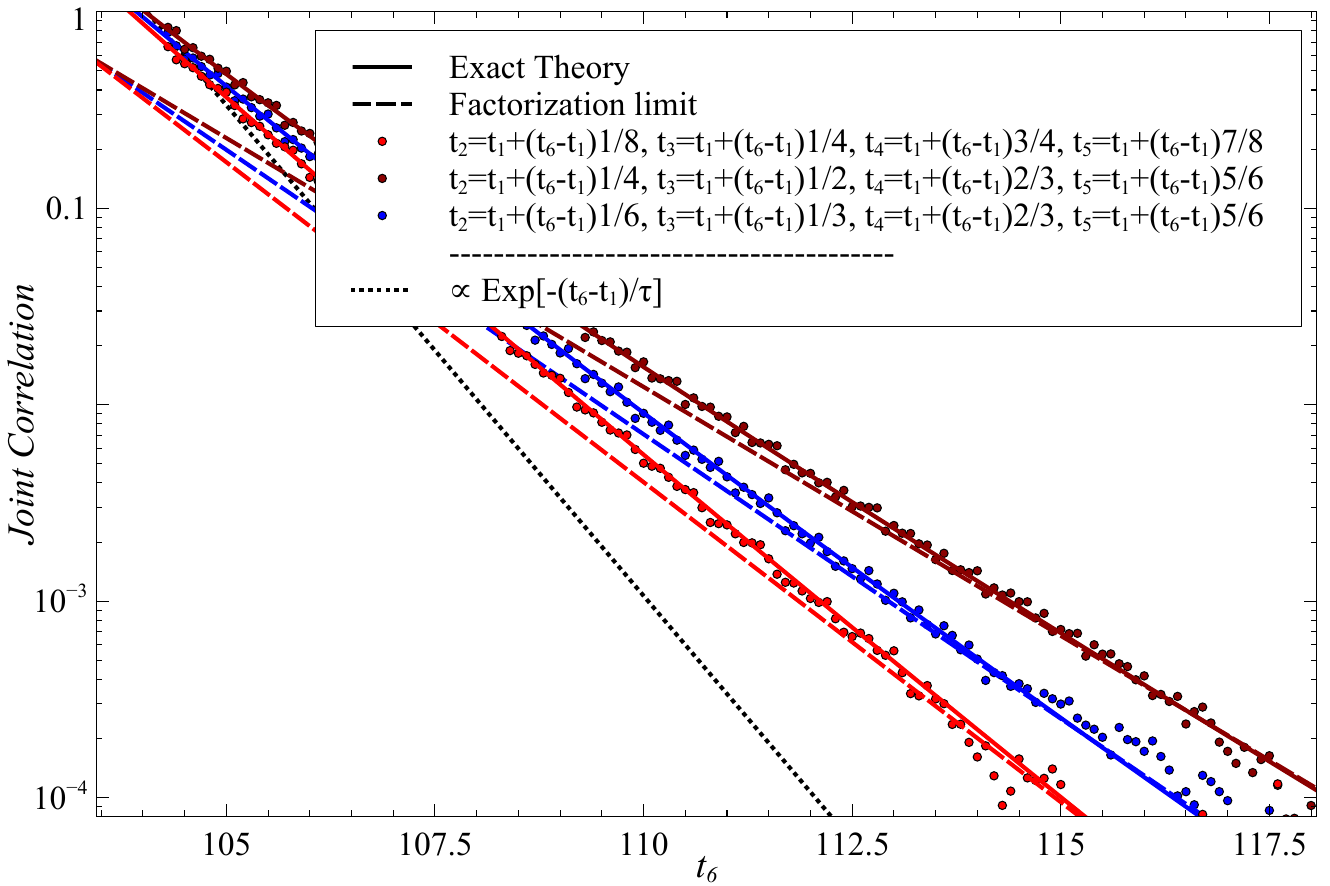}
		\caption{
			Log-plots of the 6-time correlation function for $t_1 = 100$, in the case of exponential 
			waiting times with $\tau = 1$, for the PDF of Eq.~\eqref{pFlat} (Flat PDF). 
			Dots represent the results of numerical simulations.
			Solid lines, which are nearly indistinguishable from the numerical simulations, represent the exact theoretical 
			result obtained by inserting Eq.~\eqref{pois} into Eq.~\eqref{corr6_}. 
			Dashed lines are the
			functions $\exp[-(t_2-t_1)/\tau]\exp[-(t_4-t_3)/\tau]\exp[-(t_6-t_5)/\tau]$ and represent the factorization limit:
			asymptotically, they tend to the exact theoretical results.
			The dotted line shows the 2-time correlation function, illustrating the so-called “universal limit result” which, however, fails in Poissonian cases when the PDF lacks heavy tails.}
		\label{mo6flae_T1_t1_100}
	\end{figure*}
	\begin{figure*}[h!]   
		\includegraphics[width=\textwidth]{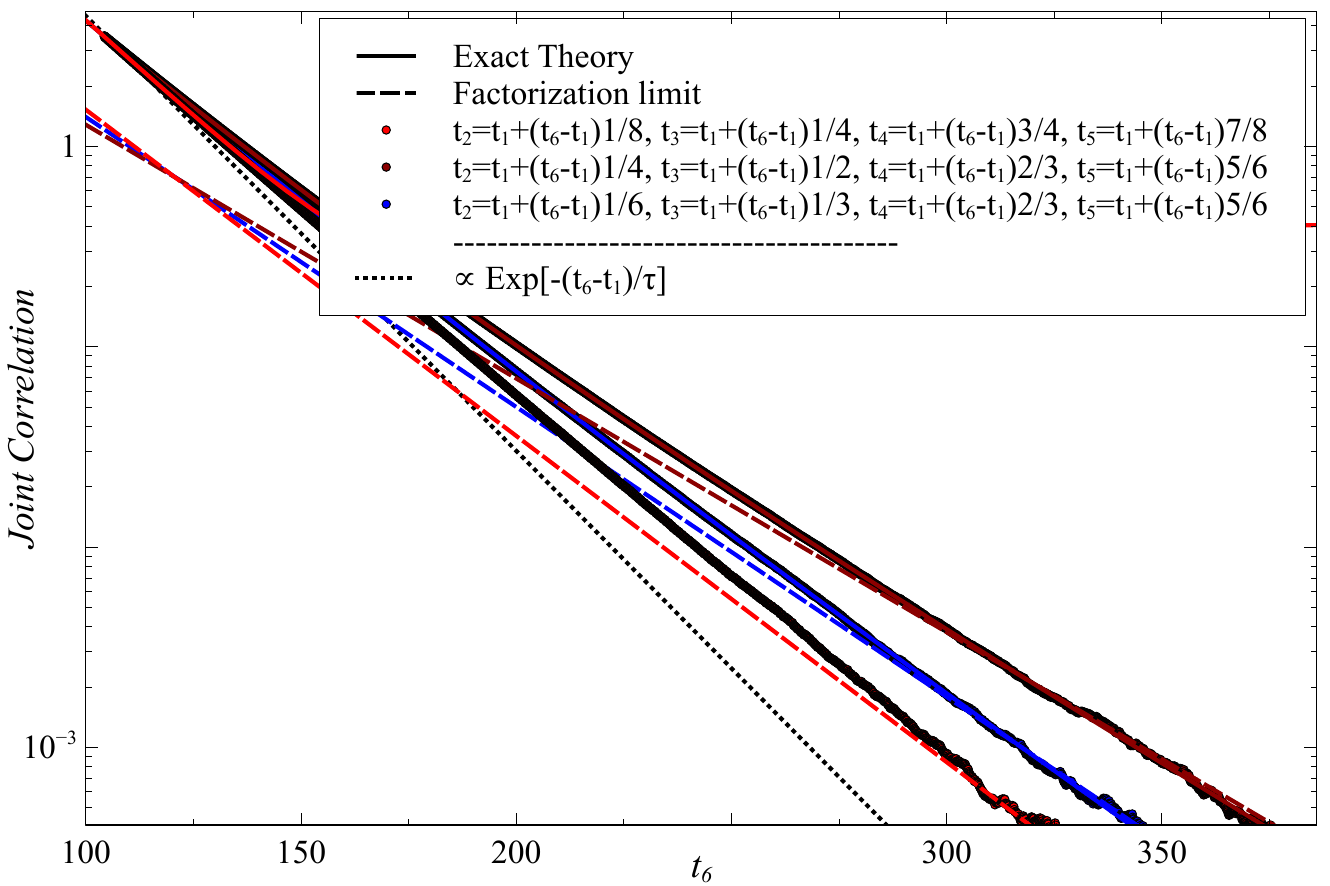}
		\caption{
			Log-plots of the 6-time correlation function for $t_1 = 100$, in the case of exponential 
			waiting times with $\tau = 20$, for the PDF of Eq.~\eqref{pFlat} (Flat PDF). 
			Dots represent the results of numerical simulations.
			Solid lines, which are nearly indistinguishable from the numerical simulations, represent the exact theoretical 
			result obtained by inserting Eq.~\eqref{pois} into Eq.~\eqref{corr6_}. 
			Dashed lines are the
			functions $\exp[-(t_2-t_1)/\tau]\exp[-(t_4-t_3)/\tau]\exp[-(t_6-t_5)/\tau]$ and represent the factorization limit:
			asymptotically, they tend to the exact theoretical results.
			The dotted line shows the 2-time correlation function, illustrating the so-called “universal limit result” which, however, fails in Poissonian cases when the PDF lacks heavy tails.}
		\label{mo6flae_T20_t1_100}
	\end{figure*}

	\begin{figure*}[h!]  
		\includegraphics[width=\textwidth]{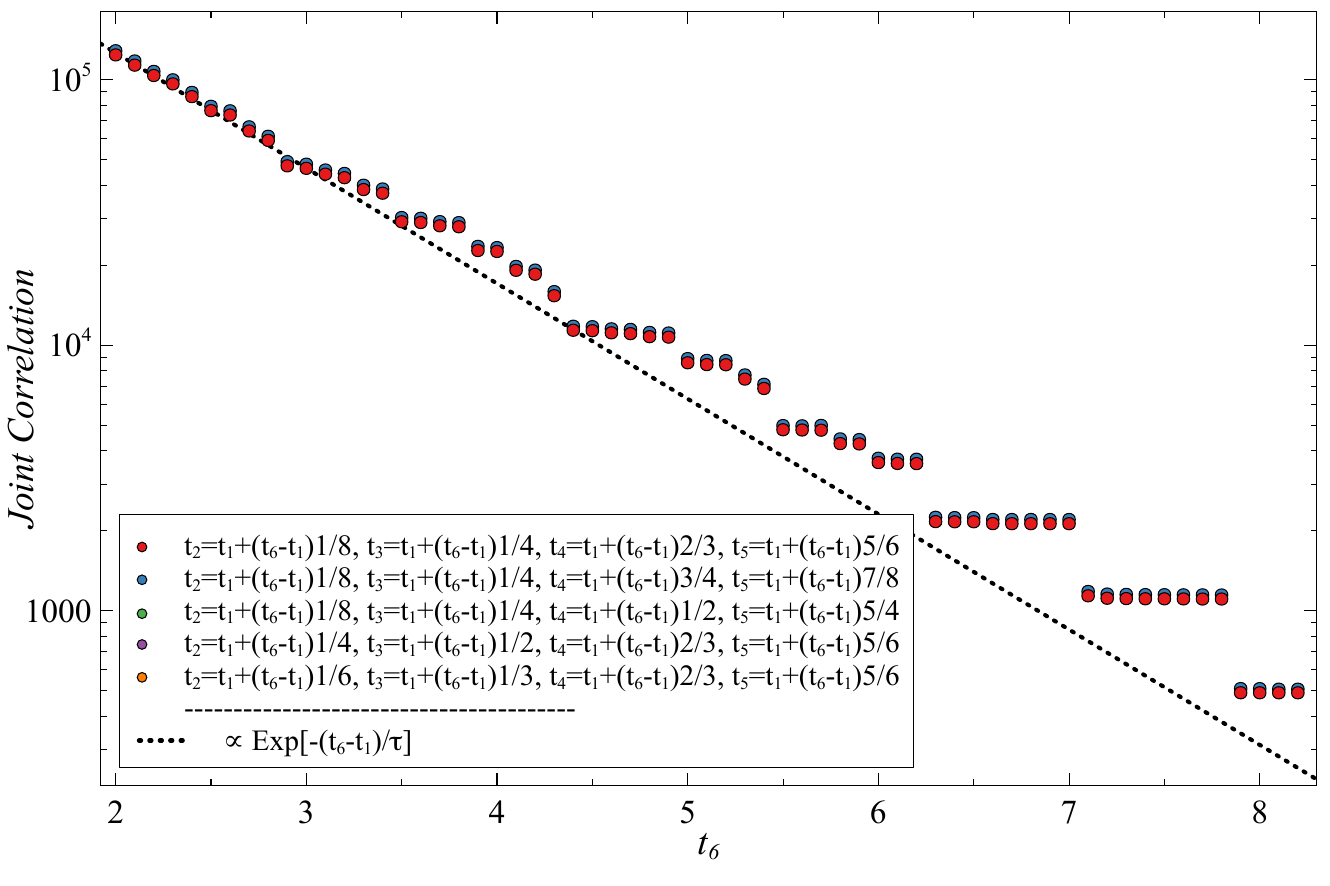}
		\caption{Log-plots of the 6-time correlation function for $t_1 = 0$, in the case of exponential 
			waiting times with $\tau = 1$, 
			for the PDF of  Eq.~\eqref{pPowerLaw} (power law PDF). Dots are
			the result of numerical simulations, for different intermediate times.
			The multi-time correlation function no longer depends on the intermediate times (dots corresponding to the different cases overlap) and it is well fitted by the universal 2-time result (the dotted line). The jump-like structure observed in the simulations is due to 
			relatively poor statistics, which is unavoidable given the divergent nature of this moment. 
			In practice, the correlation is dominated by a few trajectories in which $\xi$ is drawn 
			from the tails of the distribution. See also Fig.~\protect\ref{mo6gene_T20-t1_100} below
			where the error bars have been estimated.}
		\label{mo6gene_T1_t1_0}
	\end{figure*}
	\begin{figure*}[h!]  
		\includegraphics[width=\textwidth]{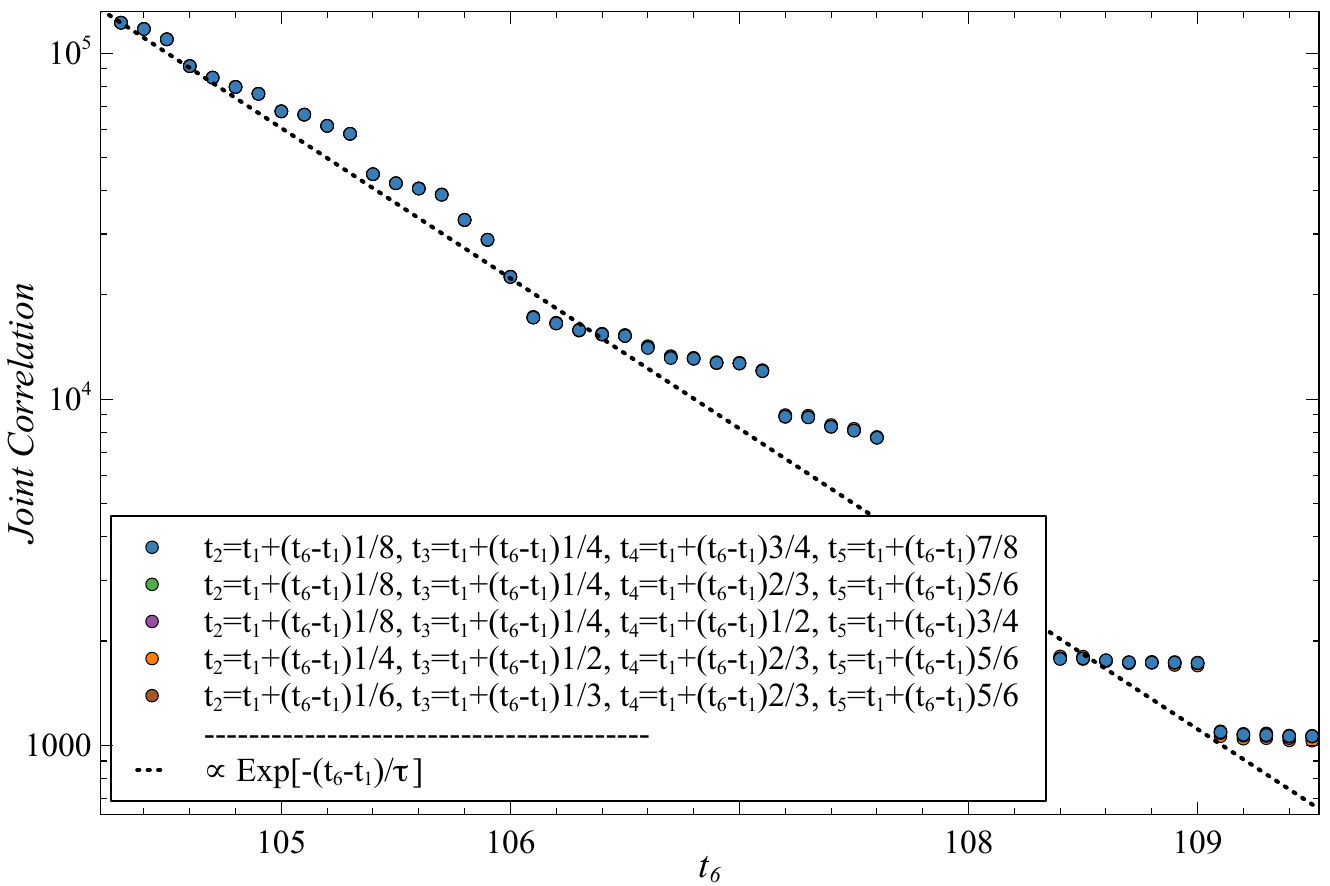}
		\caption{Log-plots of the 6-time correlation function for $t_1 = 100$, in the case of exponential 
			waiting times with $\tau = 1$,
			for the PDF of Eq.~\eqref{pPowerLaw} (power law PDF). Dots are
			the result of numerical simulations, for different intermediate times.
			The multi-time correlation function no longer depends on the intermediate times (dots corresponding to the different cases overlap) and it is well fitted by the universal 2-time result (the dotted line). The jump-like structure observed in the simulations is due to 
			relatively poor statistics, which is unavoidable given the divergent nature of this moment. 
			In practice, the correlation is dominated by a few trajectories in which $\xi$ is drawn 
			from the tails of the distribution. See also Fig.~\protect\ref{mo6gene_T20-t1_100} below
			where the error bars have been estimated.}
		\label{mo6gene_T1_t1_100}
	\end{figure*}
	\begin{figure*}[h!]   
		\includegraphics[width=\textwidth]{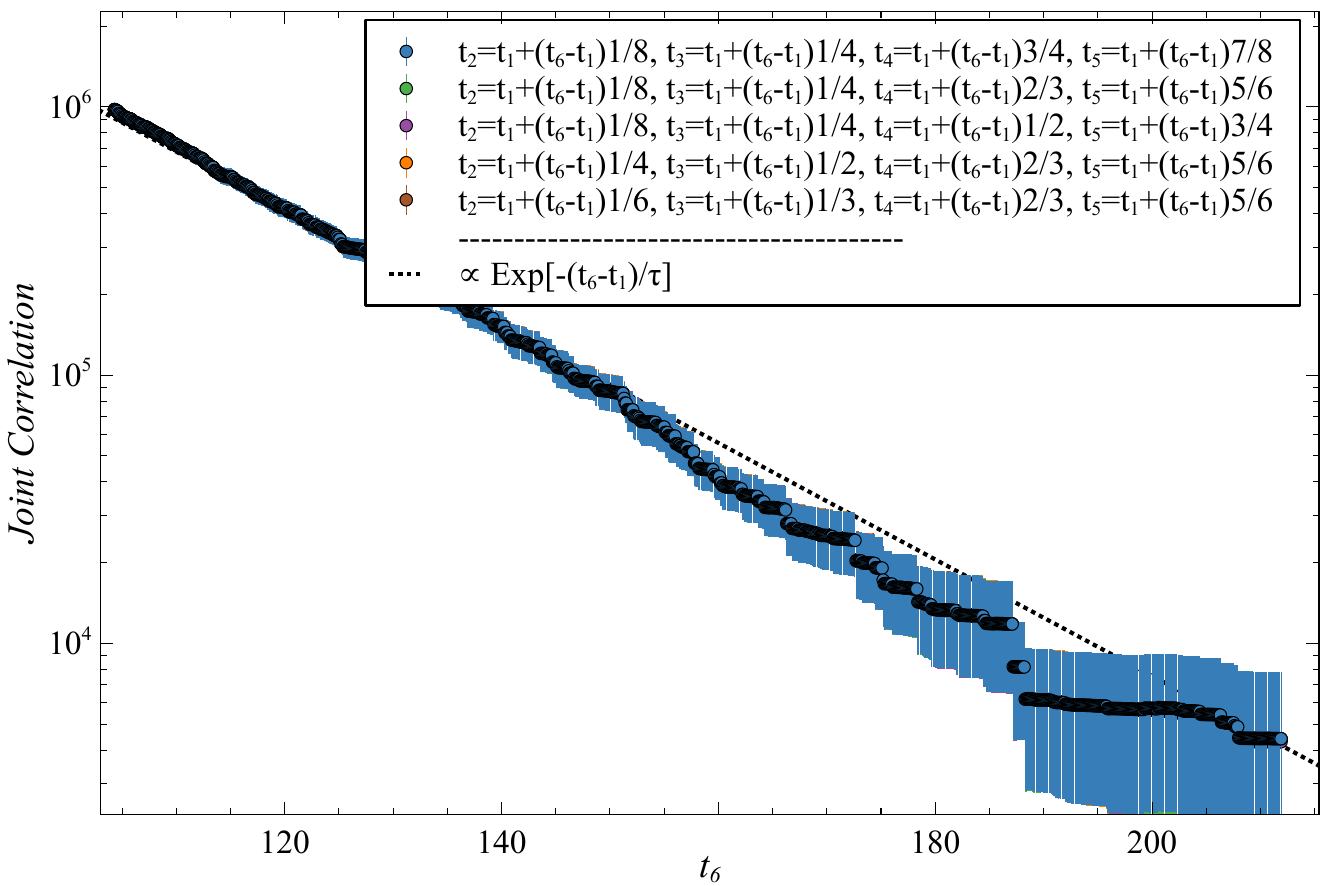}
		\caption{Log-plots of the 6-time correlation function for $t_1 = 100$, in the case of exponential 
			waiting times with $\tau = 20$,
			for the PDF of  Eq.~\eqref{pPowerLaw} (power law PDF). Dots are
			the result of numerical simulations, for different intermediate times.
			The multi-time correlation function no longer depends on the intermediate times (dots corresponding to the different cases overlap) and it is well fitted by the universal 2-time result (the dotted line). The jump-like structure observed in the simulations is due to 
			relatively poor statistics, which is unavoidable given the divergent nature of this moment. 
			In practice, the correlation is dominated by a few trajectories in which $\xi$ is drawn 
			from the tails of the distribution.  The bars are an estimate of the statistical errors. see text for details on their calculation.}
		\label{mo6gene_T20-t1_100}
	\end{figure*}

	\begin{figure*}[h!]   
		\includegraphics[width=\textwidth]{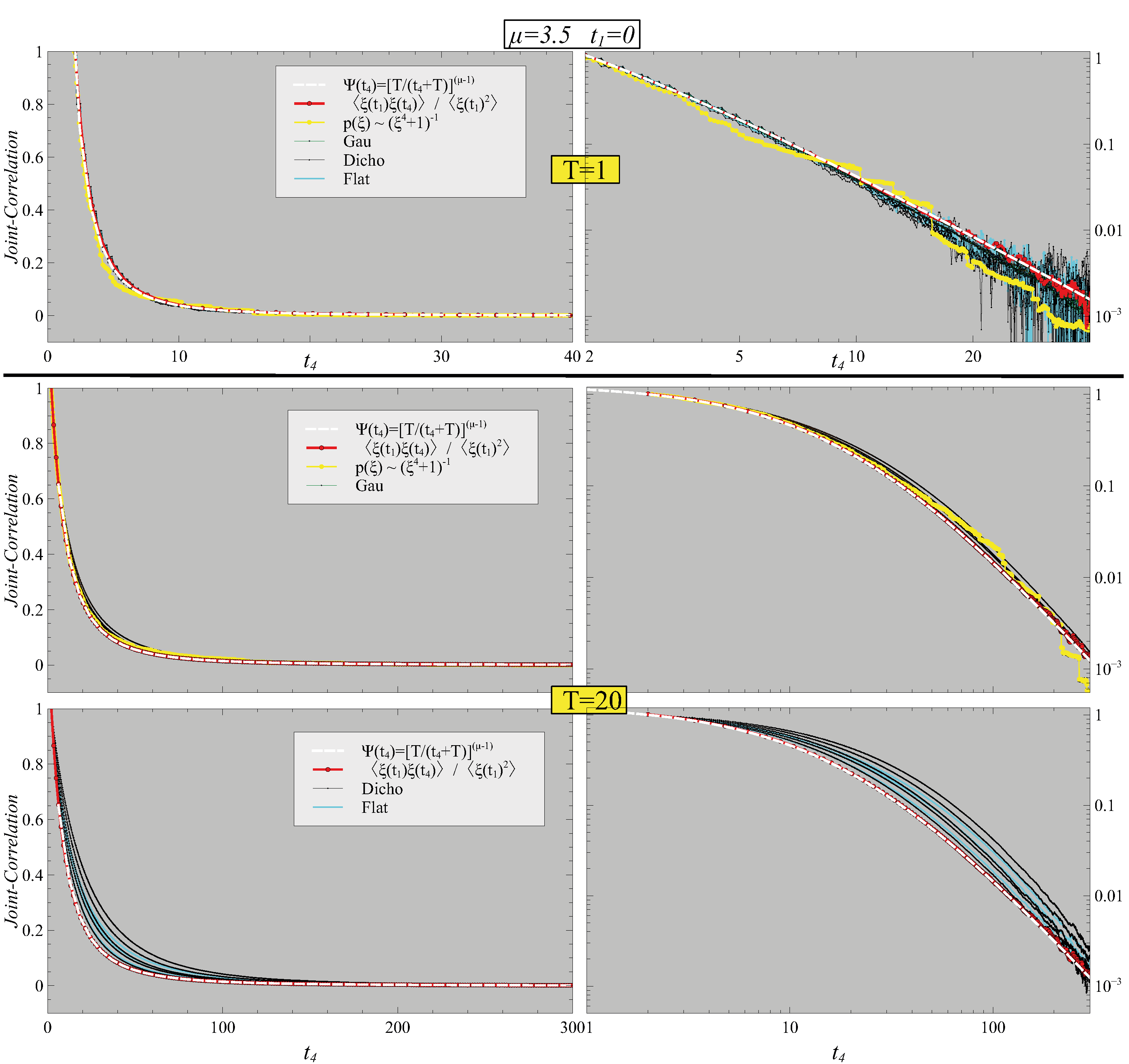}
		\caption{Comparison between the normalized universal 2-time correlation
			function (thick solid red line with small circles) from simulations and the normalized 
			4-time 
			correlation functions for  ${t_1=0}$ (thin colored lines with small circles)  from simulations done for different $\xi$  PDFs, as per the text box. The WT PDF is given by Eq.~\eqref{WTManneville},
			with ${\mu=3.5}$,  and  $T=1$, (first row) and $T=20$
			(second and third rows).
			The theoretical result for the universal 2-time correlation
			function is plotted as a dashed white line. Left panels: linear scale. Right panels:  Log-Log scale. For each $\xi$  PDF considered, six curves are plotted: the ones relative to a
			Normal and a power law PDFs collapse on a unique curve, whereas the ones relative to 
			dicothomous and flat PDFs show some spreading: see the text for a detailed explanation.
		}
		\label{Tutti_mu3p5_T_1-20_t1_0}
	\end{figure*}
	\begin{figure*}   
		\centering
		\includegraphics[width=\textwidth]{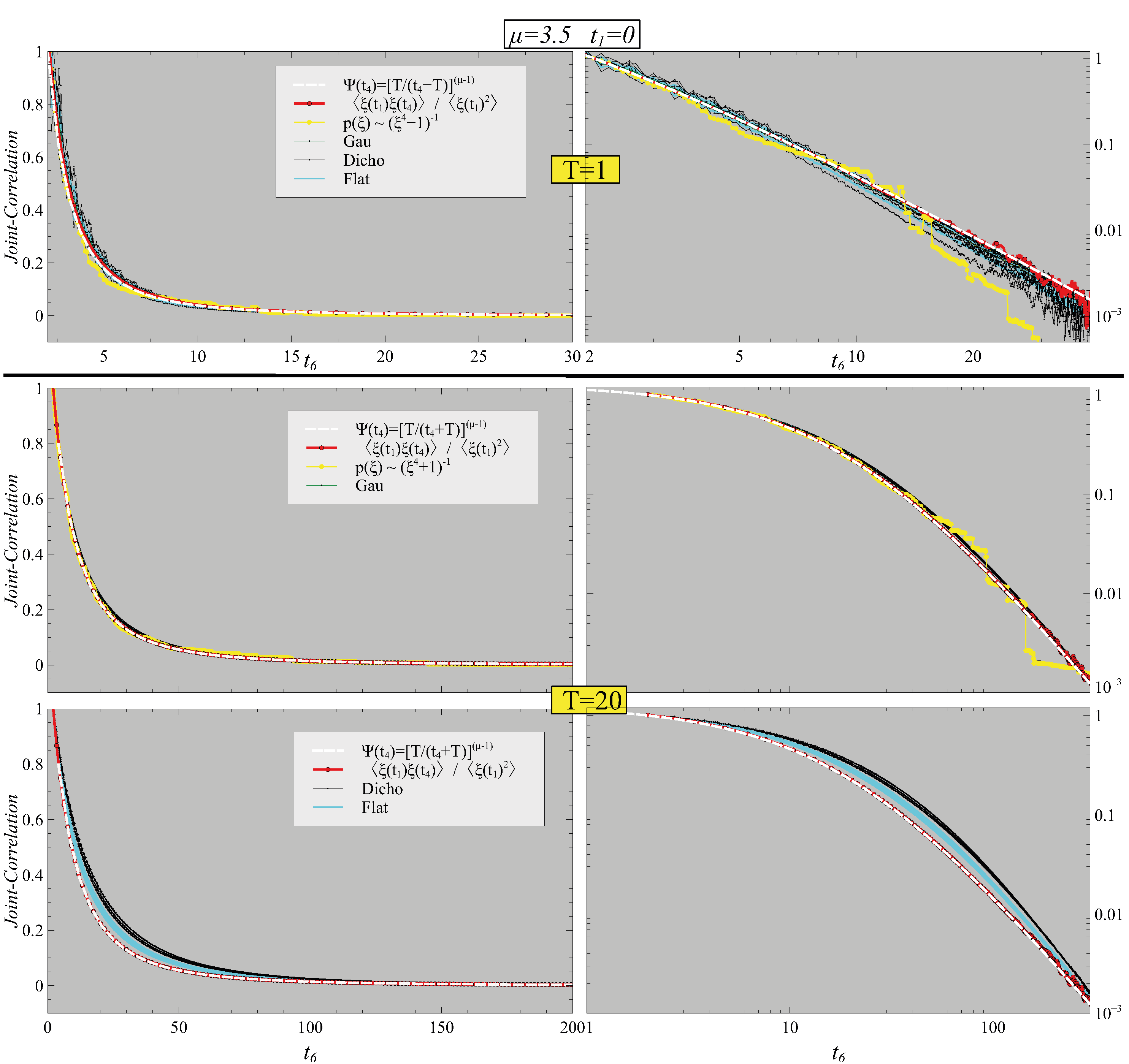}
		\caption{As in Fig.~\ref{Tutti_mu3p5_T_1-20_t1_0} but for $n=6$.
			Comparison between the normalized universal 2-time correlation
			function (thick solid red line with small circles) from simulations and the normalized 
			6-time 
			correlation functions for  ${t_1=0}$ (thin colored lines with small circles)  from simulations done for different $\xi$  PDFs, as per the text box. The WT PDF is given by Eq.~\eqref{WTManneville},
			with ${\mu=3.5}$,  and  $T=1$, (first row) and $T=20$
			(second and third rows). 
			The theoretical result for the universal 2-time correlation
			function is plotted as a dashed white line. Left panels: linear scale. Right panels:  Log-Log scale. For each $\xi$  PDF considered, six curves are plotted: the ones relative to a
			Normal and a power law PDFs collapse on a unique curve;    
			note
			that the spreading of the curves relative to the dichotomous and the flat PDS cases
			is now much reduced with respect to the case of Fig.~\ref{Tutti_mu3p5_T_1-20_t1_0}: see
			the text for a detailed explanation.}
		\label{Tutti6_mu3p5_T_1-20_t1_0}
	\end{figure*}
	\begin{figure*}  
		\centering
		\includegraphics[width=\textwidth]{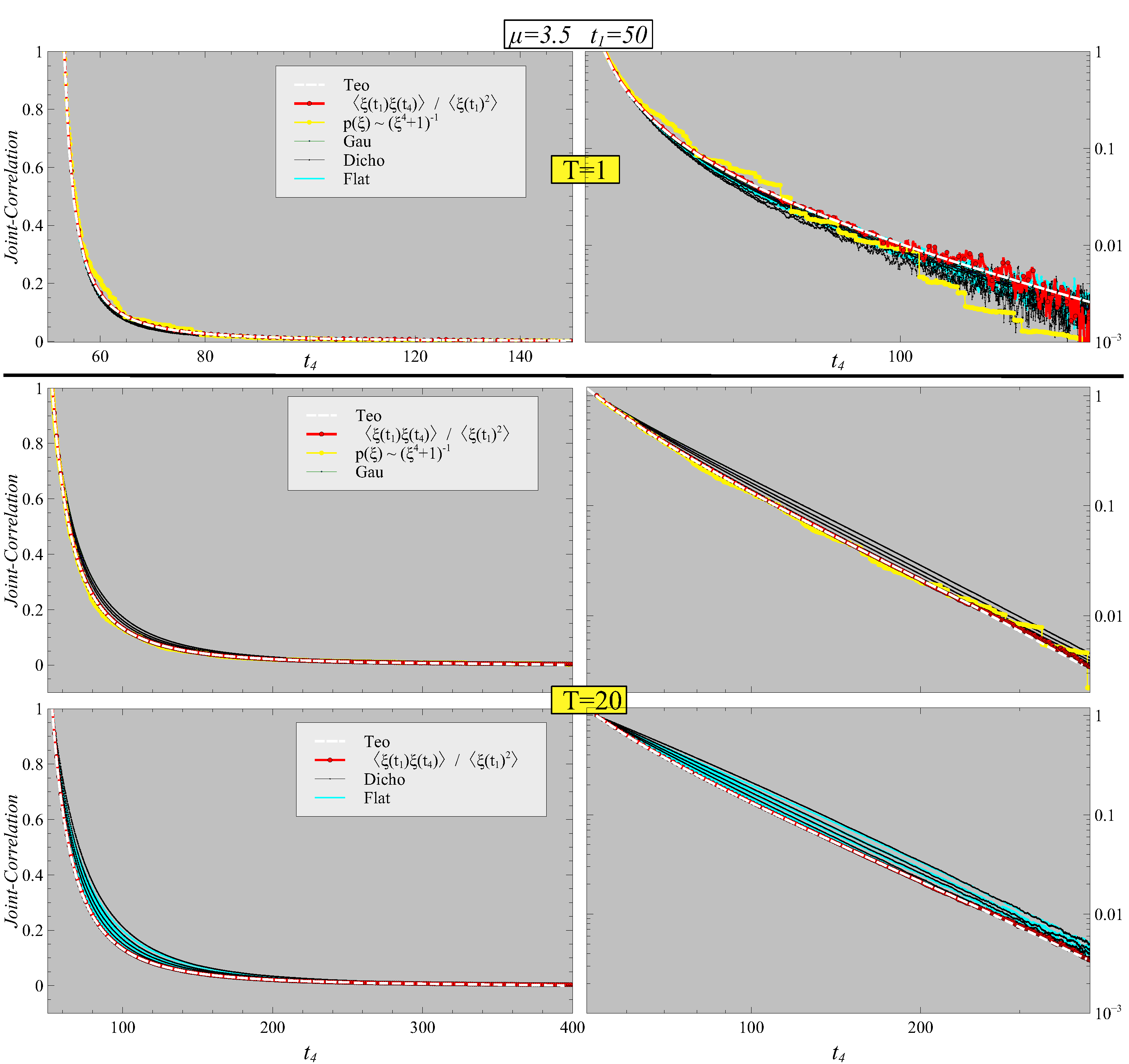}
		\caption{As in Fig.~\ref{Tutti_mu3p5_T_1-20_t1_0}, but for ${t_1=50}$.
			Comparison between the normalized universal 2-time correlation
			function (thick solid red line with small circles) from simulations and the normalized 
			4-time 
			correlation functions for  ${t_1=50}$ (thin colored lines with small circles)  from simulations done for different $\xi$  PDFs, as per the text box. The WT PDF is given by Eq.~\eqref{WTManneville},
			with ${\mu=3.5}$,  and  $T=1$, (first row) and $T=20$
			(second and third rows).
			The theoretical result for the universal 2-time correlation
			function is plotted as a dashed white line. Left panels: linear scale. Right panels:  Log-Log scale. For each $\xi$  PDF considered, six curves are plotted: the ones relative to a
			Normal and a power law PDFs collapse on a unique curve, whereas the ones relative to 
			dicothomous and flat PDFs show some spreading: see the text for a detailed explanation.
		}	
		\label{Tutti_mu3p5_T_1-20_t1_50}
	\end{figure*}
	\begin{figure*}  
		\centering
		\includegraphics[width=\textwidth]{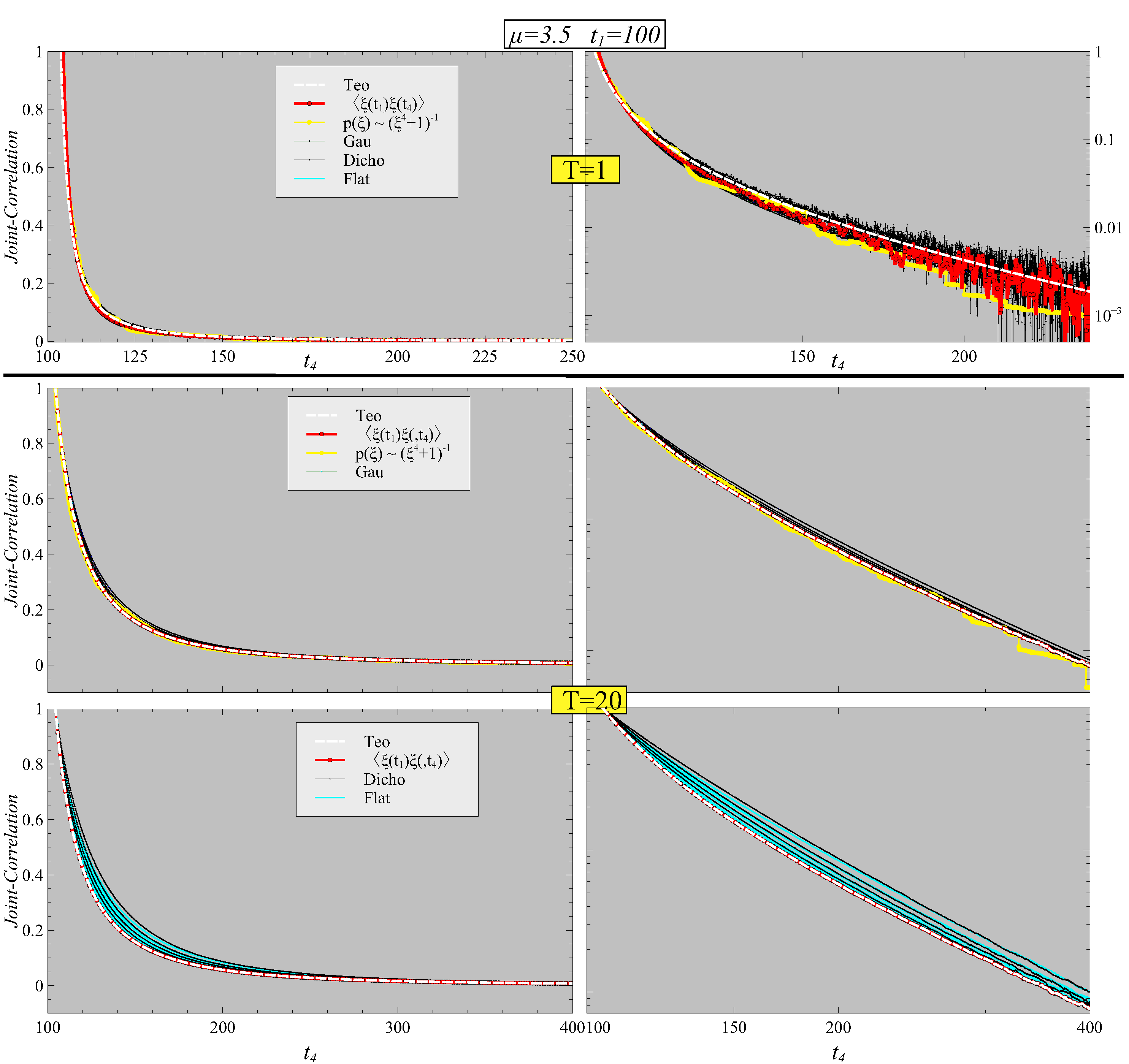}
		\caption{As in Fig.~\ref{Tutti_mu3p5_T_1-20_t1_0}, but for ${t_1=100}$.
			Comparison between the normalized universal 2-time correlation
			function (thick solid red line with small circles) from simulations and the normalized 
			4-time 
			correlation functions for  ${t_1=100}$ (thin colored lines with small circles)  from simulations done for different $\xi$  PDFs, as per the text box. The WT PDF is given by Eq.~\eqref{WTManneville},
			with ${\mu=3.5}$,  and  $T=1$, (first row) and $T=20$
			(second and third rows).
			The theoretical result for the universal 2-time correlation
			function is plotted as a dashed white line. Left panels: linear scale. Right panels:  Log-Log scale. For each $\xi$  PDF considered, six curves are plotted: the ones relative to a
			Normal and a power law PDFs collapse on a unique curve, whereas the ones relative to 
			dicothomous and flat PDFs show some spreading: see the text for a detailed explanation.}
		\label{Tutti_mu3p5_T_1-20_t1_100}
	\end{figure*}
	\begin{figure*} 
		\centering
		\includegraphics[width=\textwidth]{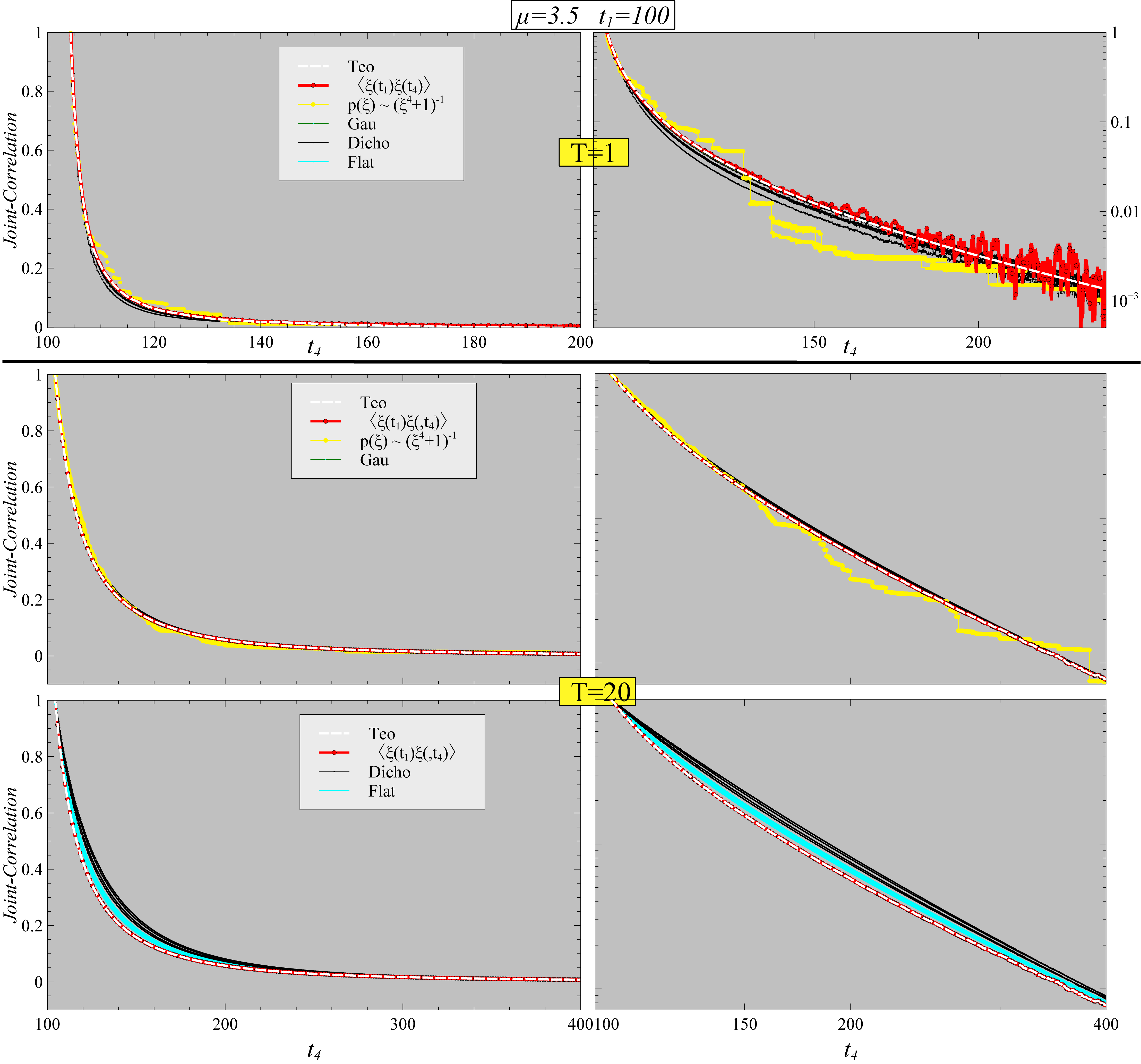}
		\caption{As in Fig.~\ref{Tutti_mu3p5_T_1-20_t1_100}, but for ${n=6}$.
			Comparison between the normalized universal 2-time correlation
			function (thick solid red line with small circles) from simulations and the normalized 
			6-time 
			correlation functions for  ${t_1=100}$ (thin colored lines with small circles)  from simulations done for different $\xi$  PDFs, as per the text box. The WT PDF is given by Eq.~\eqref{WTManneville},
			with ${\mu=3.5}$,  and  $T=1$, (first row) and $T=20$
			(second and third rows).
			The theoretical result for the universal 2-time correlation
			function is plotted as a dashed white line. Left panels: linear scale. Right panels:  Log-Log scale. For each $\xi$  PDF considered, six curves are plotted: the ones relative to a
			Normal and a power law PDFs collapse on a unique curve, whereas the spreading relative to
			the dicothomous and  the flat PDFs are now much reduced comparing
			to Fig.~\ref{Tutti_mu3p5_T_1-20_t1_100}: see the text for a detailed explanation.}
		\label{Tutti6_mu3p5_T_1-20_t1_100}
	\end{figure*}
	\begin{figure*}   
		\centering
		\includegraphics[width=\textwidth]{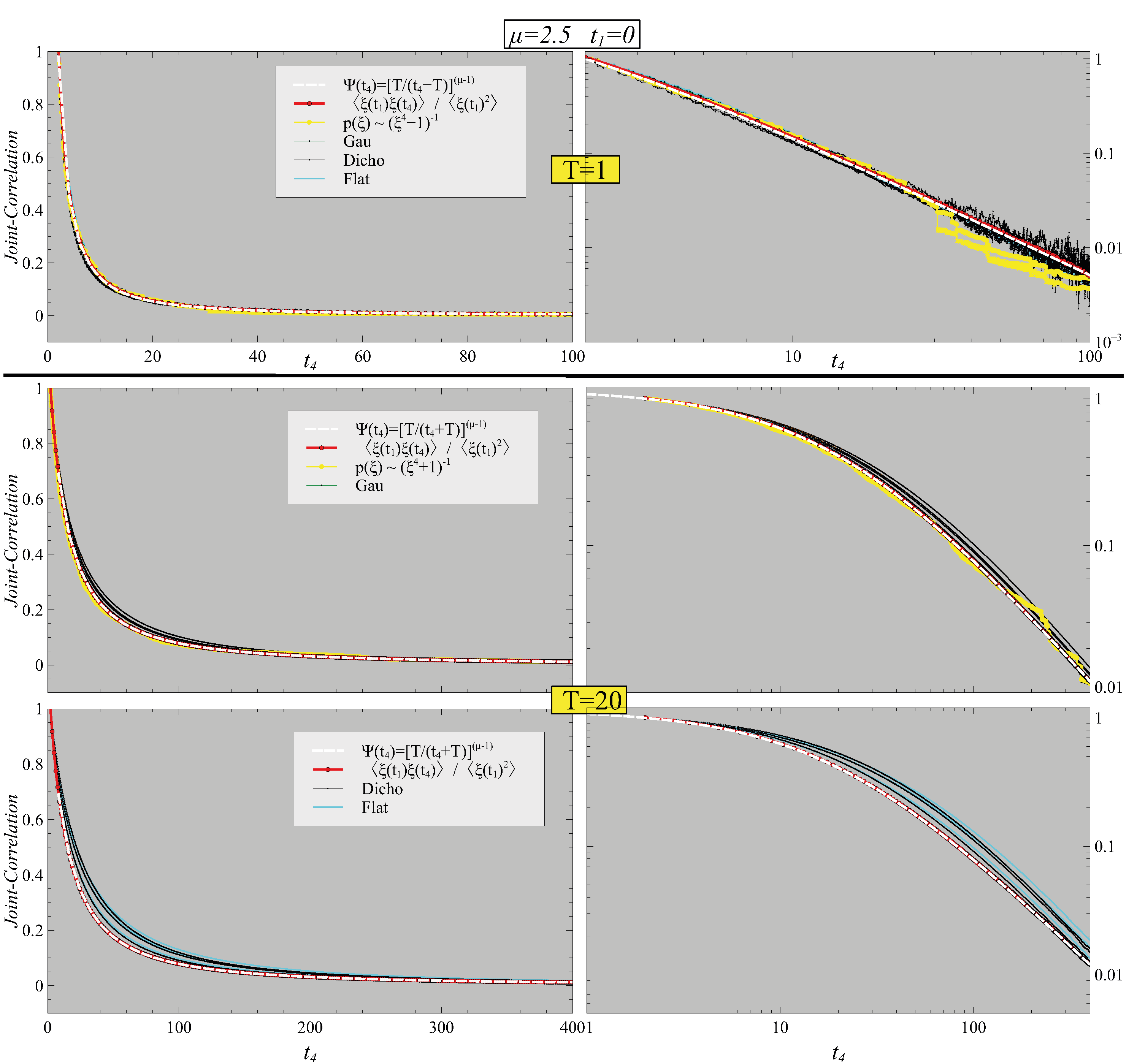}
		\caption{As in Fig.~\ref{Tutti_mu3p5_T_1-20_t1_0}, but for $\mu=2.5$. Comparison between the normalized universal 2-time correlation
			function (thick solid red line with small circles) from simulations and the normalized 
			4-time 
			correlation functions for  ${t_1=0}$ (thin colored lines with small circles)  from simulations done for different $\xi$  PDFs, as per the text box. The WT PDF is given by Eq.~\eqref{WTManneville},
			with ${\mu=2.5}$,  and  $T=1$, (first row) and $T=20$
			(second and third rows).
			The theoretical result for the universal 2-time correlation
			function is plotted as a dashed white line. Left panels: linear scale. Right panels:  Log-Log scale. For each $\xi$  PDF considered, six curves are plotted: the ones relative to a
			Normal and a power law PDFs collapse on a unique curve, whereas the ones relative to 
			dicothomous and flat PDFs show some spreading: see the text for a detailed explanation.
		}
		\label{Tutti_mu2p5_T_1-20_t1_0}
	\end{figure*}
	\begin{figure*}  
		\includegraphics[width=\textwidth]{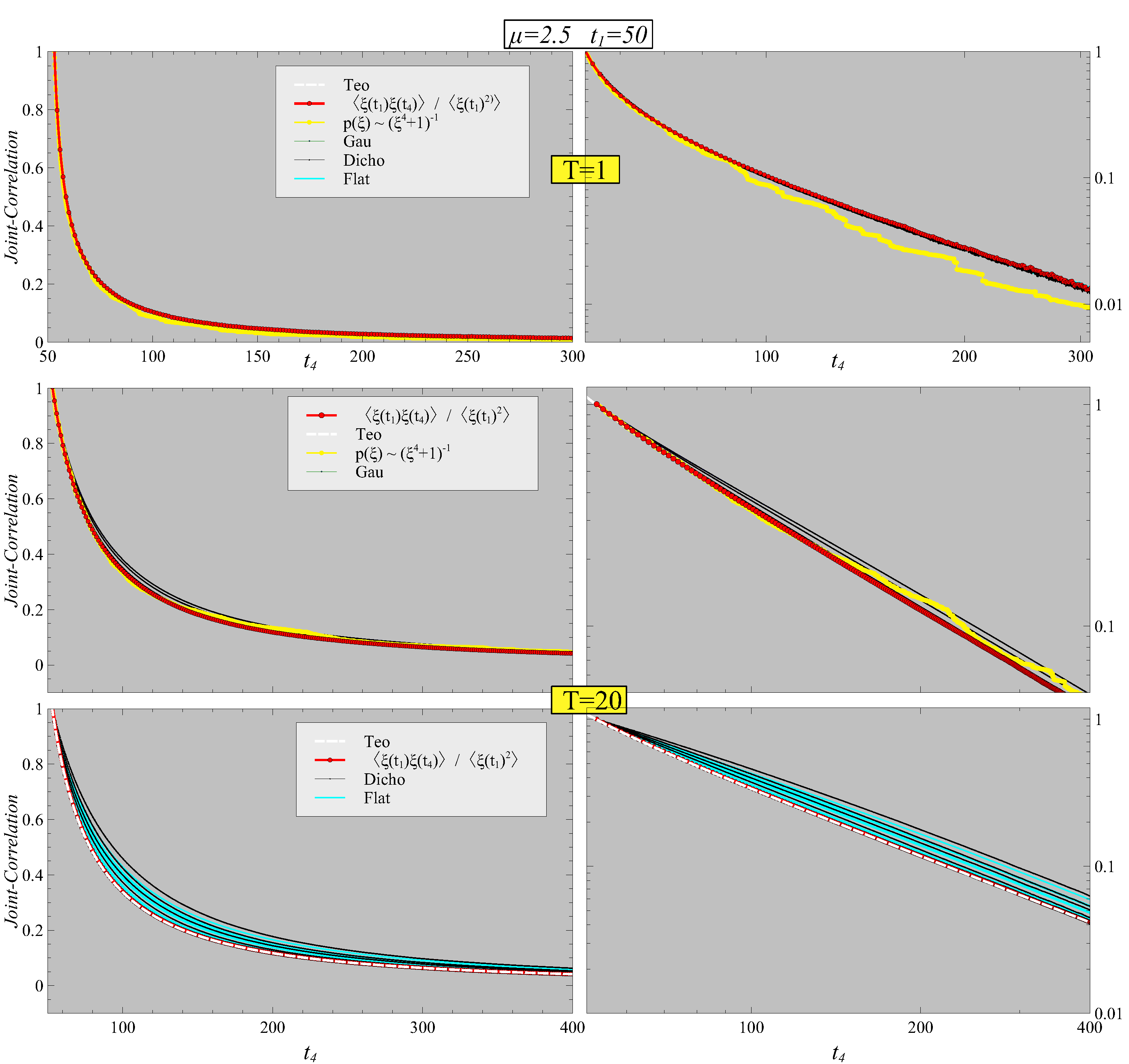}
		\caption{As in Fig.~\ref{Tutti_mu2p5_T_1-20_t1_0}, but for $t_1=50$.
			Comparison between the normalized universal 2-time correlation
			function (thick solid red line with small circles) from simulations and the normalized 
			4-time 
			correlation functions for  ${t_1=50}$ (thin colored lines with small circles)  from simulations done for different $\xi$  PDFs, as per the text box. The WT PDF is given by Eq.~\eqref{WTManneville},
			with ${\mu=2.5}$,  and  $T=1$, (first row) and $T=20$
			(second and third rows).
			The theoretical result for the universal 2-time correlation
			function is plotted as a dashed white line. Left panels: linear scale. Right panels:  Log-Log scale. For each $\xi$  PDF considered, six curves are plotted: the ones relative to a
			Normal and a power law PDFs collapse on a unique curve, whereas the ones relative to 
			dicothomous and flat PDFs show some spreading, a bit wider than in Fig.~\ref{Tutti_mu2p5_T_1-20_t1_0}: see the text for a detailed explanation.}
		\label{Tutti_mu2p5_T_1-20_t1_50}
	\end{figure*}
	\begin{figure*}  
		\centering
		\includegraphics[width=\textwidth]{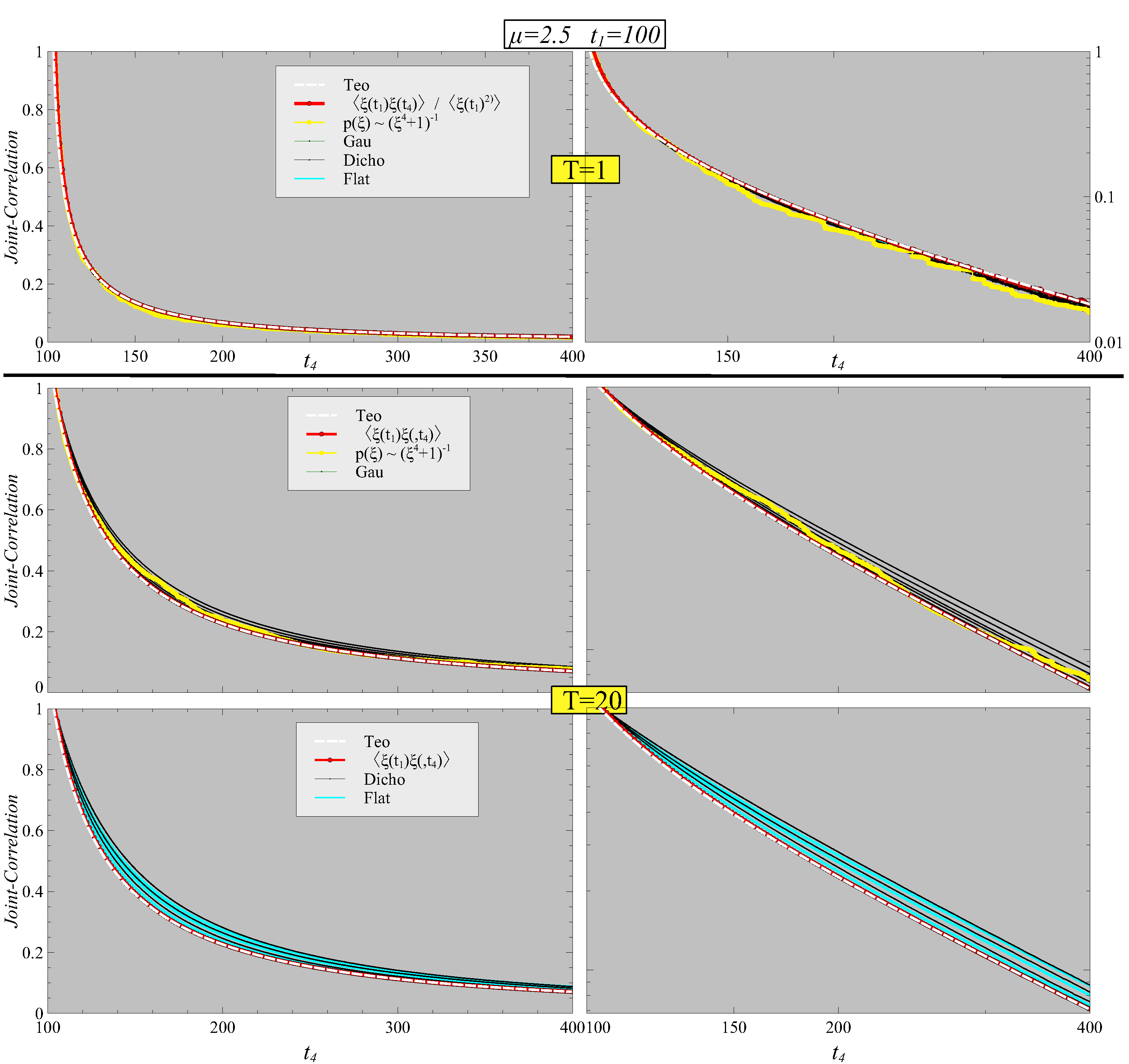}
		\caption{As in Fig.~\ref{Tutti_mu2p5_T_1-20_t1_0}, but for $t_1=100$.  Comparison between the normalized universal 2-time correlation
			function (thick solid red line with small circles) from simulations and the normalized 
			4-time 
			correlation functions for  ${t_1=100}$ (thin colored lines with small circles)  from simulations done for different $\xi$  PDFs, as per the text box. The WT PDF is given by Eq.~\eqref{WTManneville},
			with ${\mu=2.5}$,  and  $T=1$, (first row) and $T=20$
			(second and third rows).
			The theoretical result for the universal 2-time correlation
			function is plotted as a dashed white line. Left panels: linear scale. Right panels:  Log-Log scale. For each $\xi$  PDF considered, six curves are plotted: the ones relative to a
			Normal and a power law PDFs collapse on a unique curve, whereas the ones relative to 
			dicothomous and flat PDFs show some spreading: see the text for a detailed explanation.}
		\label{Tutti_mu2p5_T_1-20_t1_100}
	\end{figure*}
	%
	%
	%
	
	\begin{figure*}   
		\centering
		\includegraphics[width=\textwidth]{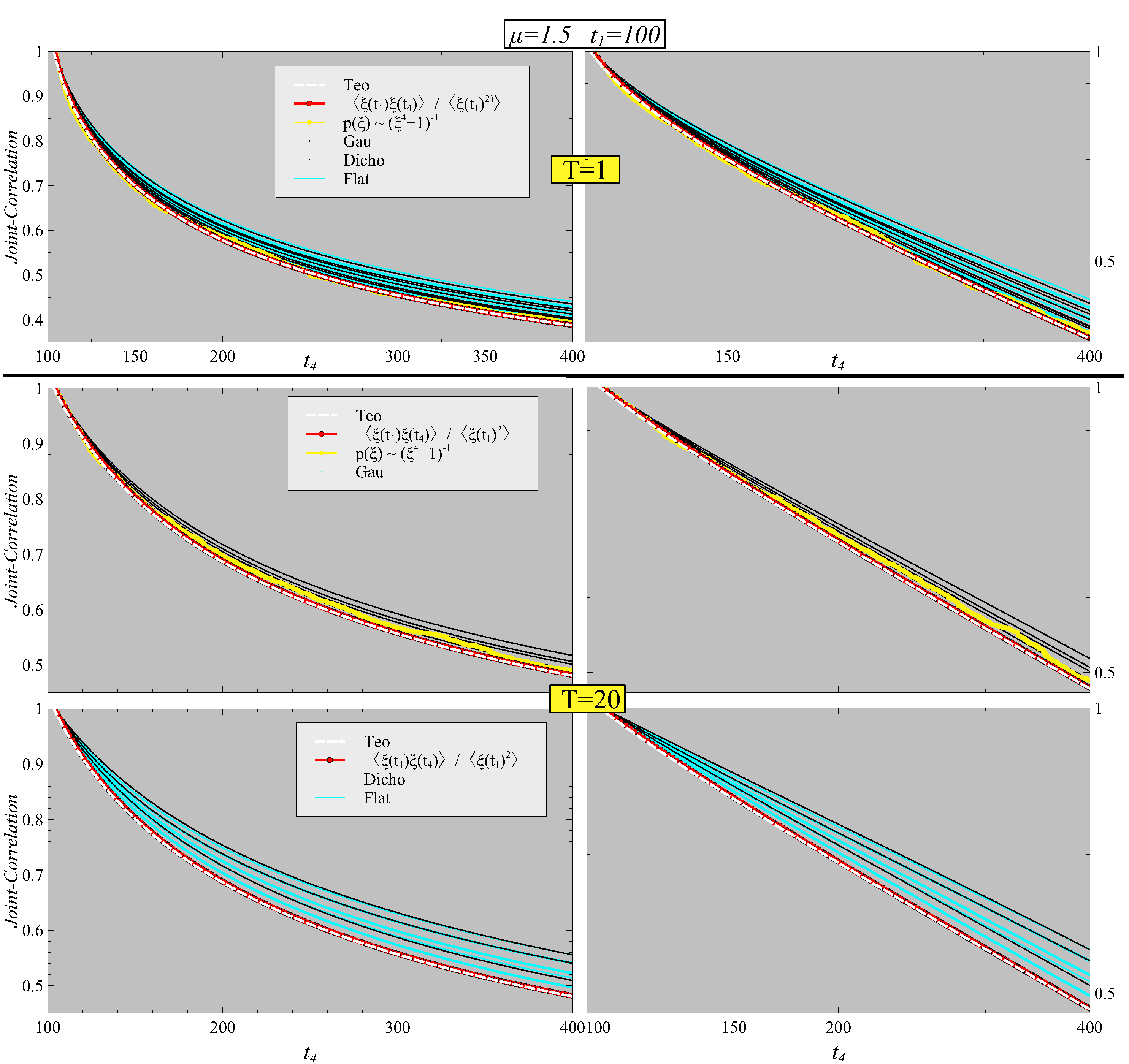}
		\caption{As in Fig.~\ref{Tutti_mu2p5_T_1-20_t1_100}, but for $\mu=1.5$. Since $\mu<2$, the conditions of Proposition~\ref{prop:universal}/Lemma~\ref{lem:2} are not met.
			The results show a clear divergence in behavior based on the $\xi$ Probability Density Function (PDF):
			Only the six correlation curves corresponding to the power-law PDF (yellow lines) agree with the 2-time correlation function (thick red line with small circles): these curves show virtually no dependence on the intermediate times. This outcome is expected, as a power-law PDF satisfies the requirements outlined in Proposition~\ref{prop:PDF_power}.
			The colored curves corresponding to the correlations computed for the other PDFs show a clear spreading or divergence when different intermediate times are considered, indicating a significant dependence on those times.}
		\label{Tutti_mu1p5_T_1-20_t1_100}
	\end{figure*}

	\begin{figure*}[h!]    
		\centering
		\includegraphics[width=\textwidth]{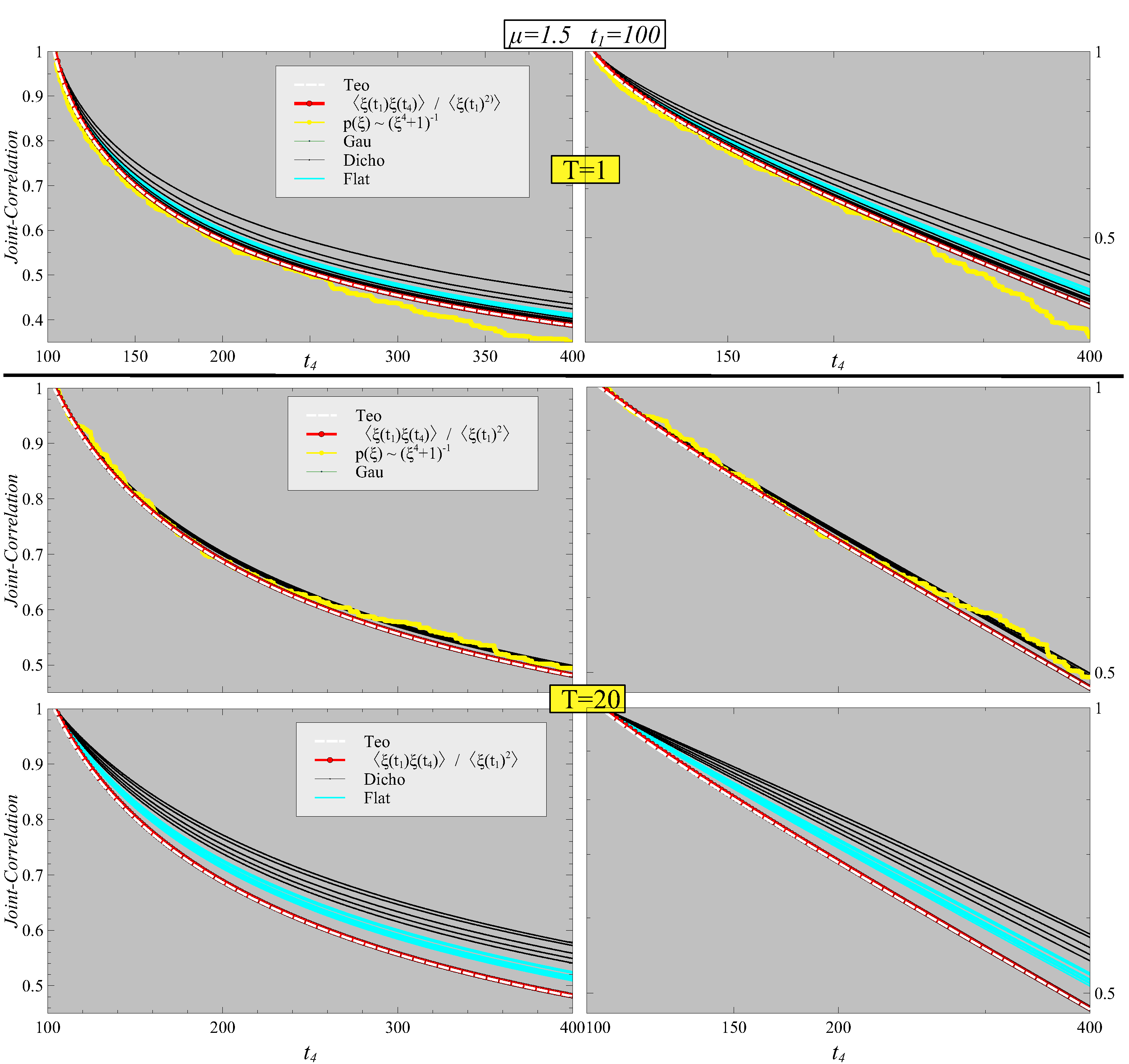}
		\caption{As in Fig.~\ref{Tutti_mu1p5_T_1-20_t1_100}, but for 6-time correlation function. 
			All the curves are now much closer to the universal 2-time correlation function (thick red line with small circles) and show a reduced spread when compared to the corresponding
			curve of Fig.~\ref{Tutti_mu1p5_T_1-20_t1_100}, with the only exception being the dichotomous case (thin black lines).
			This convergence occurs despite the fact that Proposition~\ref{prop:universal}/Lemma~\ref{lem:2} does not strictly apply.
			This happens for different reason depending on the 
			$\xi$ PDF: for the power law PDF (yellow curves) this is  explained by Proposition~\ref{prop:PDF_power}. 
			For the other $\xi$ PDFs, this is due to the condition $\overline{\xi^{n+1}} > \overline{\xi^n}$. See text for a detailed explanation.
		}
		\label{Tutti6_mu1p5_T_1-20_t1_100}
	\end{figure*}
	\begin{figure*}[h!]    
		\centering
		\includegraphics[width=0.45\textwidth]{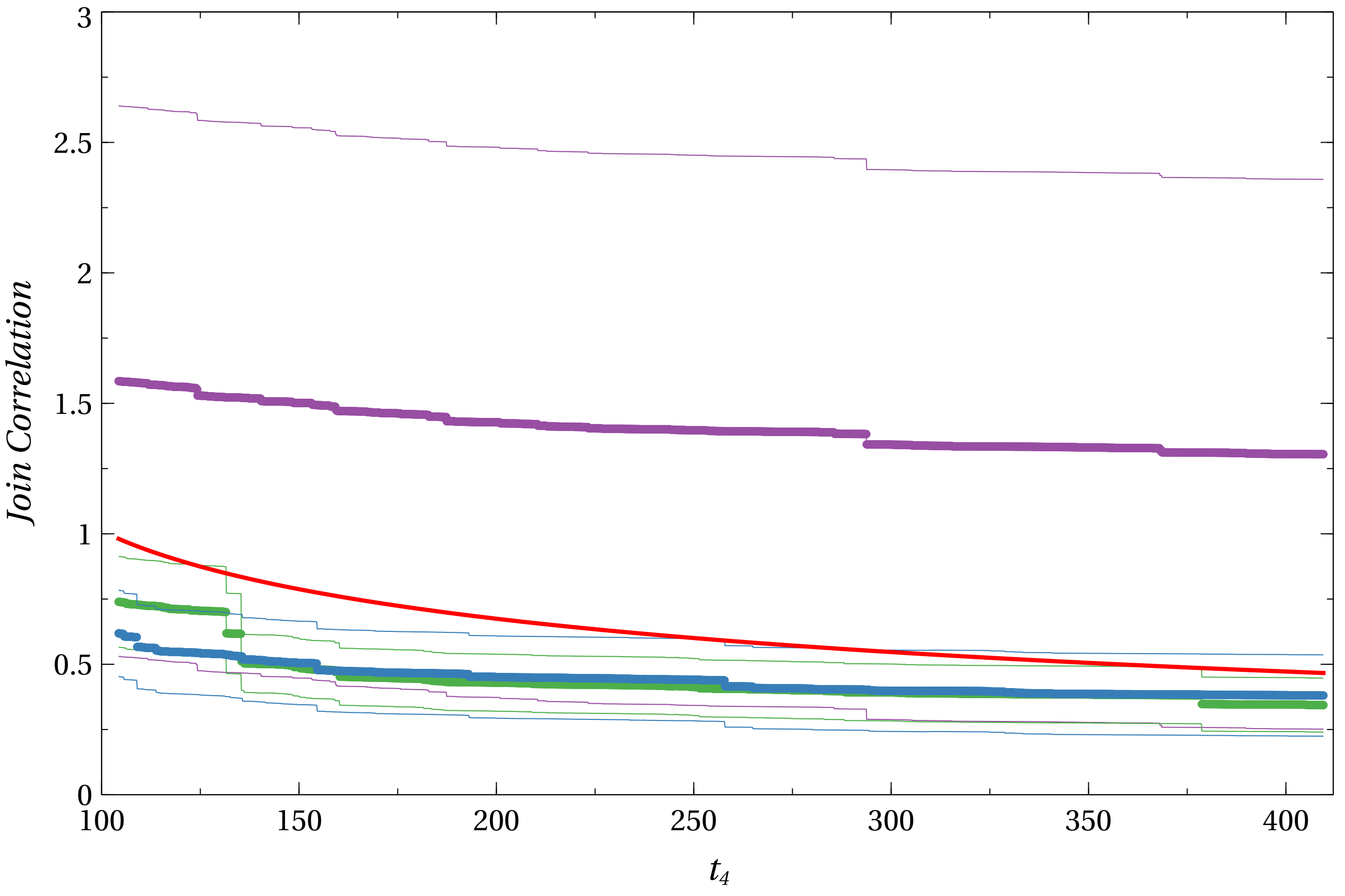}
		\includegraphics[width=0.45\textwidth]{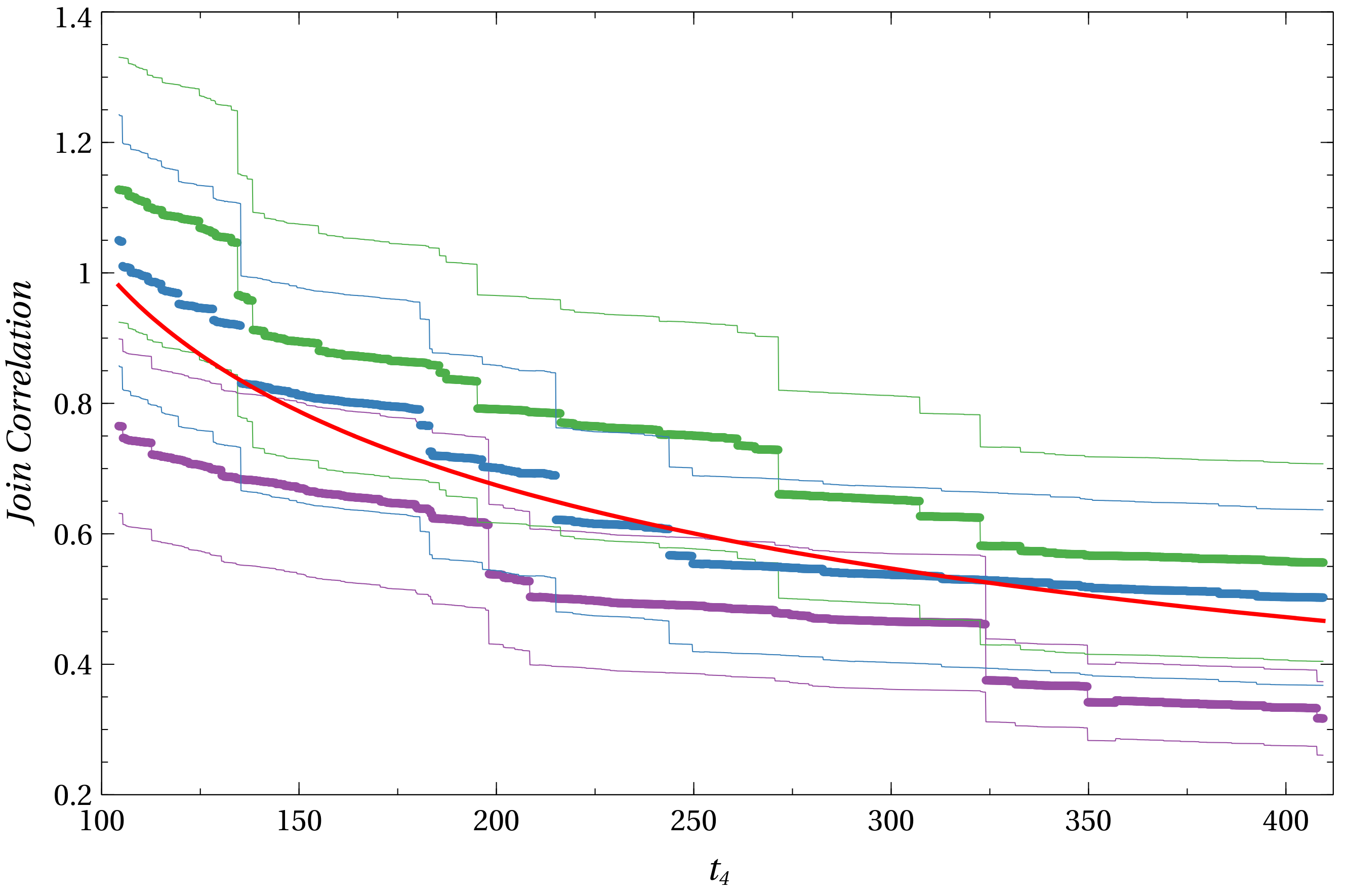} \\
		\includegraphics[width=0.45\textwidth]{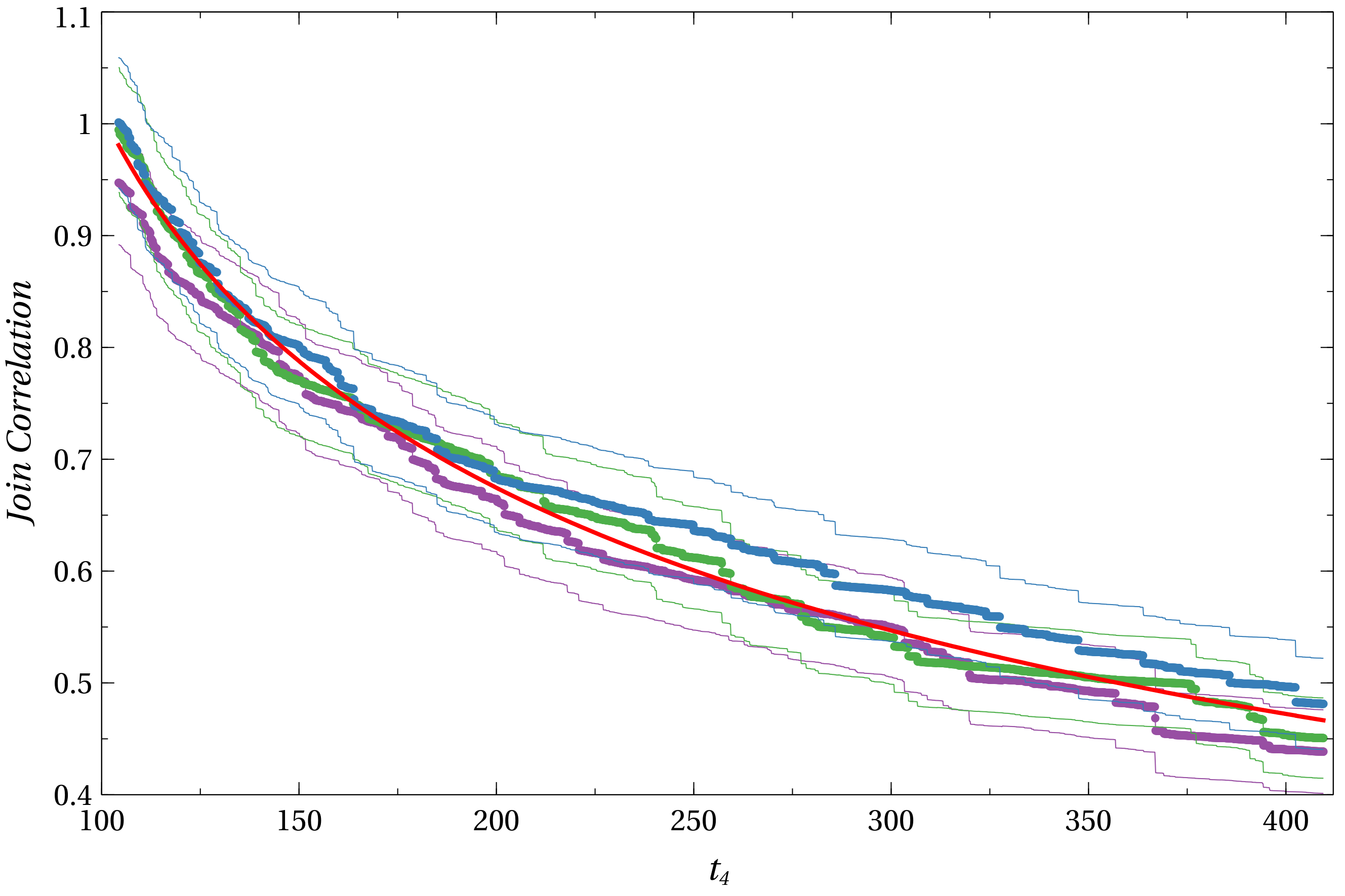}
		\caption{\added[id=1]{Simulations of the 4-time correlation function, 
				done for a $\xi$ power law PDF, a Manneville-like
				WT PDF, with $\mu=1.5$, $T=20$, $t_1=100$ and different $N$ ($\xi$ instances). 
				Top row, left: $N=2 \times 10^6$; top row, right: $N=2 \times 10^7$; bottom row: $N=2 \times 10^8$. The red line is the 2-time correlation function. Colored dots are the result of simulations for the 4-time correlation function, done for intermediate times 
				$t_2=t_1 + (t_4-t_1)/4$ and $t_3=t_1 + 3(t_4-t_1)/4$: the result of three independent runs. each done with $N$ instances, is shown in each plate with different colours. Statistical error bands on the averaged 4-time correlation function are shown as colored thin lines above and below the data points.
				Note how the simulation results  for smaller $N$ present a marked spread when different runs are considered, while they nicely converge to the 2-time correlation function as $N$ is increased. For this figure, the statistical error on the average was computed using the standard definition: $\Delta^2 \langle\Phi^{(4)}\rangle = \sum (\Phi^{(4)}-\langle\Phi^{(4)}\rangle)^2/(N(N-1))$.}
		}
		\label{NDependence}
	\end{figure*}
	
	\added[id=3]{First, we validated the numerical simulations, carrying out simulations for cases when 
		an analytic result is known: for example, we looked at the 6-time correlation $\Phi^{(6)}$
		for the dichotomous case (PDF of Eq.~\eqref{dichotomous}) with
		an exponential WT PDF: in this case, we know theoretically that the correlation function factorizes:}
	\added[id=1]{
		\begin{equation*}
			\Phi^{(6)}(t_1,t_2,t_3,t_4,t_5,t_6)  = \left(\bar{\xi}^2\right)^3 e^{-\frac{t_2-t_1}{\tau}}e^{-\frac{t_4-t_3}{\tau}}e^{-\frac{t_6-t_5}{\tau}}
		\end{equation*}
	}
	\added[id=1]{The result of the simulations is shown in Fig.~(\ref{mo6dice_T1_t1_100}), and indeed the agreement is excellent, thus validating the numerical simulations.}

	\subsection*{\added[id=1]{Numerical simulations related to proposition~\ref{prop:1} and Lemma~\ref{lem:1}}}
	Proposition~\ref{prop:1} and Lemma~\ref{lem:1} address the case where the 
	WT  PDF decays exponentially 
	with a characteristic time $\tau$, as described in Eq.~\eqref{WTpoisson}. \added[id=3]{The comparison between 
		theory and simulations is shown in Figures~(\ref{gaue_T1_t1_0})-(\ref{mo6gene_T20-t1_100}).} 
	\deleted{For this test we use} First, \added[id=3]{we consider the case where
		the $\xi$ PDF is either the} Normal \deleted{and} \added[id=3]{or} the flat distribution\deleted{s} of Eqs.~\eqref{pNormal}
	and~\eqref{pFlat}, respectively.    
	We look at the 4-time and 6-time correlation functions of 
	$\xi[ t]$, i.e., $\Phi^{(4)}(t_1,t_2,t_3,t_4)$ and $\Phi^{(6)}(t_1,t_2,t_3,t_4,t_5,t_6)$. 
	In this Poissonian case, 
	the exact theoretical results are obtained using  Eq.~\eqref{pois} in Eq.~\eqref{corr4_2} and Eq.~\eqref{corr6_}, respectively, and read:
	\begin{equation}
		\begin{aligned}
			\label{exact:poi_4}
			& \Phi^{(4)}(t_1,t_2,t_3,t_4)  =\overline{\xi^4} e^{-\frac{t_4-t_1}{\tau}} \\
			& +\left(\bar{\xi}^2\right)^2 e^{-\frac{t_2-t_1}{\tau}} 
			\left(1-e^{-\frac{t_3-t_2}{\tau}} \right)e^{-\frac{t_4-t_3}{\tau}} ,
		\end{aligned}
	\end{equation}
	\begin{equation}
		\begin{aligned}
			\label{exact:poi_6}
			& \Phi^{(6)}(t_1,t_2,t_3,t_4,t_5,t_6)  =\overline{\xi^6} e^{-\frac{t_6-t_1}{\tau}} \\
			& +\overline{\xi^4} \overline{\xi^2}\left[e^{-\frac{t_4-t_1}{\tau}} \left(1-e^{-\frac{t_5-t_4}{\tau}} \right)
			+e^{-\frac{t_2-t_1}{\tau}} \left(1-e^{-\frac{t_3-t_2}{\tau}} \right)
			e^{-\frac{t_6-t_3}{\tau}}\right] \\
			& +\left(\bar{\xi}^2\right)^3 e^{-\frac{t_2-t_1}{\tau}} \left(1-e^{-\frac{t_3-t_2}{\tau}} \right)
			e^{-\frac{t_4-t_3}{\tau}} \left(1-e^{-\frac{t_5-t_4}{\tau}} \right)e^{-\frac{t_6-t_5}{\tau}} .
		\end{aligned}
	\end{equation}
	We also analyze whether, for $\left|t_i - t_j\right| \gg \tau$, these converge to the factorization results, i.e., 
	if asymptotically we have:
	\begin{align}
		\label{asymp_fac_4}
		&\Phi^{(4)}(t_1,t_2,t_3,t_4) \approx \Phi^{(2)}(t_1,t_2)\Phi^{(2)}(t_3,t_4)
		=\left(\bar{\xi}^2\right)^2 e^{-\frac{t_2-t_1}{\tau}}e^{-\frac{t_4-t_3}{\tau}},\\
		&\text{for $n=4$ and}\nonumber \\
		\label{asymp_fac_6}
		&\Phi^{(6)}(t_1,t_2,t_3,t_4,t_5,t_6) \approx \Phi^{(2)}(t_1,t_2)\Phi^{(2)}(t_3,t_4)\Phi^{(2)}(t_5,t_6)
		\nonumber \\
		&=\left(\bar{\xi}^2\right)^3 e^{-\frac{t_2-t_1}{\tau}}e^{-\frac{t_4-t_3}{\tau}}e^{-\frac{t_6-t_5}{\tau}}\\
		&\text{for $n=6$}\nonumber;
	\end{align}
	as in the case of telegraph noise.
	
	
	%
	
	\added[id=3]{Figures~(\ref{gaue_T1_t1_0}) and~(\ref{gaue_T1_t1_100}) are relative to the 4-time correlation function for the gaussian case, and figures~(\ref{mo6flae_T1_t1_0})-(\ref{mo6flae_T20_t1_100}) are relative to the 6-time correlation function for the flat case. 
		Figures chosen are meant to be representative: similar results are found when, for instance, we consider the
		4-time correlation function for the flat case and different $t_1$.}
	In each figure, \added[id=3]{colored} dots correspond to different intermediate times choices, as indicated in 
	the \deleted{respective legends}\added[id=3]{text boxes}. 
	Solid \added[id=3]{colored lines}\deleted{, shown in different colors,} represent the exact theoretical predictions, given in 
	Eq.~\eqref{exact:poi_4} for $n=4$ and in Eq.~\eqref{exact:poi_6} for $n=6$: they are in total agreement 
	with the \added[id=3]{results of the} numerical simulations \added[id=3]{done for the corresponding intermediate times, shown with the same color of the theoretical prediction}. 
	Dashed lines correspond to the expressions given in Eq.~\eqref{asymp_fac_4} for $n=4$ and in Eq.~\eqref{asymp_fac_6} 
	for $n=6$, as predicted by the factorization property. It is evident that, for large time lags, \added[id=3]{the dashed lines
		are close to the corresponding colored solid lines: this proves that}
	the factorization 
	property is asymptotically valid, in agreement with Proposition~\ref{prop:1} and Lemma~\ref{lem:1}. \added[id=3]{Finally
		we note that correlation functions for different intermediate times are very different}.

	On the other hand, Figs.~\ref{mo6gene_T1_t1_0}--\ref{mo6gene_T20-t1_100} show that 
	\textit{when the random variable}  $\xi$ \textit{is drawn from a power-law PDF} as that in \eqref{pPowerLaw}, 
	\added[id=1]{a different scenario emerges: the multi-time correlation functions no longer depend on the intermediate
		times (simulations of the correlation done for different intermediate times, dots in the figures, overlap); } 
	the factorization property does not hold, despite the exponential WT PDF; and the multi-time correlation functions from simulations remain consistently close to the universal two-time correlation function, in full 
	agreement with Proposition~\ref{prop:PDF_power}.

	\subsection*{\added[id=1]{Numerical simulations related to Proposition~\ref{prop:universal}/Lemma~\ref{lem:2} and  Proposition \ref{prop:PDF_power}}}
	To test Proposition~\ref{prop:universal}/Lemma~\ref{lem:2} and again  Proposition \ref{prop:PDF_power},
	we turn to the Manneville-like WT  PDF
	of Eq.~\eqref{WTManneville}, with $T = 1$ and $T=20$\deleted{and for two values, greater then 2, of the exponent}. \added[id=1]{We deal first with two values of the exponent  $\mu$, both greater than 2:}
	$\mu=3.5$ (Figs.~\eqref{Tutti_mu3p5_T_1-20_t1_0}-\eqref{Tutti_mu3p5_T_1-20_t1_100}) and
	$\mu=2.5$ (Figs.~\eqref{Tutti_mu2p5_T_1-20_t1_0}-\eqref{Tutti_mu2p5_T_1-20_t1_100}). 
	We consider all the cases in Eqs.~\eqref{dichotomous}-\eqref{pPowerLaw} for the $\xi$ PDF. 
	\added[id=3]{As before,
		figures shown are illustrative, similar results are obtained when other time correlation functions or $t_1$'s are
		considered.}
	For both values of the exponent $\mu$, we compare the $n$-time correlation 
	function as a function of $t_n$ to the ``universal'' two-time correlation function: we considered both the numerically
	simulated two-time correlation function and the theoretical two-time correlation function
	from Eqs.~\eqref{corrnt}, supplied with Eqs.~\eqref{lem:2}. 
	The number of times $n$ is 4 for most cases, and  6 in some cases (as specified in the captions of figures).
	\deleted{For each  
		PDF of $\xi$ we plot 6 curves with the same color, corresponding to 6 different
		values of the intermediate times ($t_2$ and $t_3$ for $n=4$ and
		$t_2, t_3, t_4$ and $t_5$ for $n=6$).  In each figure, we fix the time $t_1$.}
	\deleted{The results are shown in Figs.~\eqref{Tutti_mu3p5_T_1-20_t1_0}-\eqref{Tutti_mu2p5_T_1-20_t1_100}.}
	\added[id=1]{The figures need some explanation. First of all, the red line in each figure is 
		the two-time correlation function obtained numerically, and the dashed white curve is the theoretical two-time correlation function. Then,
		for each $\xi$ PDF considered, we numerically
		computed the
		correlation functions for six different values of the intermediate times: these six correlation functions are plotted with curves of the same color, with the color code for
		each PDF shown in the text box of each figure. Note that in many cases it seems that
		fewer than six curves are plotted: this is because these six curves might be very close or even overlap. 
		In each figure, we fix  the time $t_1$.}
	
	For $n=4$, the intermediate times \added[id=1]{to obtain the six different correlation functions} are evaluated at ($\Delta:=t_4-t_1)$ 
	\begin{enumerate}
		\item ${t_2}={t_1}+\frac{1}{4} \Delta\;, {t_3}={t_1}+\frac{3}{4} \Delta$
		\item ${t_2}={t_1}+\frac{1}{4} \Delta\;, {t_3}={t_1}+\frac{2}{3} \Delta$
		\item ${t_2}={t_1}+\frac{1}{4} \Delta\;, {t_3}={t_1}+\frac{1}{2} \Delta$
		\item ${t_2}={t_1}+\frac{1}{2} \Delta\;, {t_3}={t_1}+\frac{2}{3} \Delta$
		\item ${t_2}={t_1}+\frac{1}{3} \Delta\;, {t_3}={t_1}+\frac{2}{3} \Delta$
		\item ${t_2}={t_1}+\frac{1}{4} \Delta\;, {t_3}={t_1}+\frac{2}{3} \Delta$
	\end{enumerate}
	
	For $n=6$, the intermediate times are evaluated at ($\Delta=({t_6}-{t_1})$):
	\begin{enumerate}
		\item ${t_2}={t_1}+\left(\frac{1}{4}-\frac{1}{8}\right)\Delta\;, {t_3}={t_1}+\frac{1}{4}\Delta\;, {t_4}={t_1}+\frac{3}{4}\Delta\;, {t_5}={t_1}+\left(\frac{3}{4}+\frac{1}{8}\right)\Delta$
		\item ${t_2}={t_1}+\left(\frac{1}{4}-\frac{1}{8}\right)\Delta\;, {t_3}={t_1}+\frac{1}{4}\Delta\;, {t_4}={t_1}+\frac{2}{3}\Delta\;, {t_5}={t_1}+\left(\frac{2}{3}+\frac{1}{6}\right)\Delta$
		\item ${t_2}={t_1}+\left(\frac{1}{4}-\frac{1}{8}\right)\Delta\;, {t_3}={t_1}+\frac{1}{4}\Delta\;, {t_4}={t_1}+\frac{1}{2}\Delta\;, {t_5}={t_1}+\left(\frac{1}{2}+\frac{1}{4}\right)\Delta$ 
		\item  ${t_2}={t_1}+\left(\frac{1}{2}-\frac{1}{4}\right)\Delta\;, {t_3}={t_1}+\frac{1}{2}\Delta\;, {t_4}={t_1}+\frac{2}{3}\Delta\;, {t_5}={t_1}+\left(\frac{2}{3}+\frac{1}{6}\right)\Delta$
		\item ${t_2}={t_1}+\left(\frac{1}{3}-\frac{1}{6}\right)\Delta\;, {t_3}={t_1}+\frac{1}{3}\Delta\;, {t_4}={t_1}+\frac{2}{3}\Delta\;, {t_5}={t_1}+\left(\frac{2}{3}+\frac{1}{6}\right)\Delta$
		\item ${t_2}={t_1}+\left(\frac{1}{4}-\frac{1}{8}\right)\Delta\;, {t_3}={t_1}+\frac{1}{4}\Delta\;, {t_4}={t_1}+\frac{2}{3}\Delta\;, {t_5}={t_1}+\left(\frac{2}{3}+\frac{1}{6}\right)\Delta$ 
	\end{enumerate}
	
	\deleted{It is evident from all the figures that the results are in excellent agreement 
		with Proposition~\ref{prop:universal}, i.e., for time lags large compared to 
		$T$, the $n$-time correlation functions are nearly independent of the 
		intermediate times and closely follow the universal two-time correlation 
		function.
		Moreover, these figures clearly show that, except in the flat and 
		dichotomous PDF cases (which, in 
		Figs.~\ref{Tutti_mu3p5_T_1-20_t1_0}--\ref{Tutti_mu2p5_T_1-20_t1_100}, for 
		$T=20$, are presented in separate panels), this universal behavior appears to 
		hold universally, that is, regardless of the time lags relative to $T$.
		In addition, this effect is more pronounced for $n=6$ than for $n=4$, 
		indicating that the universal behavior described in 
		Proposition~\ref{prop:universal} as a limiting case for large time lags may in 
		fact apply more broadly.}
	
	\added[id=3]{In all figures, the simulated two-time correlation function (red line) shows excellent agreement with the theoretical prediction (dashed white line).}
	\added[id=1]{It is also evident that the results from numerical simulations for the $n$-time correlation functions confirm Proposition~\ref{prop:universal}. Specifically, for time lags large compared to $T$ (plates on top row of each figure,
		labelled $T=1$), the $n$-time correlation functions are nearly independent of the intermediate times (the numerical data 
		for the
		$n$-time correlation functions overlap) and closely follow the universal two-time correlation function.}
	\added[id=3]{This universal behavior appears to hold more broadly (bottom plates of each figure, labelled $T=20$, where we split the comparison between theory and simulations), i.e. regardless of the time lags relative to $T$, when the case of gaussian and power law $\xi$ PDF is considered, whereas some spreading is clearly visible when a flat or a dichotomous PDF is considered. The bottom
		plates show separately these two different cases: for the gaussian and power law $\xi$ PDF, the correlation functions show little or no spreading and collapse on the two-time 
		correlation function, for a flat or a dichotomous PDF some spreading is clearly visible.}
	
	\added[id=3]{The independence on intermediate times and the collapse on the two-time correlation function 
		is more pronounced for $n=6$ than for $n=4$ correlation function 
		(compare for instance Figure~(\ref{Tutti6_mu3p5_T_1-20_t1_0}) 
		to Figure~(\ref{Tutti_mu3p5_T_1-20_t1_0}) or Figure~(\ref{Tutti6_mu3p5_T_1-20_t1_100}) 
		to Figure~(\ref{Tutti_mu3p5_T_1-20_t1_100})), suggesting that the universal behavior described in Proposition~\ref{prop:universal} as a limiting case for large time lags may, in fact, have wider applicability, in particular
		when higher $n$-time correlation functions are considered.}
	%
	%
	This fact can be easily explained by examining the general expression for the $n$-time correlation function in Eq.~\eqref{corrGen_}, as well as the illustrative examples for $n=6$ and $n=8$ in Eqs.~\eqref{corr6_} and~\eqref{corr8_}, respectively.
	
	In fact, the coefficient of the $p=1$ term, which depends on two times (and resembles a normalized two-time correlation function), is $\overline{\xi^n}$. The coefficients of the remaining terms are products of the form $\overline{\xi^{m_1}}\, \overline{\xi^{m_2}} \cdots \overline{\xi^{m_p}}$, where $m_1 + m_2 + \cdots + m_p = n$.
	
	Therefore, if the PDF is such that for a given integer $k$ we have $\overline{\xi^{k+1}} > \overline{\xi^k}$, increasing $n$ results in the dominance of the two-time correlation term, leading to the convergence described by Eq.~\eqref{corrnt-2t}, regardless of the time lags. This condition is generally satisfied by most PDFs, though notable exceptions include the dichotomous case and, to a lesser extent, the flat PDF.
	
	\subsubsection*{\added[id=1]{The case of Heavy Tails}}
	
	In relation to the last point, we also observe that the numerical simulations align with Proposition~\ref{prop:PDF_power}. In cases where the PDF has very heavy tails, such as the power-law PDF given by Eq.~\eqref{pPowerLaw} (represented by the yellow curves in the figures), convergence towards the universal two-time correlation function is achieved very rapidly simply by increasing $N$ (the number of averages of $\xi$), regardless of both the time lags and the correlation order $n$. \added[id=1]{We will discuss this point in more detail further down.}

	\subsubsection*{\added[id=1]{The $\mu<2$ case}}
	Figs.~\ref{Tutti_mu1p5_T_1-20_t1_100}--\ref{Tutti6_mu1p5_T_1-20_t1_100} concern the case
	where $\mu=1.5$, i.e.\ it is less than $2$. This implies that Proposition~\ref{prop:universal} does not
	hold. Despite this, we see that the numerical simulations for the case of a power-law PDF (the yellow curves)
	continue to agree with 
	Proposition~\ref{prop:PDF_power}, for both $n$'s considered. \added[id=1]{We will now discuss in more detail the figures.}
	
	\added[id=1]{
		The results in Fig.~\ref{Tutti_mu1p5_T_1-20_t1_100} show a clear divergence in behavior based on the $\xi$ PDF considered:}
	
		\begin{itemize}
			\item Power-Law PDF: Only the six correlation curves corresponding to the power-law PDF (yellow lines) agree with the two-time correlation function (thick red line with small circles). Crucially, these curves show virtually no dependence on the intermediate times. This outcome is expected, as a power-law PDF satisfies the requirements outlined in Proposition~\ref{prop:PDF_power}.
			\item Other PDFs: The colored curves corresponding to the correlations computed for the other $\xi$ PDFs show a clear spreading or divergence when different intermediate times are considered, indicating a significant dependence on those times.
		\end{itemize}
		
		Comparing Fig.~\ref{Tutti6_mu1p5_T_1-20_t1_100} (where $n=6$) to Fig.~\ref{Tutti_mu1p5_T_1-20_t1_100} (where $n=4$), we observe that all curves are now very close to the universal two-time correlation function (thick red line with small circles), with the only exception being the dichotomous case (thin black lines).
		This convergence occurs despite the fact that Proposition~\ref{prop:universal}/Lemma~\ref{lem:2} does not strictly apply. The reason for this close fit depends on the $\xi$ PDF:
		\begin{enumerate}
			\item For the power-law PDF (yellow curves), this behavior is explained by Proposition~\ref{prop:PDF_power}.
			\item For the other $\xi$ PDFs, this is due to the condition $\overline{\xi^{n+1}} > \overline{\xi^n}$. For a sufficiently large $n$, this inequality ensures that the term corresponding to $p=1$ in the summation of Eq.~\eqref{corrGen_gen} provides the main contribution. Since this term has the structure of a two-time correlation function, its dominance leads to the observed convergence.
		\end{enumerate}
	
	\subsubsection*{\added[id=1]{The role of $N$ in case of heavy tails}}
	
	\added[id=1]{As mentioned, in the case of heavy tails, the convergence of higher order correlation
		functions to the 2-time correlation function is expected to be particularly fast, as explained by Proposition~\ref{prop:PDF_power}. We verified
		this point carrying out simulations for the 4-time correlation function in the case of a power-law PDF for $\xi$, 
		a Manneville-like WT PDF with $\mu=1.5$, $T=20$, $t_1=100$,  
		and for different $N$. The result is
		shown in Fig.~\ref{NDependence} where we plot simulations done for three different values
		of $N$ and intermediate times $t_2=t_1 + (t_4-t_1)/4$ and $t_3=t_1 + 3(t_4-t_1)/4$.  In each plate, three statistically independent runs are plotted: the three different sets of colored dots show the three average 4-time correlation 
		functions obtained in 
		each run, with the error bar on the average shown as
		thin colored lines above and below the data points. The red line is the 2-time 
		correlation function. 
		The results for the smaller $N$ considered show a marked spread between the average
		correlations computed in different
		runs, while they nicely converge to the 2-time correlation function as $N$ increases. The 4-time correlation functions computed for the other intermediate times
		considered in this paper are indistinguishable from the ones plotted, on this scale.}

	\section{Conclusions\label{sec:conclusions}}
	
		Stochastic renewal processes are now pervasive across numerous scientific domains, underscoring
		their foundational relevance.
		
		This work has examined the important subclass of such processes, in which the random variable $\xi$
		remains constant over a random duration sampled from a waiting-time (WT) probability density
		function (PDF). These dynamics naturally arise in various contexts, such as the blinking of
		quantum dots or the velocity component in L\'evy walks with random velocities.
		
		By averaging over trajectory realizations, we derived an exact expression for arbitrary $n$-time
		correlation functions (Proposition~\ref{prop:main}). This result offers deep insight into the statistical architecture of
		renewal processes. The key findings are summarized below:
		
		\begin{itemize}
			\item
			When the WT PDF has a power-law tail with finite mean $\tau$ (i.e., $\mu > 2$), all $n$-time
			correlation functions converge, for large time lags, to the universal two-time correlation
			evaluated at the outermost times (Proposition~\ref{prop:universal}). Because this result is derived in the stationary regime, the ensemble preparation is irrelevant, as explicitly shown in Eq.~\eqref{corr2t_aged}. Notably, our formulism reproduces known results for aged systems, including the asymptotic stationary dichotomous case (e.g.,~\cite{gnzPRL54}).
			
			
			\item
			When the PDF of $\xi$ exhibits fat tails, convergence toward the two-time correlation persists
			regardless of the decay of the WT PDF. This holds even for short time intervals, provided the
			ensemble size or trajectory length is sufficiently large
			(Proposition~\ref{prop:PDF_power}). In this regime, the WT PDF may decay with $1 < \mu < 2$,
			implying infinite aging and non-stationarity. In this case, the initial ensemble preparation is relevant (see Eq.~\eqref{corrnt-2t_power}).
			\item
			If the WT PDF decays exponentially with characteristic time $\tau$, and the first $n$ moments
			of $\xi$ are finite, the $n$-time correlation function converges, for time lags larger
			than $\tau$, to that of telegraph noise. In this case, the factorization property holds
			(Proposition~\ref{prop:1}).
		\end{itemize}
		Remarkably, the former two points generalize the universality previously established for
		two-time correlations~\cite{bblmCSF196} to the full hierarchy of multi-time functions.
		All theoretical predictions are well reproduced by numerical simulations.
		
		The implications of all these results span diverse fields, including quantum physics, condensed matter, physical chemistry, atmospheric science, and non-equilibrium statistical mechanics
		(e.g.,~\cite{sanda, bon, mel, hot, lv}). In practice, any Brownian system perturbed by
		renewal-type noise—such as the stochastic differential equation in Eq.~\eqref{SDE}—inherits the
		universal statistical features of $\xi[t]$, especially in the asymptotic regime. This paves the
		way for simplified analytical treatments of complex systems, including the derivation of
		universal master equations via generalized cumulant expansions.
		
		Several extensions are currently in progress:
		
		\begin{itemize}
			\item Extending the framework to spike-type renewal processes, relevant for CTRW systems.
			\item Using Propositions~\ref{prop:universal} and~\ref{prop:PDF_power}, along with generalized
			($M$-)cumulant theory, to derive a universal master equation for L\'evy walks with
			state-dependent drift and possible multiplicative noise.
			\item Applying Propositions~\ref{prop:1} and Lemma~\ref{lem:1} to construct a universal master
			equation for Poissonian noise.
		\end{itemize}
		
		These developments will deepen our understanding of how non-Markovian fluctuations shape the
		dynamics of physical, biological, and financial systems.
	%
	%
	\section*{Code and data availability}
	Data are available upon reasonable requests. F90 codes are available at \url{https://github.com/dundacil/renewal_codes}.
	
	\section*{Acknowledgment}
	We thank the Green Data Center of University of Pisa for providing the computational 
	power needed for the present paper. 
	
	This research work was supported in part by 
	ISMAR-CNR and UniPi
	institutional funds. 
	
	D.L. acknowledges financial support from  the project 
	``ITINERIS '', under the National Recovery and 
	Resilience Plan (PNRR), Mission 4 Component 2 Investment 3.1 funded from the 
	European Union - NextGenerationEU.
	
	R.M acknowledges support from the 
	``National Centre for HPC, 
	Big Data and Quantum Computing'', under the National Recovery and 
	Resilience Plan (PNRR), Mission 4 Component 2 Investment 1.4 funded from the 
	European Union - NextGenerationEU.

	\appendix
	\FloatBarrier
	\section{Evaluation of the $n$-time correlation functions for the step noise case: the formal and general approach\label{app:A_A}}
	In this appendix we provide a rigorous derivation of the generalization of the result in Eq.~\eqref{corrGen_}. 
	This derivation is rigorous because it relies solely on algebraic manipulations of the definition 
	given in Eq.~\eqref{n_corr_def_step}.  
	
	Since the derivation is rather long and involved, we organize it into several steps, each introduced by a specific title.

	\subsection*{$a)$ The starting point}
	Because the demonstration is a little cumbersome,  
	for the sake of simplicity, let us rewrite Eq.~\eqref{n_corr_def_step} setting $t_0=0$ (a different initial
	time can always be restored by the replacement $t_i\to t_i-t_0$).   
	\begin{align}
		\label{app:n_corr_step}
		&  \langle\xi( t_1)\xi( t_2)...\xi( t_n)\rangle\nonumber \\
		&=\int 
		\sum_{i_1=0}^\infty 
		\sum_{i_2=i_1}^\infty \cdots
		\sum_{i_{n-1}=i_{n}}^\infty
		\xi_{i_1}  \xi_{i_2}\cdots
		\xi_{i_n}  \Theta\left(t_1-\sum_{k_1=0}^{i_1} \theta_{k_1}\right)  
		\Theta\left(\sum_{k_1=0}^{{i_1}+1} \theta_{k_1}-t_1\right)\nonumber \\
		&\times \Theta\left(t_2-\sum_{k_2=0}^{i_2} \theta_{k_2}\right) 
		\Theta\left(\sum_{k_2=0}^{{i_2}+1}\theta_{k_2}-t_2\right)
		\times...\nonumber \\
		&...\times \Theta\left(t_n-\sum_{k_n=0}^{i_n} 
		\theta_{k_n}\right) \Theta\left(\sum_{k_n=0}^{{i_n}+1} \theta_{k_n}-t_n\right)\nonumber \\
		&\times p_0(\xi_0)d\xi_0\prod_{q=1}^\infty\psi(\theta_q)d\theta_q\,p(\xi_q)d\xi_q.
	\end{align}
	%
	%
	
	\subsection*{$b)$ Rearrange the sums as a sum of compositions\label{app:step_b}}
	Note that
	Eq.~\eqref{app:n_corr_step} is a sums over all possible integer values (including zero) 
	of the  \(n\) indices \(i_{1}, i_{2}, \dots, i_{n}\), subject to the   constraint \(i_{1} \le i_{2} \le \dots \le i_{n}\).
	The \emph{sum} over all the indices is then in one-to-one 
	correspondence with the \emph{sum} over all the possible set of  integer indices  \(i_{1}, i_{2}, \dots, i_{n}\) with this constraint.
	We highlights the fact that some consecutive
	indices can have the same value. For example, for $n=8$, one possible set of 8  indices is: 
	$3,3,7,50,50,220,220,220$, i.e., $i_1=3,i_2=i_1,i_3=3,i_4=50,i_5=i_4, i_6=220,i_7=i_6,i_8=i_7$. 
	Thus, any list  \(i_{1}, i_{2}, \dots, i_{n}\) can be grouped in blocks 
	$\left\{m_{1}\right\}\left\{ m_{2}\right\}...\left\{ m_p\right\}$, with $\sum_{k=1}^p m_k=n$ and where
	in the $k$-th block there is a number $m_{k}$ of consecutive indices with the same value:
	\begin{align}
		\label{app:compositions}
		&   \left\{i_1,\,i_2=i_1,\,...,i_{m_{1}}=i_1\right\}
		\left\{
		i_{m_{1}+1}> i_{m_1},\,i_{m_{1}+2}=i_{m_{1}+1},...,i_{m_{1}+m_{2}}=i_{m_{1}+1}\right\}
		\nonumber \\&      
		\left\{
		i_{m_{1}+m_{2}+1}> i_{m_{1}+m_{2}},\,i_{m_{1}+m_{2}+2}
		=i_{m_{1}+m_{2}+1},...,i_{m_{1}+m_{2}+m_{3}}=i_{m_{1}+m_{2}+1}\right\}\,......
		\nonumber \\[5pt]
		&\big\{
		i_{{\underbrace{{\scriptstyle m_{1}+m_{2}+...+m_{p-1}}}_{=n-m_p}}+1}>i_{m_{1}+m_{2}+\dots+m_{p-1}},
		i_{m_{1}+m_{2}+\dots+m_{p-1}+2}=i_{m_{1}+m_{2}+\dots+m_{p-1}+1},...
		\nonumber \\
		& ...
		i_{{\underbrace{{\scriptstyle m_{1}+m_{2}+\dots+m_{p-1}+m_{p}}}_{=n}}}
		=i_{m_{1}+m_{2}+\dots+m_{p-1}+1}
		\big\}.
	\end{align}
	In other words, the same list can be written as
	\begin{align}
		\label{sets}
		\underbrace{r_1,r_1,...,r_1}_{\footnotesize \begin{array}{c}m_1\;\text{times}\\ \text{block 1}\end{array}},\underbrace{r_2,r_2,...,r_2}_{\footnotesize \begin{array}{c}m_2\;\text{times}\\ \text{block 2}\end{array}},
		...\underbrace{r_p,r_p,...,r_p}_{\footnotesize \begin{array}{c}m_p\;\text{times}\\ \text{block }p\end{array}}.
	\end{align}
	where \(r_{1}=i_1=i_2...=i_{m_1}<r_{2}=i_{m_1+1}=i_{m_1+2}=...=i_{m_1+m_2}<, \dots, <r_{p}=i_{m_1+m_2+...m_{p-1}+1}=i_{m_1+m_2+...m_{p-1}+2}=...=i_{m_1+m_2+...m_{p-1}+m_p}\), 
	is a set of \emph{strictly} ordered integers. 
	
	In the previous example with $n=8$ we have $r_1=3$ with $m_1=2$, $r_2=7$ with $m_2=1$, $r_3=50$ with $m_3=2$, and $r_4=220$ with $m_4=3$.  
	The value of the term in the $j$th position (from left to right, $0 \le j \le n$) in the list \eqref{sets} is denoted by $i_j$.  
	
	By analyzing the summand of Eq.~\eqref{app:n_corr_step}, we observe that only the times $t_j$ depend on the sub-index $j$ of $i_j$ 
	(i.e., they depend on the \textit{position} in the list \eqref{sets}), 
	whereas all the other terms depend solely on the \textit{value} of $i_j$ 
	(i.e., on the terms $r_h$, with $1 \le h \le p$, of the list \eqref{sets}).  
	Therefore, we can replace the multiple sum \eqref{app:n_corr_step} with the following multiple sum over all possible \emph{sets} of the type defined in \eqref{sets}:
	{\small
		\begin{align}
			\label{app:sums_temp}
			&	\text{\eqref{app:n_corr_step}}\to 
			\sum_{p=1}^{n} \;
			\left[
			\sum_{\{m_i\in\N\}:\sum_{i=1}^p m_i = n}  
			\int 
			\sum_{r_1=0}^\infty\sum_{r_2=r_1+1}^\infty\cdots \sum_{r_p=r_{p-1}+1}^\infty
			{\xi_{r_1}^{m_1}}\,{\xi_{r_2}^{m_2}}\,\cdots{\xi_{r_p}^{m_p}}
			\right.	 \nonumber \\
			&\begin{rcases}
				\begin{aligned}
					&	\times \Theta\left(t_1-\sum_{k_1=0}^{r_1} \theta_{k_1}\right)  
					\Theta\left(\sum_{k_1=0}^{r_{1}+1} \theta_{k_1}-t_{1}\right)
					\nonumber \\
					&\times \Theta\left(t_2-\sum_{k_1=0}^{r_1} \theta_{k_1}\right)  
					\Theta\left(\sum_{k_1=0}^{r_{1}+1} \theta_{k_21}-t_{2}\right)\times...
					\nonumber \\
					&...\times 
					\Theta\left(t_{m_1}-\sum_{k_{1}=0}^{r_1} \theta_{k_{1}}\right)  
					\Theta\left(\sum_{k_{1}=0}^{r_{1}+1} \theta_{k_{1}}-t_{m_1}\right)
				\end{aligned}
			\end{rcases}
			\quad \text{block 1}
			\nonumber \\
			&\begin{rcases}
				\begin{aligned}
					&	\times \Theta\left(t_{m_1+1}-\sum_{k_{2}=0}^{r_2} \theta_{k_{2}}\right)  
					\Theta\left(\sum_{k_{2}=0}^{r_{2}+1} \theta_{k_{2}}-t_{m_1+1}\right)
					\nonumber \\
					&\times \Theta\left(t_{m_1+2}-\sum_{k_{2}=0}^{r_2} \theta_{k_{2}}\right)  
					\Theta\left(\sum_{k_{2}=0}^{r_{2}+1} \theta_{k_{2}}-t_{m_1+2}\right)\times...
					\nonumber \\
					&...\times 
					\Theta\left(t_{m_1+m_2}-\sum_{k_{2}=0}^{r_2} \theta_{k_{2}}\right)  
					\Theta\left(\sum_{k_{2}=0}^{r_{2}+1} \theta_{k_{2}}-t_{m_1+m_2}\right)
				\end{aligned}
			\end{rcases}
			\quad \text{block 2}
			\nonumber \\
			&\times ...\nonumber \\
			&\begin{rcases}
				\begin{aligned}
					&	\times \Theta\left(t_{m_1+...+m_{p-1}+1}-\sum_{k_{p}=0}^{r_p} \theta_{k_{p}}\right)  
					\Theta\left(\sum_{k_{p}=0}^{r_{p}+1} \theta_{k_{p}}-t_{m_1+...+m_{p-1}+1}\right)
					\nonumber \\
					&\times \Theta\left(t_{m_1+...+m_{p-1}+2}-\sum_{k_{p}=0}^{r_p} \theta_{k_{p}}\right)  
					\Theta\left(\sum_{k_{p}=0}^{r_{p}+1} \theta_{k_{p}}-t_{m_1+...+m_{p-1}+2}\right)\times...
					\nonumber \\
					&...\times 
					\Theta\left(t_{m_1+...+m_{p-1}+m_p}-\sum_{k_{p}=0}^{r_p} \theta_{k_{p}}\right)  
					\Theta\left(\sum_{k_{p}=0}^{r_{p}+1} \theta_{k_{p}}-t_{m_1+...+m_{p-1}+m_p}\right)
				\end{aligned}
			\end{rcases}
			\quad \text{block p}
			\nonumber \\
			&\left.	p_0(\xi_0)d\xi_0\prod_{q=1}^\infty\psi(\theta_q)d\theta_q\,p(\xi_q)d\xi_q
			\right],
		\end{align}	
	}
	where, again $m_1+m_2+...+m_p=n$. 
	In practice, we have replaced the sum over all the non decreasing indices $i_1\le i_2\le ...\le i_n$
	with the sum over all the
	compositions (ordered partitions) where in each composition the set
	$i_1,\,i_2\,...,i_{n}$  is grouped in $p$ blocks as in \eqref{sets}.
	
	Of course,
	for any fixed number of $p$ blocks, there are  
	$N(p)=\frac{\left(n-1\right)!}{(p-1) !\left[n-p\right] !}$) possible compositions, and summing 
	for all the possible number of blocks we obtain the total number  of compositions : 
	$\sum_{p=1}^{n} N(p)=2^{n-1}$ (a block separator between any  position can be turned on or 
	off, thus we have just $2^{n-1}$ possibilities). 
	
	Now, we notice that, given the assumption
	$t_1\le t_2\le...\le t_n$, in each block in \eqref{app:sums_temp} we can disregard all the
	Heaviside theta functions, but the first and the last ones. Thus we can simplify the same equation
	in the following way:
	\begin{align}
		\label{app:n_corr_steP_{-}}
		&  \langle\xi( t_1)\xi( t_2)...\xi( t_n)\rangle\nonumber \\
		&=\sum_{p=1}^{n} \;
		\left[
		\sum_{\{m_i\in\N\}:\sum_{i=1}^p m_i = n}  
		\int 
		\sum_{r_1=0}^\infty\sum_{r_2=r_1+1}^\infty\cdots \sum_{r_p=r_{p-1}+1}^\infty
		{\xi_{r_1}^{m_1}}\,{\xi_{r_2}^{m_2}}\,\cdots{\xi_{r_p}^{m_p}}
		\right.	 \nonumber \\&
		\times \Theta\left(t_1-\sum_{k_1=0}^{r_1} \theta_{k_1}\right)  
		\Theta\left(\sum_{k_1=0}^{r_{1}} \theta_{k_1}+\theta_{r_1+1}-t_{m_1}\right)
		\nonumber \\
		&\times 
		\Theta\left(t_{m_1+1}-\sum_{k_{2}=0}^{r_{2}} \theta_{k_{2}}\right) 
		\Theta\left(\sum_{k_{2}=0}^{r_{2}} \theta_{k_{2}}
		+\theta_{r_2+1}-t_{m_1+m_2}\right)
		\nonumber \\
		&\times 
		\Theta\left(t_{m_1+m_{2}+1}-\sum_{k_{3}=0}^{r_{3}} \theta_{k_{3}}\right) 
		\Theta\left(\sum_{k_{3}=0}^{r_{3}} \theta_{k_{3}}+\theta_{r_3+1}-t_{m_1+m_2+m_3}\right)
		\times...
		\nonumber \\
		&...\times \Theta\left(t_{m_1+...+m_{p-1}+1}-\sum_{k_{p}=0}^{r_{p}} \theta_{k_{p}}\right)
		\Theta\left(\sum_{k_{p}=0}^{r_{p}} \theta_{k_{p}}+\theta_{r_p+1}-t_n\right)
		\nonumber \\
		&\times 
		\left.	p_0(\xi_0)d\xi_0\prod_{q=1}^\infty\psi(\theta_q)d\theta_q\,p(\xi_q)d\xi_q
		\right].
	\end{align}

	As a first link between this algebraic manipulation and the approach given in Proposition~\ref{prop:main}, we observe that
	the Heaviside functions in \eqref{app:n_corr_steP_{-}} leads to the following contraints 
	concerning the times of the correlation function:
	\begin{itemize}
		\item the times $t_1, t_2,...,t_{m_{1}}$  lie in the same laminar region,  bounded by the $r_1=i_1$-th 
		and the $r_1+1$-th transition events;
		\item the times  $t_{m_{1}+1},t_{m_{1}+2},...,t_{m_{1}+m_{2}}$ lie in another laminar region, 
		bounded by the $r_2=i_{m_{1}+1}$-th and the $r_2+1$-th transition events, and so on;
		\item in other words, for any block of $m_{k}$ indices there is a corresponding block of 
		$m_{k}$ times that lies in the (same) $r_k$-th laminar region. 
	\end{itemize}   

	\subsection*{$c)$ Exploiting the i.i.d. assumption in computing integrals over the PDFs of $\xi$}
	Integrating  Eq.~\eqref{app:n_corr_steP_{-}} over all the $\xi$ variables (that corresponds to averaging over all the reandom $\xi$ variables) we get:%
	\begin{align}
		\label{app:n_corr_step_3}
		&  \langle\xi( t_1)\xi( t_2)...\xi( t_n)\rangle\nonumber \\
		&=\sum_{p=1}^{n} \;
		\left[
		\sum_{\{m_i\in\N\}:\sum_{i=1}^p m_i = n}  
		\sum_{r_1=0}^\infty\sum_{r_2=r_1+1}^\infty\cdots \sum_{r_p=r_{p-1}+1}^\infty
		\overline{\xi^{m_1}}'\,\overline{\xi^{m_2}}\,\cdots\overline{\xi^{m_p}}
		\right. \nonumber \\[5pt]&
		\times \int\Theta\left(t_1-\sum_{k_1=0}^{r_1} \theta_{k_1}\right)  
		\Theta\left(\sum_{k_1=0}^{r_{1}} \theta_{k_1}+\theta_{r_1+1}-t_{m_1}\right)
		\nonumber \\
		&\times 
		\Theta\left(t_{m_1+1}-\sum_{k_{2}=0}^{r_{2}} \theta_{k_{2}}\right) 
		\Theta\left(\sum_{k_{2}=0}^{r_{2}} \theta_{k_{2}}
		+\theta_{r_2+1}-t_{m_1+m_2}\right)
		\nonumber \\
		&\times 
		\Theta\left(t_{m_1+m_{2}+1}-\sum_{k_{3}=0}^{r_{3}} \theta_{k_{3}}\right) 
		\Theta\left(\sum_{k_{3}=0}^{r_{3}} \theta_{k_{3}}+\theta_{r_3+1}-t_{m_1+m_2+m_3}\right)
		\times...
		\nonumber \\
		&\left. ...\times \Theta\left(t_{m_1+...+m_{p-1}+1}-\sum_{k_{p}=0}^{r_{p}} \theta_{k_{p}}\right)
		\Theta\left(\sum_{k_{p}=0}^{r_{p}} \theta_{k_{p}}+\theta_{r_p+1}-t_n\right)
		\prod_{q=1}^\infty\psi(\theta_q)d\theta_q
		\right].
	\end{align}
	where in $\overline{\xi^{m_1}}'$,  the prime symbol indicates that 
	$\overline{\xi^{m_1}}'=\overline{\xi^{m_1}_0}$ for $r_1=0$ and
	$\overline{\xi^{m_1}}'=\overline{\xi^{m_1}}$ for $r_1>0$.
	Given the definition of $\psi(\theta)$ as the WT PDF, it is clear that the summand on the right-hand side of Eq.~\eqref{app:n_corr_step_3} is 
	$\overline{\xi^{m_1}}'\,\overline{\xi^{m_2}} \cdots \overline{\xi^{m_p}}$
	multiplied by the probability that the times $t_1, t_2, \ldots, t_n$ are grouped as specified in the points listed at the end of the previous step. 
	Together with the fact that the multiple sums in the same equation are formally equivalent to the sum over all compositions of $n$ objects, 
	this provides a rigorous foundation for the procedure described in Proposition~\ref{prop:main}, 
	which was originally introduced through an intuitive approach based on statistical arguments.  
	
	In order to establish a formal equivalence between Eq.~\eqref{app:n_corr_step_3} and the result \eqref{corrGen_} of Proposition~\ref{prop:main}, 
	both expressions must be rewritten in terms of the rate function $R$. 
	For the latter, this is straightforward: it is sufficient to use the definition of the normalized closed correlation function given in 
	Eq.~\eqref{closed_corr}.  
	
	By contrast, rewriting Eq.~\eqref{app:n_corr_step_3} in terms of $R$ is more involved, and is carried out in the remaining steps of this Appendix.

	\subsection*{$d)$ From WT PDFs to probabilities of absolute times (or sums of waiting times)}
	Since the rate function $R$ is derived from $\psi_n$, the $n$-fold convolution of $\psi(\theta)$, 
	the purpose of this section is to clarify the role of the function $\psi_n$ in Eq.~\eqref{app:n_corr_step_3}.
	
	To this end, we perform a change of variables that transforms the description in terms of the waiting times $\theta_k$
	(i.e., the time intervals between successive events) 
	into one expressed in terms of the absolute times of the events.

	Let us begin with the sum of the waiting times preceding the first event at $t_1$, namely  
	$w = \theta_1 + \theta_2 + \cdots + \theta_{r_{1}}$. 
	Accordingly, we introduce the change of variable
	$
	\theta_1 \;\to\; w = \theta_1 + \theta_2 + \cdots + \theta_{r_{1}}, 
	\quad \text{i.e.,} \quad 
	\theta_{1} = w - \sum_{k_1=2}^{r_1} \theta_{k_1}.
	$
	
	With this substitution, the multiple integral in Eq.~\eqref{app:n_corr_step_3}, involving the variables 
	$\theta_2,\theta_3, \ldots, \theta_{r_{1}}$, becomes
	\begin{align}
		\label{psi_i_2}
		&     
		\int_0^{w} d \theta_2
		\int_0^{w-\theta_2} d \theta_3
		\int_0^{w-\theta_2-\theta_3} d \theta_4 \,\cdots\,
		\int_0^{w-\theta_2 - \cdots -\theta_{{r_1}-1}} d \theta_{r_1} \nonumber \\
		&\times
		\psi(w-\theta_2-\theta_3-\theta_4- \dots -\theta_{{i_1}})
		\psi(\theta_2)\psi(\theta_3)\psi(\theta_4) \dots \psi(\theta_{{r_1}}):=\psi_{r_1}(w)
	\end{align}
	It is straightforward to verify that Eq.~\eqref{psi_i_2} provides an alternative 
	representation of the ${r_1}$-fold convolution of the WT PDF, evaluated at the time $w$. 
	In other words, the function $\psi_{r_1}(w)$ on the right-hand side of Eq.~\eqref{psi_i_2} 
	is precisely the same as that introduced previously in Eq.~\eqref{Rtilde}.
	
	Notice that after this change of variable, the first Heaviside-function in 
	Eq.~\eqref{app:n_corr_step_3} simply restricts the upper limit of integration 
	for $w$ to $t_1$.  
	
	After performing the aforementioned change of variables and 
	\emph{after integrating over 
		the first $r_1 - 1$ waiting time variables $\theta$, }we can set 
	$\theta := \theta_{r_1+1}$ and renumber the subsequent 
	$\theta_k$ variables,  shifting head the counting: 
	$\theta_{r_1+2}\to \theta_1$, $\theta_{r_1+3}\to \theta_2,...$. 
	In other words, we reset the indexing of 
	waiting times beginning with the $(r_1+1)$-th event.
	Thus, in the end,
	Eq.~\eqref{app:n_corr_step_3} becomes 
	\begin{align}
		\label{app:n_corr_step_4}
		&  \langle\xi( t_1)\xi( t_2)...\xi( t_n)\rangle
		\nonumber \\
		&
		= \sum_{p=1}^{n} \;
		\left[\sum_{\{m_i\in\N\}:\sum_{i=1}^p m_i = n}
		\sum_{r_1=0}^\infty\sum_{r_2=r_1+1}^\infty\cdots \sum_{r_p=r_{p-1}+1}^\infty
		\overline{\xi^{m_1}}'\,\overline{\xi^{m_2}}\,\cdots\overline{\xi^{m_p}} \right.
		\nonumber \\
		& \times
		\int_0^{t_1} dw\psi_{r_1}(w)
		\int d\theta  	\Theta\left(w+\theta  -t_{m_1}\right) \psi(\theta  )
		\nonumber \\
		&\times\int\Theta \left(t_{m_1+1}-w-\theta  -\sum_{k_2=0}^{r_2-r_1-1} \theta_{k_2}\right)
		\Theta \left(w+\theta  +\sum_{k_2=0}^{r_2-r_1-1} \theta_{k_2}+\theta_{r_2-r_1}-t_{m_1+m_2}\right)
		\nonumber \\
		&\times 
		\Theta\left(t_{m_1+m_{2}+1}-w-\theta-\sum_{k_2=0}^{r_2-r_1-1} \theta_{k_2}
		-\sum_{k_{3}=r_2-r_1}^{r_{3}-r_1-1}  \theta_{k_{3}}\right) \nonumber \\&
		\times
		\Theta\left(w+\theta-\sum_{k_2=0}^{r_2-r_1-1} \theta_{k_2}
		+\sum_{k_{3}=r_2-r_1}^{r_{3}-r_1-1}  \theta_{k_{3}}+\theta_{r_{3}-r_1}-t_{m_1+m_2+m_3}\right)
		\times...\nonumber \\
		&...\times \; 
		\Theta\left(t_{m_1+...+m_{p-1}+1}-w-\theta-\sum_{k_2=0}^{r_2-r_1-1} \theta_{k_2}
		-\sum_{k_p=r_2-r_1}^{r_p-r_1-1} \theta_{k_p}\right)\nonumber \\
		&\left. 
		\times\Theta\left(w+\theta+\sum_{k_2=0}^{r_2-r_1-1} \theta_{k_2}
		+\sum_{k_p=r_2-r_1}^{r_p-r_1-1} \theta_{k_p}+\theta_{r_p-r_1}-t_n\right)
		\left(\prod_{l=1}^\infty\psi(\theta_l)d\theta_l\,\right)
		\right].
	\end{align}
	Now, for any trajectory realization, we  introduce the first time on the left ($u_{r_{1}}$) and the first time on the right 
	($u_{r_{1}+1}$) of $t_1$. I.e. $u_{r_{1}}=w$, while  
	$u_{r_{1}+1}=u_{r_{1}}+\theta$. With these definitions, the previous equation becomes:
	%
	%
	\begin{align}
		\label{app:n_corr_step_5}
		&  \langle\xi( t_1)\xi( t_2)...\xi( t_n)\rangle\nonumber \\
		&=\sum_{p=1}^{n} \;
		\left[\sum_{\{m_i\in\N\}:\sum_{i=1}^p m_i = n}
		\sum_{r_1=0}^\infty\sum_{r_2=r_1+1}^\infty\cdots \sum_{r_p=r_{p-1}+1}^\infty 
		\overline{\xi^{m_1}}'\,\overline{\xi^{m_2}}\,\cdots\overline{\xi^{m_p}} \right.
		\nonumber \\ 
		& \times
		\int_0^{t_1} du_{r_{1}}\psi_{r_1}(u_{r_{1}})
		\int du_{r_1+1}\Theta\left(u_{r_{1}+1}-t_{m_1}\right)\psi(u_{r_{1}+1}-u_{r_{1}})
		\nonumber \\
		&\times\int \Theta \left(t_{m_1+1}-u_{r_{1}+1}-\sum_{k_2=0}^{r_2-r_1-1} \theta_{k_2}\right)
		\Theta \left(u_{r_{1}+1}+\sum_{k_2=0}^{r_2-r_1-1} \theta_{k_2}+\theta_{r_2-r_1}-t_{m_1+m_2}\right)
		\nonumber \\
		&\times 
		\Theta\left(t_{m_1+m_{2}+1}-u_{r_{1}+1}-\sum_{k_2=0}^{r_2-r_1-1} \theta_{k_2}
		-\sum_{k_{3}=r_2-r_1}^{r_{3}-r_1-1}  \theta_{k_{3}}\right) \nonumber \\
		&\times
		\Theta\left(u_{r_{1}+1}-\sum_{k_2=0}^{r_2-r_1-1} \theta_{k_2}
		+\sum_{k_{3}=r_2-r_1}^{r_{3}-r_1-1}  \theta_{k_{3}}+\theta_{r_{3}-r_1}-t_{m_1+m_2+m_3}\right)
		\times...\nonumber \\
		&...\times \; 
		\Theta\left(t_{m_1+...+m_{p-1}+1}-u_{r_{1}+1}-\sum_{k_2=0}^{r_2-r_1-1} \theta_{k_2}
		-\sum_{k_p=r_2-r_1}^{r_p-r_1-1} \theta_{k_p}\right)\nonumber \\
		&\left.
		\times\Theta\left(u_{r_{1}+1}+\sum_{k_2=0}^{r_2-r_1-1} \theta_{k_2}
		+\sum_{k_p=r_2-r_1}^{r_p-r_1-1} \theta_{k_p}+\theta_{r_p-r_1}-t_n\right)
		\left(\prod_{l=1}^\infty\psi(\theta_l)d\theta_l\,\right)
		\right] .
	\end{align}
	The previous expression no longer involves the variable $w$.
	
	We now handle the waiting times inside the second pair of Heaviside functions in 
	the same way as done above for the first pair. We redefine $w$ as the time interval 
	between the first event after $t_{m_1}$ and the first event before $t_{m_1+1}$:
	$w = \sum_{k_2=0}^{r_2 - r_1 - 1} \theta_{k_2}$ and
	we make the change of variable 
	$\theta_{1} \to w = \sum_{k_2=0}^{r_2 - r_1 - 1} \theta_{k_2}$, 
	i.e., $\theta_1 = w - \sum_{k_2=1}^{r_2 - r_1 - 1} \theta_{k_2}$. Then 
	we integrate again  over all $\theta_{k_2}$ for $k_2 \in [2, r_2 - r_1 - 1]$.
	
	Once again, this change of variables and subsequent integration yield the 
	${r_2 - r_1 - 1}$-fold convolution of the WT PDF, evaluated at time $w$.
	
	To obtain again an expression that does not contain the variable $w$, we introduce, as before, the absolute times 
	$u_{r_2} = u_{r_1 + 1} + w$ and $u_{r_2 + 1} = u_{r_2} + \theta_{r_2 - r_1}$, 
	which correspond to the time of the first event before $t_{m_1+1}$ and the 
	time of the first event after $t_{m_1 + m_2}$, respectively.
	
	After these operations, the previous equation becomes:  
	\begin{align}
		\label{app:n_corr_step_5_bis}
		&  \langle\xi( t_1)\xi( t_2)...\xi( t_n)\rangle\nonumber \\
		&=\sum_{p=1}^{n} \;
		\left[\sum_{\{m_i\in\N\}:\sum_{i=1}^p m_i = n}
		\sum_{r_1=0}^\infty\sum_{r_2=r_1+1}^\infty\cdots \sum_{r_p=r_{p-1}+1}^\infty
		\overline{\xi^{m_1}}'\,\overline{\xi^{m_2}}\,\cdots\overline{\xi^{m_p}} 
		\right.
		\nonumber \\ & \times
		\int_0^{t_1} du_{r_{1}}\psi_{r_1}(u_{r_{1}})
		\int_{t_{m_1}}^{t_{m_1+1}} du_{r_1+1}\psi(u_{r_{1}+1}-u_{r_{1}})
		\nonumber \\ &
		\int_{u_{r_{1}+1}}^{t_{m_1+1}} du_{r_{2}}\, \psi_{r_2-r_1-1}(u_{r_{2}}-u_{r_{1}+1})
		\int_{t_{m_1+m_2}}^{t_{m_1+m_2+1}} du_{r_2+1}\psi(u_{r_{2}+1}-u_{r_{2}})
		\nonumber \\
		&\times \int\Theta \left(t_{m_1+m_2+1}-u_{r_{2}+1}-\sum_{k_3=0}^{r_3-r_2-1} \theta_{k_3}\right)
		\Theta \left(u_{r_{2}+1}+\sum_{k_3=0}^{r_3-r_2-1} \theta_{k_3}+\theta_{r_3-r_2}-t_{m_1+m_2+m_3}\right)
		\times...\nonumber \\
		&...\times \; 
		\Theta\left(t_{m_1+...+m_{p-1}+1}-u_{r_{2}+1}
		-\sum_{k_3=0}^{r_3-r_2-1} \theta_{k_3}-\sum_{k_p=r_3-r_2}^{r_p-r_2-1} \theta_{k_p}\right)\nonumber \\
		&\left.
		\times\Theta\left(u_{r_{2}}
		+\sum_{k_3=0}^{r_3-r_2-1} \theta_{k_p}+\sum_{k_p=r_2-r_2}^{r_p-r_2-1} \theta_{k_p}+\theta_{r_p-r_2}-t_n\right)
		\left(\prod_{l=1}^\infty\psi(\theta_l)d\theta_l\,\right)
		\right]   .
	\end{align}
	Repeating this procedure of changes of variables up to the last set of times in the same laminar
	region (i.e., up to the last couple of Heaviside functions), we end up with the following result:        
	\begin{align}
		\label{app:n_corr_step_6}
		&  \langle\xi( t_1)\xi( t_2)...\xi( t_n)\rangle\nonumber \\
		&=\sum_{p=1}^{n} \;
		\left[\sum_{\{m_i\in\N\}:\sum_{i=1}^p m_i = n}
		\sum_{r_1=0}^\infty\sum_{r_2=r_1+1}^\infty\cdots \sum_{r_p=r_{p-1}+1}^\infty 
		\overline{\xi^{m_1}}'\,\overline{\xi^{m_2}}\,\cdots\overline{\xi^{m_p}} 
		\right.
		\nonumber \\ 
		& \times
		\int_0^{t_1} du_{r_{1}}\psi_{r_1}(u_{r_{1}})
		\int_{t_{m_1}}^{t_{m_1+1}} du_{r_1+1}\psi(u_{r_{1}+1}-u_{r_{1}})
		\nonumber \\ 
		&\times
		\int_{u_{r_{1}+1}}^{t_{m_1+1}} du_{r_{2}}\, \psi_{r_2-r_1-1}(u_{r_{2}}-u_{r_{1}+1})
		\int_{t_{m_1+m_2}}^{t_{m_1+m_2+1}} du_{r_2+1}\psi(u_{r_{2}+1}-u_{r_{2}})
		\nonumber \\ 
		&\times
		\int_{u_{r_{2}+1}}^{t_{m_1+m_2+1}} du_{r_{3}}\, \psi_{r_3-r_2-1}(u_{r_{3}}-u_{r_{2}+1})
		\int_{t_{m_1+m_2+m_3}}^{t_{m_1+m_2+m_3+1}} du_{r_3+1}\psi(u_{r_{3}+1}-u_{r_{3}})
		\nonumber \\ 
		&\times...\nonumber \\
		&\times ...
		\left.
		\int_{u_{r_{1}+1}}^{t_{m_1+m_2+...m_{p-1}+1}} du_{r_{p}}\, \psi_{r_p-r_{p-1}-1}(u_{r_{p}}-u_{r_{p-1}+1})
		\int_{t_{n}}^{\infty} du_{r_p+1}\psi(u_{r_{p}+1}-u_{r_{p}})
		\right] .
	\end{align} 
	
	\subsection*{$e)$ Introducing indices for the relative distance from the ``diagonal''}
	
	To further simplify the notation,  we rewrite  \eqref{app:n_corr_step_6} by exploiting  the following change of indices 
	in the multiple sum:       
	$j_1:=r_1$ and $j_k=r_k-r_{k-1}-1$. 
	Since the variables of integration are dummy (only the limits of integration are relevant), they are not affected by this change of variables:
	\begin{align}
		\label{app:n_corr_step_7}
		&  \langle\xi( t_1)\xi( t_2)...\xi( t_n)\rangle\nonumber \\
		&=\sum_{p=1}^{n} \;
		\left[\sum_{\{m_i\in\N\}:\sum_{i=1}^p m_i = n}
		\sum_{j_1=0}^\infty\sum_{j_2=0}^\infty...\sum_{j_p=0}^\infty 
		\overline{\xi^{m_1}}'\,\overline{\xi^{m_2}}\,\cdots\overline{\xi^{m_p}} 
		\right.
		\nonumber \\ 
		& \times
		\int_0^{t_1} du_{r_{1}}\psi_{j_1}(u_{r_{1}})
		\int_{t_{m_1}}^{t_{m_1+1}} du_{r_1+1}\psi(u_{r_{1}+1}-u_{r_{1}})
		\nonumber \\ 
		&\times
		\int_{u_{r_{1}+1}}^{t_{m_1+1}} du_{r_{2}}\, \psi_{j_2}(u_{r_{2}}-u_{r_{1}+1})
		\int_{t_{m_1+m_2}}^{t_{m_1+m_2+1}} du_{r_2+1}\psi(u_{r_{2}+1}-u_{r_{2}})
		\nonumber \\ 
		&\times
		\int_{u_{r_{2}+1}}^{t_{m_1+m_2+1}} du_{r_{3}}\, \psi_{j_3}(u_{r_{3}}-u_{r_{2}+1})
		\int_{t_{m_1+m_2+m_3}}^{t_{m_1+m_2+m_3+1}} du_{r_3+1}\psi(u_{r_{3}+1}-u_{r_{3}})
		\nonumber \\ 
		&\times...\nonumber \\
		&
		\left.
		\times ...
		\int_{u_{r_{1}+1}}^{t_{m_1+m_2+...m_{p-1}+1}} du_{r_{p}}\, \psi_{j_p}(u_{r_{p}}-u_{r_{p-1}+1})
		\int_{t_{n}}^{\infty} du_{r_p+1}\psi(u_{r_{p}+1}-u_{r_{p}})
		\right].
	\end{align} 
	where now 
	$\overline{\xi^{m_1}}'=\overline{\xi^{m_1}_0}$ for $j_1=0$ and
	$\overline{\xi^{m_1}}'=\overline{\xi^{m_1}}$ for $j_1>0$.

	\subsection*{$f)$Final result via summation over the indices $j_k$}
	After summing over all the $j_k$ indices and taking into account:
	\begin{itemize}
		\item the definition of the rate function in Eq.~\eqref{Rtilde},
		\item the fact that $\psi_{0}(t) = \delta(t)$,
	\end{itemize}
	we get
	\begin{align}
		\label{app:n_corr_step_8}
		&  \langle\xi( t_1)\xi( t_2)...\xi( t_n)\rangle\nonumber \\
		&=\sum_{p=1}^{n} \;
		\left\{\sum_{\{m_i\in\N\}:\sum_{i=1}^p m_i = n}
		\overline{\xi^{m_2}}\,\cdots\overline{\xi^{m_p}} 
		\right.
		\nonumber \\ 
		& \times
		\left[\overline{\xi_0^{m_1}}+
		\overline{\xi^{m_1}}
		\int_0^{t_1} du_{r_{1}} R(u_{r_{1}})\right]
		\int_{t_{m_1}}^{t_{m_1+1}} du_{r_1+1}\psi(u_{r_{1}+1}-u_{r_{1}})
		\nonumber \\ 
		&\times
		\int_{u_{r_{1}+1}}^{t_{m_1+1}} du_{r_{2}}\, \tilde R(u_{r_{2}}-u_{r_{1}+1})
		\int_{t_{m_1+m_2}}^{t_{m_1+m_2+1}} du_{r_2+1}\psi(u_{r_{2}+1}-u_{r_{2}})
		\nonumber \\ 
		&\times
		\int_{u_{r_{2}+1}}^{t_{m_1+m_2+1}} du_{r_{3}}\, \tilde R(u_{r_{3}}-u_{r_{2}+1})
		\int_{t_{m_1+m_2+m_3}}^{t_{m_1+m_2+m_3+1}} du_{r_3+1}\psi(u_{r_{3}+1}-u_{r_{3}})
		\nonumber \\ 
		&\times...\nonumber \\
		&
		\left.
		\times ...
		\int_{u_{r_{1}+1}}^{t_{m_1+m_2+...m_{p-1}+1}} du_{r_{p}}\, \tilde R(u_{r_{p}}-u_{r_{p-1}+1})
		\int_{t_{n}}^{\infty} du_{r_p+1}\psi(u_{r_{p}+1}-u_{r_{p}}).
		\right\} .
	\end{align} 
	In the simplified case in which the  PDF of $\xi_0$ (the preparation of the ensemble) 
	is the same as the PDF of the random variable $\xi$, i.e., when $p_0(\xi_0)=p(\xi_0)$, then, 
	by exploiting the definition of the normalized closed correlation function given in 
	Eq.~\eqref{closed_corr}, we can verify that Eq.~\eqref{app:n_corr_step_8} is equivalent to 
	Eq.~\eqref{corrGen_}.
	\section{The ME for a general system driven by the telegraph noise\label{app:ME_dicho}}
	%
	%
	Here we illustrate how to derive the exact ME 
	\eqref{MEG_cnumber_PDF_appro} for the reduced probability density function (PDF)  
	of $x$ governed by the stochastic differential equation (SDE)~\eqref{SDE}, when $\xi$ is the telegraph noise.  
	There are different equivalent way to arrive to that. 
	One is starting from the following stochastic Liouville equation, equivalent to the SDE~\eqref{SDE}  (note:  
	$\partial_a := \partial/\partial a$):
	\begin{align}
		\label{stochLiouv_}
		\partial_t \rho(x,\xi(t);t) = {\mathcal{L}}_a \rho(x,\xi(t);t)
		- \partial_x I(x)\xi(t) \rho(x,\xi(t);t),
	\end{align}
	where  
	\begin{equation}
		\label{La}
		{\mathcal{L}}_a := \partial_x C(x)
	\end{equation}  
	and $\rho(x,\xi(t);t)$ is, for any trajectory realization of $\xi[t]$, the 
	time evolution of an initial ensemble $\rho(x;0)$.
	Then, we observe that an old result of~\cite{ydfJCP62} demonstrates 
	the equivalence of the Zwanzig projection approach and the
	generalized (TTO-, i.e. $G$-)cumulant method. Thus, recalling  that
	the telegraph noise is a $G$-Gaussian stochastic process~\cite{bCSF159}, 
	we have that in this case the
	Zwanzig series must stop to the second term, that exactly correspond to 
	\eqref{MEG_cnumber_PDF_appro}, provided that $P(x;t)$ is the marginal PDF of $x$,
	obtained averaging $\rho(x,\xi(t);t)$ over all the trajectories  $\xi(t)$.

	Another more direct way is taking advantage of the fact that in the simple case of 
	telegraph noise with values $\xi=\pm 1$, the
	SDE~\eqref{SDE} is equivalent to the following   
	Liouville equation for the joint PDF of $x$ and $\xi$
	\begin{align}
		\label{transitionmatrix}
		\partial_t P_{1}(x;t) &= {\mathcal{L}}_a P_{1}(x;t) 
		- \partial_x I(x) P_{1}(x;t) 
		- \lambda P_{1}(x;t) + \lambda P_{-1}(x;t), \nonumber \\
		\partial_t P_{-1}(x;t) &= {\mathcal{L}}_a P_{-1}(x;t) 
		+ \partial_x I(x) P_{-1}(x;t) 
		+ \lambda P_{1}(x;t) - \lambda P_{-1}(x;t).
	\end{align}  
	where we have indicated with $\lambda$ the transition probability per unit time.
	
	Taking the sum and the difference of these two equations, we obtain:
	\begin{align}
		\partial_t P_{+}(x;t) &= {\mathcal{L}}_a P_{+}(x;t) 
		- \partial_x \big[ I(x) P_{-}(x;t) \big], \\
		\partial_t P_{-}(x;t) &= {\mathcal{L}}_a P_{-}(x;t) 
		- \partial_x \big[ I(x) P_{+}(x;t) \big] 
		- 2\lambda P_{-}(x;t),
	\end{align}  
	where $P_{+}(x;t) := P_{1}(x;t) + P_{-1}(x;t) \equiv P(x,t)$ is the reduced PDF of $x$,  
	and $P_{-}(x;t) := P_{1}(x;t) - P_{-1}(x;t)$ is its complementary PDF.  
	
	By inserting into the first equation the expression for $P_{-}(x;t)$ obtained from  
	integrating the second one, setting $2\lambda = 1/\tau$, and using the definition  
	in Eq.~\eqref{La}, we recover  the master equation  
	\eqref{MEG_cnumber_PDF_appro}.
	
	\newcommand{\noop}[1]{}

\end{document}